\documentclass[11pt]{article}

\usepackage[usenames,dvipsnames,table]{xcolor}

\usepackage{changepage}

\usepackage{float}

\usepackage{jheppub} 

\usepackage{graphicx}
\usepackage{amsmath, amssymb}

\usepackage{manfnt}

\newcommand{\be}{\begin{eqnarray}}
\newcommand{\ee}{\end{eqnarray}}
\newcommand{\bea}{\begin{eqnarray}}
\newcommand{\eea}{\end{eqnarray}}

\newcommand{\bn}{\begin{enumerate}}
\newcommand{\en}{\end{enumerate}}

\def\SU{\mathrm{SU}}

\def\Z{{\mathbb Z}}
\def\C{{\mathbb C}}

\def\half{\frac{1}{2}}

\title{Weakly coupled conformal manifolds in $4d$}

\author[a]{Shlomo S. Razamat,}
\author[a]{Evyatar Sabag,}
\author[b,c]{and Gabi Zafrir}

\affiliation[a]{Department of Physics, Technion, Haifa, 32000, Israel}
\affiliation[b]{Kavli Institute for the Physics and Mathematics of the Universe (WPI), University of Tokyo\\
Kashiwa, Chiba 277-8583, Japan}
\affiliation[c]{Dipartimento di Fisica, Universit\`a di Milano-Bicocca \& INFN, \\ 
Sezione di Milano-Bicocca, I-20126 Milano, Italy}

\emailAdd{razamat@physics.technion.ac.il}
\emailAdd{sevyatar@campus.technion.ac.il}
\emailAdd{gabi.zafrir@unimib.it}

\abstract{We classify ${\cal N}=1$ gauge theories with simple gauge groups in four dimensions which possess a conformal manifold passing through weak coupling.
A very rich variety of models is found once one allows for arbitrary representations under the gauge group. For each such model we detail the dimension of the conformal manifold, the conformal anomalies, and the global symmetry preserved on a generic locus of the manifold. We also identify, at least some, sub-loci of the conformal manifolds preserving more symmetry than the generic locus.  Several examples of applications of the classification are discussed. In particular we consider a conformal triality such that one of the triality frames is a $USp(6)$ gauge theory with six fields in the two index traceless antisymmetric representation. We discuss an IR dual of a conformal $Spin(5)$ gauge theory with two chiral superfields in the vector representation and one in the fourteen dimensional representation. Finally, an extension of the conformal manifold of ${\cal N}=2$ class ${\cal S}$ theories by conformally gauging symmetries corresponding to maximal punctures with the addition of two adjoint chiral superfields is commented upon.
}

\begin{document} 

\maketitle
\flushbottom

\section{Introduction}

The study of supersymmetric quantum field theories in four dimensions led to numerous insights on the dynamics of strongly coupled systems. Many of these insights follow from the better control one has with supersymmetry over the renormalization group flow, thus leading to by now well established conjectures of IR equivalences of different flows as well as IR enhancements of global symmetries. Another important class of understandings is related to the existence of interacting conformal supersymmetric Lagrangians in $4d$. Such Lagrangians have tunable couplings which in particular can be taken to be small. The history of this subject is rather rich, see {\it e.g.} \cite{Sohnius:1981sn,Howe:1983wj,Parkes:1984dh}. A systematic  understanding of such conformal Lagrangians was developed first in \cite{Leigh:1995ep}. One of the highly non-trivial properties such models often have is S-duality, that is an exact equivalence of seemingly different conformal Lagrangians. The canonical example is that of ${\cal N}=4$ SYM based on gauge group $G$ with complexified coupling $\tau$ being equivalent to ${\cal N}=4$ SYM based on the Langlands dual group ${}^LG$ and coupling $-1/\tau$ \cite{Montonen:1977sn,Witten:1978mh,Osborn:1979tq,Sen:1994yi}\footnote{See for example \cite{Aharony:2013hda} for a precise statement of this duality taking into account also the spectrum of line operators. Despite these comments, for the most part when we refer to groups in this article we refer to the algebra and are cavalier about the group structure.}. These types of dualities typically relate strongly coupled regimes of one model to weakly coupled regimes of another. By now, a huge variety of such dualities with extended, ${\cal N}=4$ or ${\cal N}=2$ \cite{Seiberg:1994rs}, supersymmetry is known. Often these dualities can be understood geometrically  by compactifying the $6d$ $(2,0)$ theory on a Riemann surface \cite{Witten:1995zh,Gaiotto:2009we} which can lead to deep insights into various properties of the models, see {\it e.g.} \cite{Alday:2009aq,Gadde:2009kb}. Moreover, S-dualities are also intimately connected to deep purely mathematical ideas \cite{Kapustin:2006pk}.

However, with minimal supersymmetry in $4d$, ${\cal N}=1$, the understanding of conformal dualities is rather fragmented.  One class of examples discussed recently is that of quiver theories based on $SU(N)^k$ gauge group such that each $SU(N)$ factor sees exactly $3N$ flavors. The one loop beta function for the gauge fields vanishes and this theory admits a conformal manifold which passes through zero coupling. To deduce this the exact structure of the quiver theory is essential \cite{Gaiotto:2015usa} (see \cite{Hanany:2015pfa,Franco:2015jna,Razamat:2016dpl,Bah:2017gph} for some further developments). Moreover it was argued in \cite{Gaiotto:2015usa} that the conformal manifold has a one dimensional complex direction  such that the theory with weak coupling is dual to the same theory, modulo possible global issues, albeit with strong coupling. The full duality group should be the mapping class group of a sphere with two pairs of different punctures. Another example which was recently considered is that of $SU(3)$ SQCD with nine flavors \cite{Razamat:2018gro}. This model has a seven dimensional conformal manifold \cite{Leigh:1995ep,Green:2010da} passing through zero coupling. It was argued in \cite{Razamat:2018gro} that this manifold should admit an action of a duality group which is the mapping class group of a sphere with ten punctures. The model in four dimensions can be understood as a compactification on a ten punctured sphere  of a $(1,0)$ $6d$ SCFT described by a pure $SU(3)$ SCFT on its tensor branch \cite{Seiberg:1996qx,Bershadsky:1997sb}\footnote{Yet another intriguing example is a conformal duality between an $SU(2)^{N(N+1)}$ quiver gauge theory and a $USp(2N)^2\times SU(N+1)^2$ quiver theory discussed in \cite{Kim:2018bpg,Kim:2018lfo}. There the duality follows from geometric considerations and is related to using $5d$ dualities to construct in two different ways a compactification of $D$ type conformal matter on a torus with flux.}.  In both of these examples, as well as in the cases with extended supersymmetry, the appearance of the conformal duality  can be understood geometrically by constructing the models at hand as compactifications of six dimensional theories on Riemann surfaces. The duality group is then the mapping class group of the Riemann surface. It is natural to wonder whether any conformal Lagrangian with a non trivial conformal manifold would admit an action of a duality group. That is the conformal manifold will have loci describable by weakly coupled conformal gauge theories\footnote{A  more general scenario would be that the conformal manifold might have loci, cusps, which are described by a conformal weak gauging of a global symmetry of some SCFT, which might by itself be strongly-coupled \cite{Argyres:2007cn,Razamat:2020gcc}.}.

A first step when trying to answer such a question is to compile a list of four dimensional supersymmetric theories which admit a description in terms of a conformal Lagrangian. Partially achieving this goal is the purpose of the current paper. As ${\cal N}=1$ models have a rather rich structure allowing to choose a gauge group, matter content, and superpotential; classifying the most general model with a weakly coupled Lagrangian conformal description is a rather intricate problem. Here we will focus on Lagrangians based on a simple gauge group $G$. However, we will keep the matter content and the interactions generic, guided only by the demand that the theory should have a conformal manifold passing through zero couplings. Given the list of conformal gauge theories with simple gauge groups one can start building theories with non-simple groups by coupling the currents of a (sub)group of the global symmetry to dynamical vector fields. In that respect the models we classify here are a set of building blocks for more general conformal Lagrangian theories with weakly coupled conformal manifolds\footnote{In some cases conformally gauging a subgroup of the global symmetry of a free gauge theory with a simple gauge group that {\it does not} have a conformal manifold can lead to a non trivial conformal manifold. Such a theory must have a vanishing beta function for this to happen. Thus, the full set of building blocks of conformal gauge theories include also free gauge theories with simple gauge groups.}.

The classification detailed here consists of two main simple steps. As we are interested in conformal Lagrangians we take all the matter chiral superfields to have the free conformal R-symmetry assignment of $\frac23$. First we need to find all the possible choices of matter content such that all the one loop beta functions of gauge couplings vanish, which is a rather trivial exercise to perform.  The second step is to figure out whether the theory with this matter content, and with assignment of free R charges to the matter fields, has a conformal manifold, that is, it is not IR free. This step amounts to computing certain K\"ahler quotients \cite{Green:2010da} (see also \cite{Kol:2002zt,Kol:2010ub}), which is, though mostly straightforward, a laborious task to perform. We will find a rich variety of conformal Lagrangian theories from which the well known cases of $SU(3)$ SQCD with nine flavors, ${\cal N}=4$ SYM, and ${\cal N}=2$ $SU(N)$ SQCD with $2N$ flavors are just few examples. For each such model we will detail the conformal anomalies, the dimension of the conformal manifold, the symmetry group preserved on a generic locus of the manifold, as well as (at least some) directions of the conformal manifold preserving larger sub-groups of the global symmetry of the free point\footnote{Let us here mention some other classification programs of various properties of supersymmetric QFTs. First, the conformal ${\cal N}=2$ Lagrangians with arbitrary gauge groups were classified in \cite{Bhardwaj:2013qia}. The $6d$ tensor branch Lagrangian descriptions were analyzed in \cite{Bhardwaj:2015xxa}, and there is a rapidly growing literature on classifying $6d$ and $5d$ SCFTs (see \cite{Heckman:2018jxk} for a recent review of some aspects of the $6d$ classification program, and \cite{Jefferson:2017ahm,Jefferson:2018irk,Hayashi:2018lyv,Apruzzi:2018nre,Closset:2018bjz,Apruzzi:2019vpe,Apruzzi:2019opn,Bhardwaj:2019jtr,Apruzzi:2019enx,Bhardwaj:2019xeg,Apruzzi:2019syw,Bhardwaj:2020wgz} plus references within for some recent approaches to the $5d$ classification program). Another strand of recent works deals with the analysis of possible protected operators, relevant and marginal in particular, of supersymmetric QFTs in various dimensions \cite{Cordova:2016xhm,Cordova:2016emh}. Yet another strand of recent works deals with the various properties of ${\cal N}=2$ theories: classification of $4d$ ${\cal N}=2$ SCFTs using their Coulomb branch geometries, culminating in a proposed classification of all $4d$ rank 1 ${\cal N}=2$ SCFTs (see \cite{Argyres:2020pmm} for a recent review, and references within); 
the chiral algebra program \cite{Beem:2013sza}; the relation to quantum mechanical integrable models \cite{Nekrasov:2009rc}; and many facets of BPS/CFT correspondence, see {\it e.g.}  \cite{Nekrasov:2015wsu,Nekrasov:2016ydq,Alday:2009aq}.  The connection between the integrable models and SCFTs persists also for minimal supersymmetric cases through 
various index computations \cite{Gaiotto:2012xa,Gaiotto:2015usa,Maruyoshi:2016caf,Razamat:2018zel,Nazzal:2018brc}. All the studies listed here are deeply interrelated.    
}.

The space of conformal gauge theories can be thought of as defining a special slice, an interface, in the space of quantum field theories. Theories with larger matter content, but same gauge group, are IR free (though often UV completed by other gauge theories \cite{Seiberg:1994pq}). Whereas theories with smaller matter content are asymptotically free.\footnote{Note that by smaller and larger we mean removing or adding representations, not merely changing the dimension of the representation under the gauge group. There are interesting examples of gauge theories with non-simple gauge groups such that some of the simple gauge factors are asymptotically free and some are IR free. The latter gauge couplings are actually dangerously irrelevant leading to, conjecturally,  interacting SCFTs in the IR. See \cite{Kim:2018bpg} for some recent examples.
 } We thus classify a sub-slice of this interface given by simple gauge groups. There are several ways in which our analysis can be generalized given this picture. One can either complete the classification of this interface by considering non simple gauge groups, or one can start exploring the bulk of the asymptotically  free theory space by turning on some relevant deformations. In fact the only interesting relevant deformations of the conformal theories on the boundary are mass terms or vacuum expectation values and thus we expect it will be not too hard to analyze. This is to be contrasted with the models farther away from the interface in the theory space which might admit a large variety of relevant deformations, see {\it e.g.} \cite{Maruyoshi:2018nod} for a recent discussion. 

The results we derive here can be exploited in a variety of ways. For example, one can try to seek conformal dualities between different ${\cal N}=1$ theories in four dimensions. A way to systematically do so was discussed in \cite{Razamat:2019vfd}. The main idea is that once we restrict the discussion to conformal Lagrangians, the dimension of the gauge group and the dimension of the representation are fixed by the conformal anomalies leaving a finite set of theories which might, or might not, contain a dual of a given model. Many examples of conformal dualities were obtained in this manner in \cite{Razamat:2019vfd,Razamat:2020gcc}. It would be very interesting to study such conformal dualities using a variety of techniques involving various types of supersymmetric indices \cite{Kinney:2005ej,Romelsberger:2005eg,Dolan:2008qi,Benini:2011nc,Razamat:2013opa,Kels:2017toi,Closset:2017bse} and going beyond these, see {\it e.g.} \cite{Komargodski:2020ved}.
 As a concrete motivation for our program, and also a detailed example of computations of conformal manifolds, in section \ref{sec:prologue} we discuss an example of a putative conformal triality between three different conformal gauge theories. Some of the conformal Lagrangian theories might have an alternative UV complete non-conformal description which involves an RG flow. We will discuss a particular example of this in section \ref{sec:irduality}: thus, one can not just go inside the bulk of the theory space away from the conformal interface by RG flows but also go back to the interface  from the bulk. Another interesting question is whether conformal manifolds have any interesting geometric interpretation. In addition to examples of this that we have already mentioned, let us mention one more interesting case. In  \cite{Razamat:2019vfd} it was shown that certain conformal quiver theories with $SU(3)$ gauge nodes with nine flavors each have conformal manifolds which can be understood as parametrizing the space of complex structure moduli of genus $g$ Riemann surfaces as well as the space of flat connections of $E_8$ on these surfaces. This is again related to the fact that these models can be conjecturally obtained by compactifying the $(1,0)$ $6d$ rank one E-string theory on a genus $g$ Riemann surface (see \cite{Kim:2017toz,Razamat:2019ukg,Pasquetti:2019hxf} for related works). As a concrete example of an application of the classification program in this direction we discuss in section \ref{sec:classs} an extension of the conformal manifold of class ${\cal S}$ \cite{Gaiotto:2009we,Gaiotto:2009hg} theories corresponding to the compactification of the $A$ type $(2,0)$ theory on a genus $g$ Riemann surface with $s$ maximal punctures. The extension is done by conformally gauging the maximal puncture symmetries with the addition of two chiral superfields in the adjoint representation. 

The paper is organized as follows. We start in section \ref{sec:prologue} with a detailed example of a derivation of a conformal triality which illustrates the set of techniques we will use to classify theories with non-trivial conformal manifolds. In section \ref{sec:algorithm} the classification procedure is detailed and in the following sections we apply it to various classical gauge groups. ${\cal N}=1$ conformal  theories with simple unitary, section \ref{sec:unitary}, symplectic, section \ref{sec:symplectic},  orthogonal, section \ref{sec:orthogonal}, and exceptional, section \ref{sec:exceptional}, gauge groups are  considered. In section \ref{sec:extended} we discuss conformal theories with simple gauge group which have extended supersymmetry at zero coupling. Finally, in section \ref{sec:epilogue} we discuss a couple of physical applications of our results. Several appendices complement the bulk with additional technical details and definitions.  Specifically, Appendix \ref{app:confmancalc} has numerous worked out examples of computations illustrating the various subtleties in analyzing the structure of the conformal manifolds.

\section{Prologue: a conformal triality}
\label{sec:prologue}

We start our discussion with a simple example of a conformal triality. We claim that three different looking conformal gauge theories describe different weakly coupled cusps of the same $21$ dimensional conformal manifold. The derivation of this triality proceeds by first  considering the theory in triality frame $A$, which is  a $USp(6)$ conformal gauge theory with six fields in the ${\bf 14}$, the two index antisymmetric traceless representation. We argue that this theory is conformal and that on a generic locus of the conformal manifold there is no global (non-R) symmetry. 
The model has $21$ vector fields and $6\times 14=84$ chiral superfields. All the 't Hooft anomalies for symmetries that are not broken on the conformal manifold must agree between the different triality frames. For the case at hand, the only symmetry that is preserved on the conformal manifold is the $U(1)$ R-symmetry, whose anomalies can be expressed in terms of the $a$ and $c$ conformal anomalies. For conformal gauge theories, these are expressible in terms of the dimension of the gauge group ($\text{dim}\,{\frak G}$) and the dimension of the representation of the chiral fields under it ($\text{dim}\,{\frak R}$),
\be
a=\frac3{16}\text{dim}\,{\frak G}+\frac1{48}\text{dim}\,{\frak R}\,,\qquad\qquad c=\frac1{8}\text{dim}\,{\frak G}+\frac1{24}\text{dim}\,{\frak R}\,.
\ee 
This implies \cite{Razamat:2019vfd} that any conformal dual gauge theory to this model has to have the same dimension of the gauge group and the same dimension of representations of matter.
We then proceed to find two models that fit this bill and analyze their conformal manifolds to claim that in fact they are dual to the theory in frame $A$. The discussion concretely illustrates the various methods  to analyze the conformal manifolds which we use in the next sections of the paper to classify conformal gauge theories with simple gauge groups.

\subsection{Frame A}

Let us consider a $USp(6)$ gauge theory with matter comprised of six chiral fields in the ${\bf 14}$, which is the two index traceless antisymmetric representation of $USp(6)$. The Dynkin index of this matter representation is $2$ and of the adjoint is $4$ and thus, as $4+6\times 2\times (\frac23-1)=0$, the one loop beta function for the gauge coupling vanishes. This matter content does not have a Witten anomaly \cite{Witten:1982fp}.  We also note that the ${\bf 14}$ has a non-trivial totally symmetric cubic invariant and thus the theory has $8\times 7=56$ marginal operators, which transform in the three index symmetric representation of the $SU(6)$ global symmetry group of the free point. As we will discuss shortly the dimension of the conformal manifold is $21$ on a generic point of which the $SU(6)$ global symmetry is completely broken. We can also count the number of supersymmetric relevant operators and we find $21$ of these corresponding to the symmetric square of the matter fields. 
\begin{figure}[htbp]
	\centering
  	\includegraphics[scale=0.3]{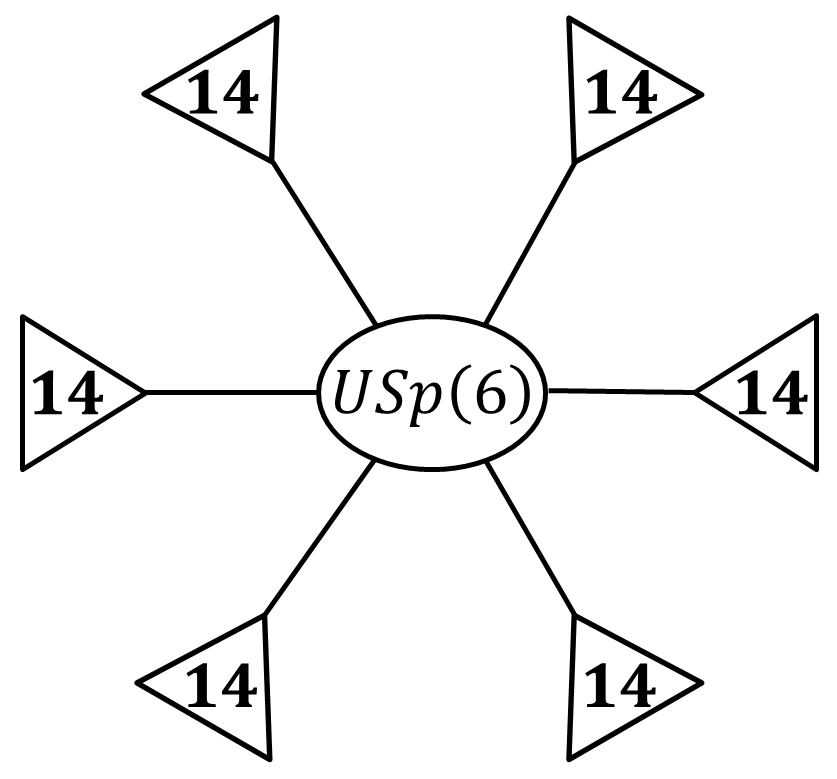} 
    \caption{Triality frame A.}
    \label{F:frameA}
\end{figure}

To determine the dimension of the conformal manifold we will follow here, and in most similar computations in the paper, the procedure developed in \cite{Green:2010da}. In this reference, analyzing the general structure of the beta functions of ${\cal N}=1$ theories in four dimensions (as well as theories with four supercharges in lower dimensions), it was determined that to compute the dimension of the conformal manifold one could proceed in two simple steps. We start from some SCFT and as a first step we list all the supersymmetric marginal couplings, the set of  which we will denote by $\{\lambda_i\}$. The theory can be a free UV Lagrangian, an IR end point of an RG flow, or even an abstractly defined strongly-coupled SCFT.
The SCFT has some global symmetry which we will denote by $G$.  Note that if the SCFT has weakly coupled vector fields, we should in principle also consider the gauge couplings and include the anomalous symmetries in $G$.  In the second step we compute the K\"ahler quotient of the space given by dividing the marginal couplings by the complexified global symmetry group,
\be
\{ \lambda_i\}/G_{\mathbb C}\,.
\ee 
The dimension of the K\"ahler quotient is the dimension of the conformal manifold, {\it i.e.} the space of {\it exactly marginal} couplings, of the theory. The intuition one might have about this is that the only way a marginal coupling might not be exactly marginal, in fact marginally irrelevant, is if it recombines with a component of a conserved current multiplet. In particular there are no marginally relevant supersymmetric deformations. We note here, that this fact can be also neatly observed \cite{Beem:2012yn} by computing the supersymmetric index of a theory \cite{Romelsberger:2005eg,Kinney:2005ej}.

Let us apply this procedure for the case at hand by computing the relevant K\"ahler quotient. The non-anomalous symmetry of the free theory is $SU(6)$ and the marginal operators $\lambda_{ijk}$ form the ${\bf 56}$ which is the three index symmetric representation. There is an anomalous $U(1)$ symmetry under which all the matter fields have the same charge and there is one (complexified) gauge coupling $\tau$.  The K\"ahler quotient which we need to compute then is,
\be
{\cal M}_c\sim \{\lambda_{ijk},\, \tau\}/SU(6)_\C\times U(1)_\C=\{{\bf 56}\}/SU(6)_\C\,.
\ee Here we have used the fact that the anomalous $U(1)$ transformation can be always absorbed by  appropriate factors depending on $\tau$.
By $\sim$ in the above equation we mean the equality to hold locally near the weakly coupled point.
To explicitly compute the quotient we first note that it is easy to construct invariants of $SU(6)$ and thus show that the K\"ahler quotient is not empty leading to a conformal manifold. For example let us define,
\be
f_{i_3j_3k_3s_3m_3n_3}=\epsilon^{i_1j_1k_1s_1m_1n_1}\epsilon^{i_2j_2k_2s_2m_2n_2}\lambda_{i_1i_2i_3}\lambda_{j_1j_2j_3}\lambda_{k_1k_2k_3}\lambda_{s_1s_2s_3}\lambda_{m_1m_2m_3}\lambda_{n_1n_2n_3}\,,
\ee which is symmetric in all the indices and then,
\be
\prod_{i=1}^6 f_{u^i_1u^i_2u^i_3u^i_4u^i_5u^i_6} \epsilon^{u^1_iu^2_iu^3_iu^4_iu^5_iu^6_i}\,,
\ee is an $SU(6)$ invariant. 

We can also consider a maximal $SU(3)$ subgroup of $SU(6)$ such that the fundamental representation of $SU(6)$ is mapped to the second rank symmetric representation of $SU(3)$, the ${\bf 6}$. Then the ${\bf 56}$ decomposes as ${\bf 1}\oplus {\bf 27}\oplus {\bf 28}$. Moreover, the adjoint of $SU(6)$, ${\bf 35}$, decomposes as ${\bf 27}\oplus {\bf 8}$. This implies that the theory has a one dimensional sublocus of the conformal manifold on which the $SU(3)$ is preserved. Here the exactly marginal operator parametrizing this one dimensional sublocus is given by the singlet in the decomposition of the ${\bf 56}$. The marginal operators in the ${\bf 27}$ of the preserved $SU(3)$ are actually marginally irrelevant, as these are eaten by the conserved currents enhancing the $SU(3)$ to $SU(6)$ that are now no longer conserved \cite{Green:2010da}. 
As a result, when on a generic point of this one dimensional sublocus, the marginal operators are in the ${\bf 1}\oplus {\bf 28}$, where the ${\bf 28}$ is the six index symmetric representation of $SU(3)$. The singlet just moves us along the sublocus, but we can insert the marginal operator in the ${\bf 28}$ to move away from it.  

We can next decompose $SU(3)$ into $SU(2)$ maximal subgroup such that ${\bf 3}$ goes to ${\bf 3}$. Then ${\bf 28}$ decomposes as ${\bf1 }\oplus {\bf 5}\oplus {\bf 9}\oplus{\bf 13}$ and the adjoint ${\bf 8}$ decomposes as ${\bf 3}\oplus {\bf 5}$. Thus we have an additional one dimensional manifold on which $SU(3)$ is farther broken to $SU(2)$ with marginal operators being in ${\bf 1}\oplus {\bf 1}\oplus{\bf 9}\oplus {\bf 13}$. We can then break $SU(2)$ to its $U(1)$ Cartan using the $U(1)$ singlets in either the ${\bf 9}$ or ${\bf 13}$. This gives an additional two dimensional manifold preserving only the $U(1)$. On this subspaces we have, besides the four singlets parametrizing it, also $18$ marginal operators charged under the $U(1)$. As this originated from $SU(2)$ representations, they come in pairs with opposite charges, giving additional $18-1=17$ directions breaking the $U(1)$. Overall, we end with a $21$ dimensional conformal manifold, preserving no symmetry on a generic locus of it, but with a one dimensional subspace preserving $SU(3)$, a two dimensional subspace preserving $SU(2)$ and a four dimensional subspace preserving $U(1)$.

The conformal anomalies are fixed by $\text{dim}\,{\frak G}=\text{dim}(USp(6))=21$ and $\text{dim}\,{\frak R}=6\times 14=84$. We can also compute some quantities  of this theory which do not depend on the location on the conformal manifold  (We will call them ${\cal M}_c$ invariants following \cite{Razamat:2019vfd}), and which thus should match in any duality frame . For example the superconformal index \cite{Kinney:2005ej, Romelsberger:2005eg,Dolan:2008qi} is given by (for definitions see appendix \ref{app:indexdef}),

\be
{\cal I}_A &=&\frac{(q;q)^3(p;p)^3}{48} \oint \frac{dz_1}{2\pi i z_1} \oint \frac{dz_2}{2\pi i z_2} \oint \frac{dz_3}{2\pi i z_3}\frac{\left(\Gamma_e((qp)^{\frac13})^2 \prod_{i\neq j}\Gamma_e((qp)^{\frac13} z_i^{\pm1} z_j^{\pm1})\right)^6}{\prod_i \Gamma_e(z_i^{\pm2}) \prod_{i\neq j}\Gamma_e(z_i^{\pm1} z_j^{\pm1})}\,.\nonumber\\
\ee Here $z_i$ are $USp(6)$ fugacities, that is the character of the fundamental is $z_1+\frac1{z_1}+z_2+\frac1{z_2}+z_3+\frac1{z_3}$. This can be generalized to other partition functions, {\it e.g.} lens index \cite{Benini:2011nc,Razamat:2013opa,Kels:2017toi}.

\subsection{Frame B}

Let us now consider an $SU(2)^7$ gauge theory with matter comprised by a single bifundamental chiral field between every pair of $SU(2)$ gauge factors. Each $SU(2)$ gauge group has then twelve fundamentals and thus has a vanishing one loop beta function. The marginal operators correspond to the triangles in the quiver, the number of which is $\begin{pmatrix}7\\ 3\end{pmatrix}=35$. The free theory has $21-7=14$ non-anomalous $U(1)$ symmetries all of which are broken on the conformal manifold as we will soon show. We thus get a $21$ dimensional conformal manifold with no symmetry on a generic point. The relevant operators are built from quadratic gauge invariants of the fields, the number of which is given by the number of the edges in the quiver which is $21$.
\begin{figure}[htbp]
	\centering
  	\includegraphics[scale=0.3]{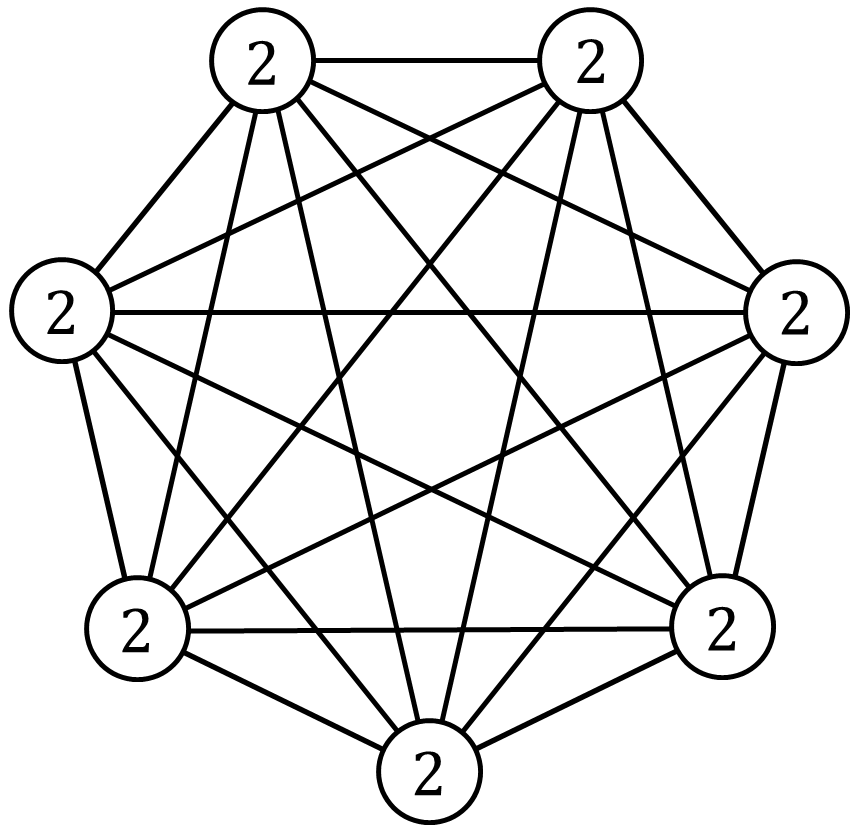} 
    \caption{Triality frame B.}
    \label{F:frameB}
\end{figure}

Let us count explicitly the exactly marginal deformations. Here it is convenient to perform the analysis using the methods developed by Leigh and Strassler \cite{Leigh:1995ep}. The basic idea is that in supersymmetric theories the beta functions of superpotential interactions are given as a linear combination of the anomalous dimensions of the fields participating in the superpotential term (see for example \cite{West:1984dg}). Moreover, the beta function of a gauge interaction is proportional to a linear combination of anomalous dimensions of the fields transforming non-trivially under the gauge group \cite{Novikov:1985rd,Shifman:1991dz}. Thus, demanding that the beta functions vanish gives a set of linear equation for anomalous dimensions with the number of independent variables given by the couplings. Then if one can show that a solution for this set exists (say at zero coupling in our case) and that the number of couplings is larger than the number of equations one can deduce the existence of a conformal manifold and compute its dimension. This procedure is particularly straightforward if the theory has only abelian global symmetries as is the case for the theory at hand.

 Let us denote the anomalous dimensions of bifundamental chiral fields between the {\it i}th and {\it j}th gauge group ($i\neq j$) as $\gamma_{ij}$. Then the demand that all the beta functions vanish translates to,
\be
\forall\; i\neq j\neq k\;\;\;  \gamma_{ij}+\gamma_{jk}+\gamma_{ki}=0\,,\qquad \forall  \; j\;\;\; \sum_{i\neq j} \gamma_{ij}=0\,.
\ee 
The first type of equations are beta functions for superpotential couplings and the second is the NSVZ beta function  \cite{Novikov:1985rd,Shifman:1991dz}. The only solution for these equations is that all $\gamma_{ij}$ vanish. We have thus $21$ equations for functions depending on $42$ variables (the $35$ superpotential couplings and $7$ gauge couplings), which gives us a $21$ dimensional space of  solutions, that is $\text{dim} {\cal M}_c =  21$. 

It is not hard to construct some of the invariants explicitly.  We have $21$ $U(1)$ symmetries associated to the edges of the quiver, $7$ of which are anomalous. Let us define the fugacity of the $U(1)$ corresponding to the edge between $i$th and $j$th gauge group as $s_{ij}$. Then the non-anomalous symmetries satisfy for any $i$ $\prod_{j\neq i} s_{ij}=1$. The fugacity associated to the superpotential term of the vertices $i$, $j$, and $k$ ($i\neq j\neq k$) is $\lambda_{ijk}=s_{ij}s_{jk}s_{ki}$. Let us by abuse of notation also denote by $\lambda_{ijk}$ the coupling of the superpotential term. Then for example $\prod_{j,k\neq i; j\neq k}\lambda_{ijk}$ are invariants for any $i$ out of all the nonanomalous symmetries and thus correspond to exactly marginal deformations.

The conformal anomalies are fixed by $\text{dim}\,{\frak G}=7\times dim(SU(2)) =21$ and $\text{dim}\,{\frak R}=\frac{7\times 6}2 \times 4 =84$. The index is given by,

\be
{\cal I}_B &=&\frac{(q;q)^7(p;p)^7}{2^7} \oint \prod_{i=1}^7\frac{dz_i}{2\pi i z_i} \frac{\prod_{i\neq j} \Gamma_e((qp)^{\frac13} z_i^{\pm1} z_j^{\pm1})}{\prod_{i=1}^7 \Gamma_e(z_i^{\pm2})}\,.
\ee  The  $z_i$ are fugacities for the seven $SU(2)$ groups, that is the character of fundamental of each is $z_i+\frac1{z_i}$.

\subsection{Frame C}

Let us consider an $SU(2)^2\times SU(4)$ gauge theory with matter comprised of three bifundamental chiral fields between each $SU(2)$ and the $SU(4)$ gauge group, and six chiral fields in the two index antisymmetric representation, ${\bf 6}$, of the $SU(4)$. Each $SU(2)$ gauge group has twelve fundamentals and thus has vanishing one loop beta function. The $SU(4)$ has six fundamentals, six antifundamentals, and six antisymmetrics. The Dynkin index of the latter representation is $1$ and thus as $4+6\times 1\times (\frac23-1)+(6+6)\times \frac12 \times  (\frac23-1)=0$ the one loop beta function for the $SU(4)$ gauge group also vanishes. There are marginal operators made from two bifundamentals and an antisymmetric, where the indices for the $SU(2)\times SU(4)$ gauge groups are contracted using epsilon tensors for both groups. We have one such operator for every choice of $SU(2)$ group, combination of bifundamentals, and antisymmetric chiral giving a total of $2\times 6\times 6=72$ marginal operators. Here we note that the bifundamental pair must be symmetric, leading to six different symmetric combinations of three bifundamentals. There are also $21$ relevant deformations built from the symmetric squares of the $SU(4)$ antisymmetrics.
\begin{figure}[htbp]
	\centering
  	\includegraphics[scale=0.3]{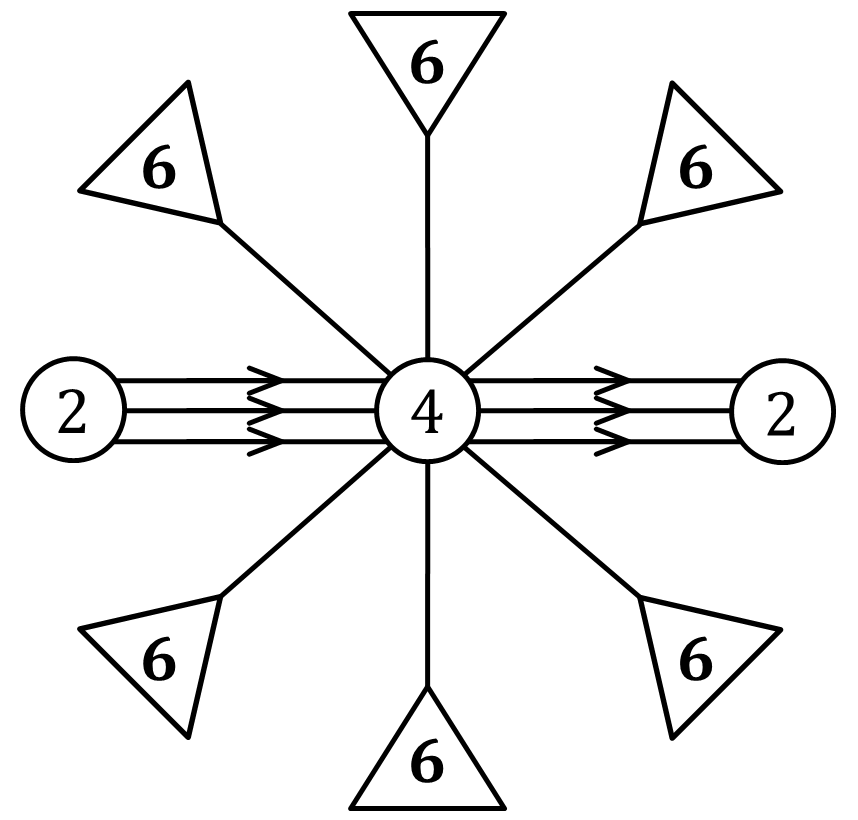} 
    \caption{Triality frame C.}
    \label{F:frameC}
\end{figure}

The symmetry group of the free theory is $SU(3)_1\times SU(3)_2\times SU(6)$ and the marginal operators are in the $({\bf 1},{\bf  \bar 6},{\bf 6})\;\oplus\; ({\bf  \bar 6},{\bf 1},{\bf 6})$ where the representations are of $(SU(3)_1,SU(3)_2,SU(6))$. Let us analyze the conformal manifold in more detail. We denote the two types of marginal operators as $\lambda^{i}_{\alpha}$ and $\lambda^{i}_{\dot\alpha}$ where $i$ are $SU(6)$ indices, $\alpha$ $SU(3)_1$ indices in its symmetric representation, and $\dot \alpha$ $SU(3)_2$ indices also in its symmetric representation. First, it is very easy to see that the K\"ahler quotient is not empty and indeed we have a conformal manifold. For instance for $\lambda^{i}_{\alpha}$ we can build the invariant
\be
\epsilon^{i j k l m n}\epsilon^{\alpha \beta \gamma \delta \mu \nu} \lambda^{i}_{\alpha}\lambda^{j}_{\beta}\lambda^{k}_{\gamma}\lambda^{l}_{\delta}\lambda^{m}_{\mu}\lambda^{n}_{\nu}\,,
\ee
 and likewise for $\lambda^{i}_{\dot\alpha}$. 

This means that there must be at least a two dimensional conformal manifold corresponding to turning on both of these operators. Let us first go along the subspace corresponding to inserting one of these to the superpotential. Doing so will break the symmetry to the subgroup that the chosen operator is invariant under. In our case the breaking is very similar to that of an $SU(N) \times SU(N)$ bifundamental where one breaks to the diagonal $SU(N)$. Likewise here we are going to break to the diagonal between the $SU(3)$ and an $SU(3)$ subgroup of the $SU(6)$ such the fundamental of $SU(6)$ becomes the symmetric ${\bf 6}$ of the $SU(3)$. Under this decomposition the operator in the $({\bf  \bar 6},{\bf 6})$ of $SU(3)\times SU(6)$ is decomposed into ${\bf  \bar 6} \otimes{\bf 6} = {\bf 1}\oplus {\bf 8}\oplus {\bf 27}$. The singlet is just the exactly marginal operator we turned on, which is indeed a singlet of the preserved symmetry. The remaining operators are additional marginal operators breaking the preserved $SU(3)$, however, these turn out to be marginally irrelevant. To see this recall that when turning on this operator we have broken the $SU(3)\times SU(6)$ part of the global to a diagonal $SU(3)$, and so the broken conserved currents must eat marginal operators to become long multiplets. The eaten marginal operators must transform in the same representation as those of the broken conserved currents under the preserved diagonal $SU(3)$, otherwise these cannot be eaten and this decomposition is impossible. Indeed under the $SU(3)$ subgroup of $SU(6)$ that we used we have that the adjoint of $SU(6)$ decomposes to ${\bf 8}\oplus {\bf 27}$. These are precisely the representations appearing besides the singlet in the decomposition of the marginal operator, and so they combine with them to form long multiplets. 

So overall we see that there is a one dimensional subspace of the conformal manifold along which the $SU(3)\times SU(3)\times SU(6)$ global symmetry is broken to $SU(3)\times SU(3)$. There are in fact two such subspaces depending on which of the two operators, $\lambda^{i}_{\alpha}$ or $\lambda^{i}_{\dot\alpha}$, we use. We can also turn on both operators. As we previously determined, once one operator is turned on then besides its singlet component under the preserved symmetry, the other components become marginally irrelevant. However we still have the second operator which is now in the $({\bf  \bar 6} ,{\bf 6})$ of the $SU(3)\times SU(3)$ global symmetry that is preserved on this one dimensional subspace. We can then insert it into the superpotential. This will break the $SU(3)\times SU(3)$ to the diagonal and we again have that ${\bf  \bar 6} \otimes{\bf 6} = {\bf 1}\oplus {\bf 8}\oplus {\bf 27}$. The singlet is the operator we are inserting. Since we break one of the $SU(3)$ global symmetry groups, the ${\bf 8}$ is actually marginally irrelevant. The ${\bf 27}$, though, remains marginal. So we conclude that there is a two dimensional subspace preserving a diagonal $SU(3)$ of the $SU(3)\times SU(3)\times SU(6)$ global symmetry group, spanned by the two singlets we get from $\lambda^{i}_{\alpha}$ and $\lambda^{i}_{\dot\alpha}$. On this subspace we still have a marginal operator in the ${\bf 27}$ that we can turn on. Note that  ${\bf 27}$ is the tensor representation which is traceless with two symmetric upper indices and two symmetric lower indices, let us denote the couplings as $\widetilde \lambda^{\alpha\beta}_{\gamma\delta}$. We can now turn on these couplings and break the symmetry completely. It is easy to build invariants and classify them, {\it e.g.} $\widetilde \lambda^{\alpha\beta}_{\gamma\delta} \widetilde \lambda^{\gamma\delta}_{\alpha\beta}$, $\widetilde \lambda^{\alpha\beta}_{\gamma\delta}\widetilde \lambda_{\alpha\beta}^{\rho\sigma}\widetilde \lambda_{\rho\sigma}^{\gamma\delta}$, $\widetilde \lambda^{\alpha\beta}_{\gamma\delta}\widetilde \lambda_{\alpha\rho}^{\gamma\sigma}\widetilde \lambda_{\beta\sigma}^{\delta\rho}$ {\it etc.} This gives a $21$ dimensional conformal manifold on a generic point of which all the symmetry is broken. 

The conformal anomalies are fixed by $\text{dim}\,{\frak G}=2\;dim(SU(2))+dim(SU(4))=21$ and $\text{dim}\,{\frak R}=6\times 6+8\times 6=84$. The index is given by,

\be
{\cal I}_C & = & \frac{(q;q)^5(p;p)^5}{2^2 4!}\oint \prod_{i=1}^3\frac{dz_i}{2\pi i z_i}\oint\frac{du}{2\pi i u}\oint\frac{dv}{2\pi i v}\times \nonumber\\
&& \frac{\left(\prod_{i=1}^4(\Gamma_e((qp)^{\frac13} u^{\pm1} z_i)\Gamma_e((qp)^{\frac13} v^{\pm1} z_i^{-1})\right)^3\left(\prod_{i\neq j}\Gamma_e((qp)^{\frac13} (z_iz_j)^{\pm1})\right)^6}{\Gamma_e(u^{\pm2})\Gamma_e(v^{\pm2})\prod_{i\neq j} \Gamma_e(z_i/z_j)}\,.
\ee 
Here $z_i$ are $SU(4)$ fugacities and satisfy $\prod_{i=1}^4z_i=1$, that is the character of the fundamental is $z_1+z_2+z_3+\frac1{z_1 z_2 z_3}$.

\subsection{Triality}

Note that each of the three gauge theories we have discussed are conformal, and have $\text{dim}\,{\frak G}=21$ and $\text{dim}\,{\frak R}=84$ and thus the same conformal anomalies. Also all three have a $21$ dimensional conformal manifold passing through zero coupling on a generic point of which there is no flavor symmetry, and $21$ supersymmetric relevant operators. We thus can conjecture that these models describe different weakly coupled cusps of the same conformal manifold, that is they are dual to each other. In particular this implies that the index of all the models should be equal,

\be
 {\cal I}_A={\cal I}_B={\cal I}_C\,.
 \ee The indices are given by rather different expressions, for example the number of integrations is different. We have verified that in expansion in fugacities they indeed agree when computed using any one of the three dual models. 
Expanding the indices  we obtain the following result in the three duality frames,
\be
&&1 + 21 (qp)^{\frac23} + 21 qp + 21(q+p) (qp)^{\frac23} + 
 231 (qp)^{\frac43} +(21 q^2 + 21 p^2 + 378 qp) (qp)^{\frac23} \nonumber\\
 &&+  420(q+p) (q p)^{\frac43} + 
1750 (qp)^2+21(q^3+p^3)(qp)^{\frac23}+441(q+p)(qp)^{\frac53}+651(q^2+p^2)(q p)^{\frac43}\nonumber\\
&&+3573 (q p)^{\frac73}+4599(q+p) (q p)^2+\cdots\,.
\ee 
 
 Note also that all three models have the same choice of global structure for the gauge group. Namely, $USp(4)$ or $USp(4)/\Z_2$, $SU(2)^7$ or $SU(2)^7/\Z_2$, and $SU(4)\times SU(2)^2$ or $(SU(4)\times SU(2)^2)/\Z_2$. This is noteworthy since the existence of these choices implies the existence of one-form symmetries \cite{Aharony:2013hda,Gaiotto:2014kfa}, either electric or magnetic, in these theories. These are hard to break and so must also agree on all sides of the duality, and indeed the presence of this choice on all three sides guarantees this\footnote{In principle one can probe \cite{Razamat:2013opa} the map between various choices of global structure for the gauge group by computing the lens index \cite{Benini:2011nc,Razamat:2013opa,Kels:2017toi}.}. Note also that although we gave some evidence for the validity of the triality we did not discuss the precise mapping between the $21$ exactly marginal couplings in each triality frame. This is a very interesting question which should be addressed but it goes beyond what we can comment on now.  
 
 Note that in triality frames $B$ and $C$ each one of the gauge groups by themselves are IR free. What makes these theories conformal is the fact that several symmetries are gauged which changes the global symmetry of the theory, and this alters the computation of the conformal manifold as a quotient by global symmetry. This fact stresses that if one is to classify most general conformal gauge theories, one should also keep IR free gauge theories (such that the one loop gauge beta function vanishes) as building blocks for more complicated theories.
 
 \
 
 The construction of the triality illustrates the basic technology we will use in classifying conformal theories in what follows as well as the physical motivation behind looking for such models. With this discussion we are ready to divert our attention to the classification program.

\section{The classification program}\label{sec:algorithm}

We shall begin by specifying the general approach of our classification program. As previously stated, we are interested in classifying all ${\cal N}=1$ gauge theories with simple gauge groups in four dimensions which possess a conformal manifold passing through weak coupling. We are then lead to consider a non-abelian gauge group $G$ with some number of chiral fields in representations $\bold{R}_i$. The conditions that this gauge theory is an interacting SCFT at the weak coupling point are\footnote{We shall only consider cases where the gauge theory itself is conformal at weak coupling, and ignore cases where the gauge theory flows to an interacting SCFT with a different weak coupling description.}:

\begin{enumerate}
  \item The theory must be gauge anomaly free. This is necessary for the consistency of the gauge theory. Since in $4d$ the gauge anomalies are all cubic, they depend on the cubic Casimir which only exist for $G=SU(N)$ with $N>2$. As a result this condition is only non-trivial for this case, and reads:
	
	\be
	\sum_i C(\bold{R}_i) = 0, \label{AnomalyC}
	\ee
	where we use $C(\bold{R}_i)$ for the cubic index of the representation $\bold{R}_i$. We note for convenience that it obeys $C(\bold{R}_i) = - C(\overline{\bold{R}_i})$, and as such vanishes for real and pseudo-real representations.    
		\item The one loop $\beta$-function must vanish with no anomalous dimensions for any field. This ensures that the theory can be conformal at weak coupling. This leads to the condition:
		
\be
\sum_i T(\bold{R}_i) = 3h^{\vee}_G, \label{BetaZero}
\ee
where we use $T(\bold{R}_i)$ for the Dynkin index of the representation $\bold{R}_i$, and $h^{\vee}_G$ for the dual Coxeter number of the group $G$.
\item The theory must have a conformal manifold. This ensures that the theory is indeed an interacting SCFT, that is, it remains conformal away from zero coupling for some appropriate combination of couplings. This means that, first, there must be appropriate cubic superpotentials, so as to be marginal at weak coupling, that one can turn on. Furthermore, the K$\ddot{a}$hler quotient of the space of marginal couplings by the complexified global symmetry group of the free point must be non-zero \cite{Green:2010da}.
\end{enumerate}  

For a given theory we need to verify that these three conditions are satisfied. The first two are given by trivial algebra. The third one, the computation of the K\"ahler quotient, can be rather tricky. In Appendix \ref{app:confmancalc} we detail an algorithm to compute such a quotient and the various subtleties involved, as well as work out quite a few concrete examples. In what follows we will detail the final result of such computations along with comments which should be instrumental in rederiving the quoted results. 

We next consider each simple group in turn, and enumerate our results. We shall consider only cases that are intrinsically $\mathcal{N}=1$ at weak coupling. We shall return to consider cases with extended supersymmetry in section \ref{sec:extended}. For each case we specify the dimension of the conformal manifold, the symmetry of the free point, the symmetry on the generic point of the conformal manifold. The conformal manifolds might contain subspaces which preserve larger symmetry than the generic point. {\it We identify some of such subspaces, though we do not claim that we identify all of them, as these types of computations tend to be rather involved.}

\section{$G=SU(N)$}\label{sec:unitary}

We shall first consider the case of group $G=SU(N)$.

\subsection{Generic cases}

For generic $N$, condition $\ref{BetaZero}$ can only be satisfied by representations containing at most two indices. These are the adjoint, symmetric, antisymmetric, fundamental and their conjugates. We have summarized the relevant group theory data on these in table \ref{SUGenRep}. For sufficiently low $N$, some other representations are possible. We shall refer to the cases involving them as exotic cases, and discuss them in the next section. Cases made from the above mentioned representations will be dubbed generic cases, and considered in this section, even if some of them are only conformal for small $N$.

\begin{table}[h!]
\begin{center}
\begin{tabular}{|c|c|c|c|c|c|}
  \hline 
   Symbol & Label & Dimension& Index & \mancube\; Casimir & Reality  \\
 \hline
 $F, \overline{F}$  & $(1,0,0,...,0),$ & $\bold{N}$  & $\frac{1}{2}$ & $1, -1$ & Complex \\
 & $(0,0,...,0,1)$ &  &  &  &  \\
\hline
 $AS, \overline{AS}$ & $(0,1,0,...,0),$  & $\bold{\frac{N(N-1)}{2}}$ & $\frac{N-2}2$ & $N-4, -(N-4)$ & Complex \\
 & $(0,...,0,1,0)$ &  &  &  & (Real for $N=4$) \\
\hline
 $S, \overline{S}$  & $(2,0,0,...,0),$ & $\bold{\frac{N(N+1)}{2}}$ & $\frac{N+2}2$ & $N+4, -(N+4)$ & Complex \\
 & $(0,0,...,0,2)$ &  &  &  &  \\
\hline
 $Ad$ & $(1,0,...,0,1)$  & $\bold{N^2-1}$ & $N$ & $0$ & Real \\
\hline
 \end{tabular}
 \end{center}
\caption{Various group theory data for $SU(N)$ representations associated with tensors with at most two indices. Here the label entry stands for the standard Dynkin label of the representation, and the symbol entry provides the shorthand symbol that will be employed for that representation throughout this article. The remaining entries list the dimension, Dynkin index ({\it a.k.a.} quadratic Casimir), cubic Casimir, and the reality properties of each representation.} 
\label{SUGenRep}
\end{table}

An important part in the classification is the possible superpotential terms one can add, which will be discussed forthwith. First adjoints can be coupled using a cubic superpotential. This can be done with an antisymmetric coupling for any group and with a symmetric coupling for $N\geq 3$. The antisymmetric coupling is the one used in $\mathcal{N} = 4$ super Yang-Mills, but will not play a role here as we always considered cases with less than $3$ adjoint chirals. We shall here only concentrate on cases where $N\geq 3$, and so can always use the symmetric coupling. These comes about as for $SU(2)$, the only interesting cases are the ones with extended supersymmetry. 

Additionally all two index representations can be coupled to two fundamentals via cubic superpotentials. For symmetrics and antisymmetrics this is to two anti-fundamentals, where the coupling is respectively symmetric or antisymmetric, and similarly for the conjugate representations. Adjoints couple to a fundamental and anti-fundamental, and similarly can be coupled also to any other pair of a representation and its conjugate. These are just the superpotential types that exist for $\mathcal{N}=2$ theories. There is also a superpotential involving all three two-index representations: a symmetric, conjugate antisymmetric and adjoint, and likewise for the complex conjugate.

For generic $N$, these are the only superpotentials, but for low values of $N$ additional superpotential terms are possible. We next review some of these special cases. 

\begin{enumerate}
  \item $N=3$: several special things occur for $N=3$. First the antisymmetric is the same as the anti-fundamental, so there are less choices in terms of matter multiplets. However, the fundamentals and anti-fundamentals can be coupled with an antisymmetric cubic superpotential given by the baryons. Also the symmetrics and their conjugates can be coupled with a symmetric cubic superpotential given by the determinant. As a results almost every representation can be coupled to itself making conformal manifolds quite common for $SU(3)$ gauge theories.
	
	\item $N=4$: the special feature of $N=4$ is that the antisymmetric representation becomes real. As a result there are more cubic superpotentials for theories with antisymmetrics as if these can couple to a given  representation, they can also couple to the conjugate one. We also note that the $\mathcal{N}=2$ type coupling between two antisymmetrics and the adjoint is antisymmetric. 
	
	\item $N=5$: here one can build a baryonic invariant from two antisymmetrics and a fundamental, and likewise for the complex conjugate. This invariant is symmetric.
	
	\item $N=6$: here one can build a baryonic invariant from three antisymmetrics, and likewise for the complex conjugate. This invariant is symmetric.
	
\end{enumerate}

We next list the possible theories and their properties. For ease of presentation we shall break the possibilities into cases.

\subsubsection{Cases with two adjoints}

We first consider the cases with two chiral fields in the adjoint representation. The possible solutions to conditions (\ref{AnomalyC}) and (\ref{BetaZero}) for generic $N$ are:

\begin{enumerate}

\item - $N_F=N_{\overline{F}}=N$

\item - $N_{AS}=1, \quad N_F=3 , \quad N_{\overline{F}}=N-1$

\item - $N_{AS}=N_{\overline{AS}}=1, \quad N_F=N_{\overline{F}}=2$ 

\end{enumerate}

Here we have suppressed the two adjoints for brevity, but we remind the reader that they are also part of the matter content. Also, for chiral choices there are two possibilities given by complex conjugation and we have only written one, as they are physically the same differing merely by redefining the $SU(N)$ generators.   

Besides these, there are several solutions that exist only for small $N$:

\begin{enumerate}

\item - $N_{AS}=2, \quad N_F=6-N , \quad N_{\overline{F}}=N-2, \quad N=5,6$

\item - $N_{AS}=3, \quad N_F=N_{\overline{F}}=1, \quad N=4$

\item - $N_{AS}=4, \quad N=4$

\item - $N_{AS}=2, \quad N_{\overline{AS}}=1, \quad N_{\overline{F}}=1, \quad N=5$

\end{enumerate}

Here we have only listed cases that do not reduce to one of the generic families. Finally, we need to consider the possible superpotentials and perform the K$\ddot{a}$hler quotient. We list the cases where this is non-trivial, together with some of their properties in table \ref{suNtwoAd}. In all the tables in the paper we use the following notations. ${\text dim}{\cal M}$ stands for the dimension of the conformal manifold. $G_F^{free}$ and $G_F^{gen}$ are the symmetries preserved at the free point and at a generic point of the conformal manifold respectively. The symbol $\emptyset$ denotes the situation when no symmetry is preserved.  Under $G_F^{gen}$ we also list some of the larger symmetries preserved on sub-loci of the conformal manifold. Finally, we  list for reference the $a$ and $c$ anomalies.

\begin{table}[h!]
\begin{center}
\begin{adjustwidth}{-.8cm}{}
\begin{tabular}{|c|c|c|c|c|c|}
  \hline 
 &  Matter & $dim\, {\cal M}$ & $G_F^{free}$ & $G^{gen}_F$  & $a$, $c$  \\
 \hline
$1$&  $N_F=N_{\overline{F}}=N$ & $N+1$ & $U(1)^2\times SU(2) $ & $U(1)^N$  & $a=\frac{13N^2-11}{48},$ \\
&	 & ($5$ for & $\times SU(N)^2$ & ($U(1)^2$ for $N=3$) & $c=\frac{7N^2-5}{24}$ \\
&	&  $N=3$) &  & has a $1d$ subspace &  \\
&	&  &  & preserving $U(1)^2\times SU(N)$ &  \\
 \hline
$2$& $N_{AS}=1, N_F=3,$ & $7$ & $U(1)^3\times SU(2) $ & $\emptyset$  & $a=\frac{28}{3},$ \\
& $N_{\overline{F}}=5, N=6$ &  & $\times SU(3)\times SU(5) $ & has a $2d$ subspace  & $c=\frac{119}{12}$ \\
 & &  &  & preserving $U(1)\times SU(2)$ &  \\
&	&  &  & $\times SU(3)$ &  \\
 \hline
$3$& $N_{AS}=1, N_F=3,$ & $4$ & $U(1)^3\times SU(2) $ & $U(1)^2$  & $a=\frac{103}{16},$ \\
 & $N_{\overline{F}}=4, N=5$ &  & $\times SU(3)\times SU(4) $ & has a $2d$ subspace  & $c=\frac{55}{8}$ \\
&  &  &  & preserving $U(1)^2\times SU(2)^2$ &  \\
	\hline
$4$&	$N_{AS}=1, N_F=3,$ & $2$ & $U(1)^3\times SU(2) $ & $U(1)^2 \times SU(2)^2$  & $a=\frac{65}{16},$ \\
&  $N_{\overline{F}}=3, N=4$ &  & $\times SU(3)^2 $ &  & $c=\frac{35}{8}$ \\
	\hline
$5$&	$N_{AS}=N_{\overline{AS}}=1,$ & $5$ & $U(1)^4\times SU(2)^3 $ & $U(1)^2$ ($U(1)$ for $N=6$)  & $a=\frac{12N^2+3N-11}{48},$ \\
&  $N_F=N_{\overline{F}}=2$ & ($6$ for & ($U(1)^3\times SU(2)^4$ & ($\emptyset$ for $N=5$)  & $c=\frac{6N^2+3N-5}{24}$ \\
 &  &  $N=6$) &  for $N=4$) & has a $2d$ subspace &  \\
&	& ($7$ for &  & preserving $U(1) \times SU(2)^2$ &  \\
&	&  $N=5$) &  & also has a $1d$ subspace &  \\
&	&  &  & preserving $U(1)^3 \times SU(2)$ &  \\
&	&  &  & ($U(1)^2\times SU(2)^2$ &  \\
&	&  &  & for $N=4$) &  \\
	\hline
$6$&	$N_{AS}=4, N=4$ & $3$ & $U(1)\times SU(2) $ & $SU(2)^2$  & $a=\frac{63}{16},$ \\
 & &  & $\times SU(4) $ &  has a $1d$ subspace & $c=\frac{33}{8}$ \\
 &	&  &  & preserving $U(1)\times USp(4)$ &  \\
	\hline
$7$&	$N_{AS}=2, N_{\overline{F}}=4$ & $3$ & $U(1)^2\times SU(2)^2 $ & $SU(2)^2$  & $a=\frac{439}{48},$ \\
& $N=6$ &  & $\times SU(4) $ &  has a $1d$ subspace & $c=\frac{229}{24}$ \\
&	&  &  & preserving $U(1)\times USp(4)$ &  \\
	\hline
\end{tabular}
\end{adjustwidth}
 \end{center}
\caption{} 
\label{suNtwoAd}
\end{table}
 
Let us make several specific comments about the various cases.

\begin{itemize}
\item Case 1 can be conformaly gauged with the diagonal $SU(N)$. The Cartan of this group is preserved on generic points. The other preserved $U(1)$ groups are the baryon $U(1)$, which is preserved generically, and some combination of the other $U(1)$ group and the Cartan of the $SU(2)$, which is preserved only on a special line.
\item In case 2 the preserved $SU(3)$ is the diagonal of the intrinsic $SU(3)$ and an $SU(3)$ subgroup of $SU(5)$, while the $SU(2)$ is the complement of this diagonal $SU(3)$ in $SU(5)$. The $U(1)$ is a combination of the commutant $U(1)$ in $SU(5)$ and the intrinsic $U(1)$ groups. 
\item In case 3 we have the breaking of $SU(3)\rightarrow U(1)\times SU(2)$ and $SU(4)\rightarrow U(1)\times SU(2)^2$, where the preserved $SU(2)^2$ is the diagonal of the $SU(2)$ in $SU(3)$ and one of the $SU(2)$ groups in $SU(4)$, while the other $SU(2)$ is its commutant in $SU(4)$. The two $U(1)$ groups are combinations of the intrinsic $U(1)$ groups and the $U(1)$ commutants in the non-abelian groups. One of these cannot be broken and is one of those preserved generically, while the second generically preserved $U(1)$ is a combination of the other $U(1)$ and Cartans of the $SU(2)$ groups.
\item  In case 4 we have the breaking $SU(3)\rightarrow U(1)\times SU(2)$ for both $SU(3)$ groups, where the $SU(2)^2$ is preserved as well as two combinations of the commutant and intrinsic $U(1)$ groups. As both of these $SU(2)$ groups see effectively $4$ doublet chiral fields, this can easily be used to build linear quivers using for instance $\mathcal{N}=2$ $SU(2)+4F$. 
\item In case 5 the two $SU(2)$ groups rotating the flavor symmetries can be preserved along a $2d$ subspace together with some combination of the intrinsic $U(1)$ groups. This allows one to gauge them, which can also be used to build quivers for small $N$, like $N=4$. Additionally there is a $1d$ subspace, generated by the cubic adjoint superpotential and the $\mathcal{N}=2$ superpotentials for the fundamentals and antisymmetrics, where these $SU(2)$ groups are broken to the diagonal, but one preserves the two baryonic symmetries as well as another $U(1)$ which is a combination of the intrinsic $U(1)$ groups and the Cartan of the adjoint $SU(2)$. for $N=4$, one of the baryonic symmetries enhances to $SU(2)$, which is then also preserved on this subspace. Generically, the Cartan of the diagonal fundamental $SU(2)$ and some combination of the baryonic symmetries is preserved, except for $N=6$ where all baryonic symmetries can be broken and $N=5$ where all symmetries can be broken.
\item  In case 6 we first have the breaking $SU(4)\rightarrow USp(4)$, with an additional preserved $U(1)$ being a combination of the intrinsic $U(1)$ and the Cartan of the $SU(2)$. We can further completely break the $U(1)$ and break $USp(4)$ to $SU(2)^2$. 
\item Case 7 features similar breaking as in the previous case, with first breaking $SU(4)\rightarrow USp(4)$, with an additional preserved $U(1)$ being a combination of the intrinsic $U(1)$ and the Cartan of the antisymmetric $SU(2)$. We can further completely break the $U(1)$ and break $USp(4)$ to $SU(2)^2$.
\end{itemize}

\subsubsection{Cases with one adjoint}

We next consider cases with one chiral field in the adjoint representation, that only have $\mathcal{N}=1$ supersymmetry. The possible solutions to conditions (\ref{AnomalyC}) and (\ref{BetaZero}) for generic $N$ are then:

\begin{enumerate}

\item - $N_S=N_{\overline{S}}=1, \quad N_{AS}=1, \quad N_F=1, \quad N_{\overline{F}}=N-3$

\item - $N_{S}=1, \quad N_{AS}=1, \quad N_{\overline{F}}=2N$

\item - $N_{S}=1, \quad N_{\overline{AS}}=1, \quad N_{F}=N-4, \quad N_{\overline{F}}=N+4$

\item - $N_{S}=1, \quad N_{\overline{AS}}=2, \quad N_{F}=N-5, \quad N_{\overline{F}}=7$

\item - $N_{S}=1, \quad N_{F}=N-3, \quad N_{\overline{F}}=2N+1$

\item - $N_{AS}=2, \quad N_{\overline{AS}}=1, \quad N_F=5 , \quad N_{\overline{F}}=N+1$

\item - $N_{AS}=2, \quad N_F=6 , \quad N_{\overline{F}}=2N-2$

\item - $N_{AS}=1, \quad N_F=N+3 , \quad N_{\overline{F}}=2N-1$ 

\end{enumerate}

Here we have suppressed the adjoint for brevity, but we remind the reader that it is also part of the matter content. Also, for chiral choices there are two possibilities given by complex conjugation and we have only written one, as they are physically the same differing merely by redefining the $SU(N)$ generators. We also note that while for generic $N$ these models have only $\mathcal{N}=1$ supersymmetry, for some this is enhanced to $\mathcal{N}=2$ for low values of $N$. 

Besides these, there are several solutions that exist only for small $N$:

\begin{enumerate}

\item - $N_{S}=1, \quad N_{\overline{AS}}=3, \quad N_F=N-6 , \quad N_{\overline{F}}=10-N, \quad N=6, 7, 8, 9, 10$

\item - $N_{AS}=7, \quad N_F=N_{\overline{F}}=1, \quad N=4$

\item - $N_{AS}=5, \quad N_F=15-3N , \quad N_{\overline{F}}=2N-5, \quad N=4, 5$

\item - $N_{AS}=4, \quad N_{\overline{AS}}=2, \quad N_{\overline{F}}=2, \quad N=5$

\item - $N_{AS}=4, \quad N_{\overline{AS}}=1,  \quad N_F=1 , \quad N_{\overline{F}}=4, \quad N=5$

\item - $N_{AS}=4, \quad N_F=12-2N , \quad N_{\overline{F}}=2N-4, \quad N=5, 6$

\item - $N_{AS}=3, \quad N_{\overline{AS}}=2, \quad N_F=7-N , \quad N_{\overline{F}}=3, \quad N=5, 6, 7$

\item - $N_{AS}=3, \quad N_{\overline{AS}}=1, \quad N_F=8-N , \quad N_{\overline{F}}=N, \quad N=5, 6, 7, 8$

\item - $N_{AS}=3, \quad N_F=9-N , \quad N_{\overline{F}}=2N-3, \quad N=5, 6, 7, 8, 9$

\end{enumerate}

Here we have only listed cases that do not reduce to one of the generic families. Finally, we need to consider the possible superpotentials and perform the k$\ddot{a}$hler quotient. We list the cases where this is non-trivial, together with some of their properties in tables \ref{suNoneAd1} and \ref{suNoneAd2}.

\begin{table}[h!]
\begin{center}
\begin{adjustwidth}{-.9cm}{}
\begin{tabular}{|c|c|c|c|c|c|}
  \hline 
 &  Matter & $dim\, {\cal M}$ & $G_F^{free}$ & $G^{gen}_F$  & $a$, $c$  \\
 \hline
$1$&$N_S=N_{\overline{S}}=1,$ & $2$ & $U(1)^5$ & $\emptyset$  & $a=\frac{23}{6},$ \\
&	$N_{AS}=1, N_F=1,$ &  &  & has a $1d$ subspace & $c=\frac{47}{12}$ \\
&$N_{\overline{F}}=1, N=4$	&  &  & preserving $U(1)^2$ &  \\
 \hline
 $2$& $N_S=N_{\overline{S}}=1,$ & $2$ & $U(1)^5\times SU(2) $ &  $U(1)^2$ & $a=\frac{295}{48},$ \\
&	$N_{AS}=1, N_F=1,$ &  &  & has a $1d$ subspace & $c=\frac{151}{24}$ \\
&$N_{\overline{F}}=2, N=5$	&  &  & preserving $U(1)^2\times SU(2)$ &  \\
 \hline
$3$&$N_S=N_{\overline{S}}=1,$ & $4$ & $U(1)^5\times SU(3) $ & $\emptyset$  & $a=\frac{431}{48},$ \\
&	$N_{AS}=1, N_F=1,$ &  &  & has a $1d$ subspace & $c=\frac{221}{24}$ \\
&$N_{\overline{F}}=3, N=6$	&  &  & preserving $U(1)^2\times SU(2)$ &  \\
 \hline
$4$&$N_S=1, N_{\overline{AS}}=1,$ & $\frac{N-2}{2}$ for & $U(1)^4 $ & $U(1)^{\frac{N-2}{2}} \times SO(8)$ for $N$ even & $a=\frac{13N^2-10}{48},$ \\
&	$N_F=N-4,$ &  $N$ even,  & $\times SU(N-4)$ & $U(1)^{\frac{N-1}{2}} \times SO(8)$ for $N$ odd & $c=\frac{7N^2-4}{24}$ \\
&$N_{\overline{F}}=N+4$	& $\frac{N-3}{2}$ for & $\times SU(N+4)$ & for $N$ even has $1d$ subspace &  \\
&	&  $N$ odd & ($U(1)^3\times SU(8)$ & preserving $U(1)$ &  \\
&	&  & for $N=4$) & $\times USp(N-4)\times SO(N+4)$ &  \\
&	&  &  & for $N$ odd has $1d$ subspace &  \\
&	&  &  & preserving $U(1)^2$ &  \\
&	&  &  & $\times USp(N-5)\times SO(N+3)$ &  \\
 \hline
$5$&$N_S=1, N_{\overline{F}}=7,$ & $15$ & $U(1)^2\times SU(7)$ & $\emptyset$  & $a=\frac{107}{48},$ \\
&	$N=3$ &  &  & has $1d$ subspace preserving & $c=\frac{59}{24}$ \\
&	&  &  & $SO(7)$ &  \\
 \hline
$6$&$N_{AS}=2, N_{\overline{AS}}=1,$ & $11$ & $U(1)^4\times SU(2)$ & $U(1)$  & $a=\frac{467}{48},$ \\
&	$N_{F}=5, N_{\overline{F}}=7,$ &  & $\times SU(5) \times SU(7)$ & has $1d$ subspace preserving & $c=\frac{257}{24}$ \\
&$N=6$	&  &  & $U(1)^2\times SU(2) \times SU(5)$ &  \\
&   &  &  & and $1d$ subspace preserving & \\
&   &  &  & $U(1)^{2}\times USp(4)\times USp(6)$ & \\
 \hline
$7$& $N_{AS}=2, N_{\overline{AS}}=1,$ & $28$ & $U(1)^4\times SU(2)$ & $\emptyset$ & $a=\frac{325}{48},$ \\
&	$N_{F}=5, N_{\overline{F}}=6,$ & & $\times SU(5) \times SU(6)$ & has $1d$ subspace preserving & $c=\frac{181}{24}$ \\
& $N=5$	&  &  & $U(1)^3\times SU(2) \times SU(4)$ &  \\
&	&  &  & and $1d$ subspace preserving &  \\
&	&  &  & $U(1)^2\times USp(4) \times USp(6)$ &  \\
 \hline
$8$& $N_{AS}=3,$ & $30$ & $U(1)^3\times SU(3)$ & $\emptyset$  & $a=\frac{13}{3},$ \\
&	$N_{F}=N_{\overline{F}}=5,$ & & $\times SU(5)^2$ & has $1d$ subspace preserving & $c=\frac{59}{12}$ \\
& $N=4$	&  &  & $U(1)\times SU(2) \times USp(4)$ &  \\
 \hline
$9$ & $N_{AS}=2, N=6$ & $3$ & $U(1)^{3}\times SU(2)$ & $U(1)\times SU(6)\times SU(2)^{2}$ & $a=\frac{119}{12},$ \\
& $N_{F}=6, N_{\overline{F}}=10$ &  & $\times SU(6)$ & has $1d$ subspace preserving & $c=\frac{133}{12}$ \\
&  &  & $\times SU(10)$ & $U(1)^{2}\times SU(6)\times USp(4)$ &  \\
 \hline                
\end{tabular}
\end{adjustwidth}
 \end{center}
\caption{} 
\label{suNoneAd1}
\end{table}

\begin{table}[h!]
\begin{center}
\begin{adjustwidth}{-0.4cm}{}
\begin{tabular}{|c|c|c|c|c|c|}
  \hline 
   &  Matter & $dim\, {\cal M}$ & $G_F^{free}$ & $G^{gen}_F$  & $a$, $c$  \\
 \hline
$10$&$N_{AS}=2,$ & $6$ & $U(1)^3\times SU(2)\times$ & $U(1)^{2}\times SU(3)$ & $a=\frac{55}{8},$ \\
&	$N_{F}=6, N_{\overline{F}}=8,$ &  & $SU(6)\times SU(8)$ & has $1d$ subspace preserving & $c=\frac{31}{4}$ \\
&$N=5$	&  &  & $U(1)^3\times SU(4) \times USp(4)$ &  \\
 \hline
$11$&$N_{AS}=1, N=6,$ & $1$ & $U(1)^3\times SU(9)$ & $U(1)\times SU(2)$  & $a=\frac{485}{48},$ \\
&	$N_{F}=9, N_{\overline{F}}=11,$ & & $\times SU(11)$ & $\times SU(9)$ & $c=\frac{275}{24}$ \\
 \hline
$12$& $N_{AS}=1, N=5,$ & $1$ & $U(1)^3\times SU(8)$ & $U(1)^2\times SU(2)$  & $a=\frac{335}{48},$ \\
&	$N_{F}=8, N_{\overline{F}}=9,$ & & $\times SU(9)$ & $\times SU(7)$ & $c=\frac{191}{24}$ \\
 \hline
$13$& $N_{AS}=1,$ & $3$ & $U(1)^3\times SU(7)^2$ & $U(1)\times SU(2)^3$  & $a=\frac{53}{12},$ \\
&	$N_{F}=N_{\overline{F}}=7,$ &  & & has $1d$ subspace preserving & $c=\frac{61}{12}$ \\
&$N=4$	&  &  & $U(1)\times USp(6)$ &  \\
& &  & & and $1d$ subspace preserving &  \\
&	&  &  & $U(1)^2\times SU(2)^2\times SU(5)$ &  \\
 \hline
$14$&$N_{AS}=5, N=4,$ & $1$ & $U(1)^3\times SU(3)^2$ & $U(1)^2\times SU(2)^2$  & $a=\frac{17}{4},$ \\
&	$N_{F}=N_{\overline{F}}=3,$ & & $\times SU(5)$ & $\times USp(4)$ & $c=\frac{19}{4}$ \\
 \hline
$15$&$N_{AS}=4,$ & $53$ & $U(1)^2\times SU(4)$ & $\emptyset$  & $a=\frac{229}{24},$ \\
&	$N_{\overline{F}}=8, N=6$ & & $\times SU(8)$ & has $1d$ subspace preserving & $c=\frac{31}{3}$ \\
 &&  &  & $U(1)\times USp(8)$ &  \\
 \hline
$16$&$N_{AS}=3, N_{\overline{AS}}=2,$ & $10$ & $U(1)^4\times SU(2)$ & $U(1)\times SU(2)$  & $a=\frac{449}{48},$ \\
&	$N_{F}=1, N_{\overline{F}}=3,$ & & $\times SU(3)^2$ & has $1d$ subspace preserving & $c=\frac{239}{24}$ \\
&$N=6$ &  &  & $U(1)^2\times SU(2)^2$ &  \\
 \hline
$17$ & $N_{AS}=3, N_{\overline{AS}}=2,$ & $19$ & $U(1)^4\times SU(2)^2$ & $\emptyset$  & $a=\frac{105}{16},$ \\
&	$N_{F}=2, N_{\overline{F}}=3,$ & & $\times SU(3)^2$ & has $1d$ subspace preserving & $c=\frac{57}{8}$ \\
&$N=5$ &  &  & $U(1)^3\times SU(2)^2$ &  \\
 \hline
$18$& $N_{AS}=3, N_{\overline{AS}}=1,$ & $23$ & $U(1)^4\times SU(2) \times$ & $\emptyset$  & $a=\frac{229}{24},$ \\
&	$N_{F}=2, N_{\overline{F}}=6,$ & & $SU(3) \times SU(6)$ & has $1d$ subspace preserving & $c=\frac{31}{3}$ \\
& $N=6$ &  &  & $U(1)\times SU(2)\times USp(6)$ &  \\
 \hline
$19$ &$N_{AS}=3, N_{\overline{AS}}=1,$ & $31$ & $U(1)^4\times SU(3)^2$ & $\emptyset$  & $a=\frac{20}{3},$ \\
&	$N_{F}=3, N_{\overline{F}}=5,$ & & $\times SU(5)$ & has $1d$ subspace preserving & $c=\frac{22}{3}$ \\
& $N=5$ &  &  & $U(1)^4\times USp(4)$ &  \\
 \hline
$20$& $N_{AS}=3,$ & $12$ & $U(1)^3\times SU(3)^2$ & $U(1)^2\times SU(3)$ & $a=\frac{467}{48},$ \\
&	$N_{F}=3, N_{\overline{F}}=9,$ &  & $\times SU(9)$ & has $1d$ subspace preserving & $c=\frac{257}{24}$ \\
& $N=6$ &  &  & $U(1)^2\times SU(3)\times USp(6)$ &  \\
 \hline
$21$ &$N_{AS}=3,$ & $42$ & $U(1)^3\times SU(3)\times$ & $\emptyset$  & $a=\frac{325}{48},$ \\
&	$N_{F}=4, N_{\overline{F}}=7,$ & & $SU(4)\times SU(7)$ & has $1d$ subspace preserving & $c=\frac{181}{24}$ \\
& $N=5$ &  &  & $U(1)^3\times SU(2)\times USp(6)$ &  \\
 \hline                       
\end{tabular}
\end{adjustwidth}
 \end{center}
\caption{} 
\label{suNoneAd2}
\end{table}
 Let us make several specific comments about the various cases.

\begin{itemize}
\item

In case 3 the $SU(2)$ is embedded in $SU(3)$ as $U(1)\times SU(2) \subset SU(3)$. 
\item In case 5 the symmetry is embedded as $SO(7)\subset SU(7)$. Incidentally, it can be further broken to $G_2$ along a 2d subspace. 
\item In case 6 the $SU(2)$ is embedded inside $U(1)\times SU(2) \times SU(5)\subset SU(7)$ and the $SU(5)$ is the diagonal one of the flavor $SU(5)$ and the one inside $SU(7)$. The $USp(4)$ and $USp(6)$ are embedded inside $U(1)\times USp(4) \subset SU(5)$ and $U(1)\times USp(6) \subset SU(7)$, respectively.
\item  In case 7 the symmetries are embedded as follows. In the first case we have $SU(4)\times SU(2)\subset SU(6)$ and $SU(4)\times U(1)\subset SU(5)$, where the $SU(4)$ group is the diagonal one. In the second case we have $USp(6)\subset SU(6)$ and $USp(4)\times U(1)\subset SU(5)$. In both cases, the preserved $U(1)$ groups are combinations of the intrinsic $U(1)$ groups, the Cartan of the $SU(2)$, and the $U(1)$ commutants in the non-abelian groups.
\item  In case 8 the $SU(2)$ is embedded inside $U(1)\times SU(2) \subset SU(3)$, and the $USp(4)$ is in the diagonal $SU(5)$, where the embedding is such that there is a $U(1)$ commutant.
\item In case 9 the $USp(4)$ is embedded inside $SU(6)\times USp(4)\subset SU(10)$ and the $SU(6)$ is the diagonal one of the flavor $SU(6)$ and the one inside $SU(10)$.
\item In case 10 the breaking is as follows: $SU(4)\times U(1)^2\subset SU(6)$ and $SU(4)\times USp(4)\subset SU(8)$, where the $SU(4)$ group is the diagonal one. Generically the diagonal $SU(4)$ is further broken to $SU(3)$. The preserved $U(1)$ groups are combinations of the intrinsic $U(1)$ groups, the Cartan of the $SU(2)$, and the $U(1)$ commutants in the non-abelian groups. 
\item In case 11 we break $SU(11)\rightarrow SU(9)\times SU(2)\times U(1)$, where the preserved $SU(9)$ is the diagonal one and the $U(1)$ is a combination of the intrinsic $U(1)$ groups and the $U(1)$ commutant in $SU(11)$.
\item  In case 12 we break $SU(9)\rightarrow SU(7)\times SU(2)\times U(1)$ and $SU(8)\rightarrow SU(7)\times U(1)$, where the preserved $SU(7)$ is the diagonal one and the $U(1)$ groups are combinations of the intrinsic $U(1)$ groups and the $U(1)$ commutants in the non-abelian groups. 
\item In case 13 for the first $1d$ subspace we first break both $SU(7)\rightarrow SU(6)\times U(1)$ and then break $SU(6)\rightarrow USp(6)$. The preserved $USp(6)$ is then the diagonal one and the $U(1)$ is a combination of the intrinsic $U(1)$ groups and the $U(1)$ commutants in the non-abelian groups. For the second $1d$ subspace we first break both $SU(7)\rightarrow SU(5)\times SU(2)\times U(1)$. The preserved $SU(5)$ is then the diagonal one. In this subspace, the generically preserved $SU(2)^3$ is embedded as $SU(2)^2\rightarrow SU(2)_{diagonal}$, $SU(5)\rightarrow U(1)\times USp(4)\rightarrow U(1)\times SU(2)^2$. 
\item In case 14 we break both $SU(3)\rightarrow SU(2)\times U(1)$ and break $SU(5)\rightarrow USp(4)\times U(1)$. The preserved $U(1)$ groups are combinations of the intrinsic $U(1)$ groups and the $U(1)$ commutants in the non-abelian groups. 
\item In case 15 the $USp(8)$ group is embedded in $SU(8)$ and the $U(1)$ is a combination of the intrinsic $U(1)$ groups and the Cartans of $SU(4)$. 
\item In case 16 the symmetries are embedded as follows: first we break both $SU(3)$ groups to $U(1)\times SU(2)$. The preserved $SU(2)^2$ are then the diagonal combination of the intrinsic $SU(2)$ with $SU(2)\subset SU(3)_{AS}$ and the $SU(2)\subset SU(3)_{\overline{F}}$. The former is further broken on generic points. The $U(1)$ groups are combinations of the intrinsic $U(1)$ groups and the $U(1)$ commutants in the non-abelian groups.
\item In case 17 the symmetries are embedded as follows: first we break both $SU(3)$ groups to $U(1)\times SU(2)$. The preserved $SU(2)^2$ are then the diagonal combination of the intrinsic $SU(2)$ acting on the conjugate antisymmetrics with $SU(2)\subset SU(3)_{AS}$ and the $SU(2)\subset SU(3)_{\overline{F}}$. the $U(1)$ groups are combinations of the intrinsic $U(1)$ groups and the $U(1)$ commutants in the non-abelian groups. 
\item In case 18 the $SU(6)$ is broken to $USp(6)$ and the $SU(3)$ is broken to a $U(1)$ which combines with the intrinsic ones.
\item In case 19 the $SU(5)$ is broken to $USp(4)\times U(1)$ and the $U(1)$ groups are combinations of the intrinsic $U(1)$ groups and the $U(1)$ commutants in the non-abelian groups.
\item  In case 20 we first break $SU(9)\rightarrow U(1)\times SU(3)\times USp(6)$. The preserved $SU(3)$ is then the diagonal one with the $SU(3)$ rotating the fundamentals. This $SU(3)$ cannot be broken on the conformal manifold.
\item  In case 21 we break $SU(7)\rightarrow U(1)\times USp(6)$, $SU(4)\rightarrow U(1)\times SU(3)$ and $SU(3)\rightarrow U(1)\times SU(2)$. The preserved $SU(2)$ is then the diagonal one of $SU(2)\subset SU(3)$ and $SO(3)\subset SU(3)\subset SU(4)$.
\end{itemize}

\subsubsection{Cases with symmetrics but no adjoints}

We next consider the cases where there is at least one chiral field in the symmetric representation, but no chiral fields in the adjoint representation. The number of solutions to   conditions (\ref{AnomalyC}) and (\ref{BetaZero})  is very large and we list all the cases in Appendix \ref{appendix:symmnoadj}. Among those there are only a few which also have a non trivial conformal manifold once generic superpotentials are turned on.
We  consider the possible superpotentials and perform the K$\ddot{a}$hler quotient. We list the cases where this is non-trivial, together with some of their properties in table \ref{suNSym1}.

\begin{table}[h!]
\begin{center}
\begin{adjustwidth}{-0.5cm}{}
\begin{tabular}{|c|c|c|c|c|c|}
  \hline 
   &  Matter & $dim\, {\cal M}$ & $G_F^{free}$ & $G^{gen}_F$  & $a$, $c$  \\
 \hline
$1$&$N_{S}=N_{\overline{S}}=1,$ & $3$ & $U(1)^3\times SU(4)^2$ & $SU(2)^2$  & $a=\frac{9}{4},$ \\
&	$N_{F}=N_{\overline{F}}=4,$ & & & has $1d$ subspace preserving & $c=\frac{5}{2}$ \\
&$N=3$	&  &  & $SU(2)^4$ &  \\
 \hline 
$2$&$N_{S}=1, N_{F}=3,$ & $76$ & $U(1)^2\times SU(3)$ & $SU(3)$  & $a=\frac{39}{16},$ \\
&	$N_{\overline{F}}=10,$ & & $\times SU(10)$ & has $1d$ subspace preserving & $c=\frac{23}{8}$ \\
&$N=3$	&  &  & $U(1)\times SU(3)\times SO(10)$ &  \\
&	&  &  &  and $1d$ subspaces preserving &  \\
&	&  &  & $SU(3)^{M+1}\times SO(10-3M)$ &  \\
&	&  &  & for $M=1,2,3$ &  \\
 \hline
$3$& $N_{S}=1, N_{AS}=2,$ & $1$ & $U(1)^3\times SU(2)$ & $U(1)^2\times SU(2)\times SO(14)$  & $a=7,$ \\
&	$ N_{F}=3, N_{\overline{F}}=14,$ & & $\times SU(3)$ & & $c=8$ \\
&$N=5$	&  & $\times SU(14)$ & &  \\
 \hline
$4$ & $N_{S}=1, N_{\overline{AS}}=1,$ & $1$ & $U(1)^3$ & $U(1)^2\times USp(2N-4)$  & $a=\frac{14N^2-9}{48},$ \\
&	$ N_{F}=2N-4,$ & & $\times SU(2N-4)$ & $\times SO(2N+4)$ & $c=\frac{8N^2-3}{24}$ \\
& $N_{\overline{F}}=2N+4,$	&  & $\times SU(2N+4)$ & &  \\
&$N \ne 3$	&  &  & &  \\
 \hline
$5$&$N_{S}=1, N_{AS}=3,$ & $2$ & $U(1)^2\times SU(3)$ & $U(1)\times SO(16)$  & $a=\frac{159}{16},$ \\
&	$ N_{\overline{F}}=16,$ & & $\times SU(16)$ & has $1d$ subspace preserving & $c=\frac{89}{8}$ \\
&$N=6$	&  &  & $U(1)^3\times SO(16)$ &  \\
 \hline                         
\end{tabular}
\end{adjustwidth}
 \end{center}
\caption{} 
\label{suNSym1}
\end{table}
 Let us make several specific comments about the various cases.

\begin{itemize}
\item
In case 1 the breaking is $SU(4)\rightarrow SO(4)$ for both groups. We can then further break each $SO(4)$ to $SO(3)$. 
\item In case 2 the $SU(3)$ rotating the fundamentals is unbroken while the one rotating the anti-fundamentals is broken to $SO(10-3M)\times SU(3)^{M}\times U(1)^M$. For $M=0$ one does not need the anti-baryon superpotential, and an additional $U(1)$ can be preserved. It is interesting to not that the $SO(7)\times SU(3)^2$ case can be further broken to $G_2\times SU(3)^2$ on a $2d$ subspace. 
\item In case 3 the symmetries are embedded as follows: $SO(14)\subset SU(14)$ and the $SU(2)$ is the diagonal of the antisymmetric $SU(2)$ and $SO(3)\subset SU(3)$.
\end{itemize}

\subsubsection{Cases with only antisymmetrics and fundamentals}

We next consider the cases where there are only chiral fields in the antisymmetric and fundamental representation, or their conjugate.  We list the theories solving (\ref{AnomalyC}) and (\ref{BetaZero})  in \ref{appendix:asymmnoadjnosymm}. Most of these are IR free. However once superpotentials are turned on some have interacting conformal manifolds.
 We list the cases where this is non-trivial, together with some of their properties in tables \ref{suNASF1} and \ref{suNASF2}.

\begin{table}[h!]
\begin{center}
\begin{adjustwidth}{-0.5cm}{}
\begin{tabular}{|c|c|c|c|c|c|}
  \hline 
   &  Matter & $dim\, {\cal M}$ & $G_F^{free}$ & $G^{gen}_F$  & $a$, $c$  \\
 \hline
$1$&$N_{AS}=N_{\overline{AS}}=3,$ & $23$ & $U(1)^3\times SU(3)^2$ & $U(1)^2$  & $a=\frac{159}{16},$ \\
&	$N_{F}=N_{\overline{F}}=6,$ & & $\times SU(6)^2$ & has $1d$ subspace preserving & $c=\frac{89}{8}$ \\
&$N=6$	&  &  & $U(1)^2\times USp(6)^2$ & \\
 \hline
$2$&$N_{AS}=N_{\overline{AS}}=3,$ & $73$ & $U(1)^3\times SU(3)^2$ & $\emptyset$  & $a=7,$ \\
&	$N_{F}=N_{\overline{F}}=6,$ & & $\times SU(6)^2$ & has $1d$ subspace preserving & $c=8$ \\
&$N=5$	&  &  & $SU(2)$ &  \\
 \hline
$3$&$N_{AS}=3, N_{\overline{AS}}=2,$ & $28$ & $U(1)^3\times SU(2)$ & $SU(2)^2$ & $a=\frac{81}{8},$ \\
&	$N_{F}=7, N_{\overline{F}}=9,$ & & $\times SU(3)\times $ & has $1d$ subspace preserving & $c=\frac{23}{2}$ \\
&$N=6$	&  & $SU(7)\times SU(9)$ & $SU(2)^3$ &  \\
 \hline
$4$&$N_{AS}=3, N_{\overline{AS}}=2,$ & $67$ & $U(1)^3\times SU(2)$ & $\emptyset$ & $a=\frac{341}{48},$ \\
&	$N_{F}=7, N_{\overline{F}}=8,$ & & $\times SU(3)\times $ & has $1d$ subspace preserving & $c=\frac{197}{24}$ \\
&$N=5$	&  & $SU(7) \times SU(8)$ & $U(1)^{5}\times SU(2)^{3}\times USp(4)$ &  \\
 \hline
$5$&$N_{AS}=N_{\overline{AS}}=2,$ & $11$ & $U(1)^3\times SU(2)^2$ & $SU(2)^{10}$  & $a=\frac{165}{16},$ \\
&	$N_{F}=N_{\overline{F}}=10,$ & & $\times SU(10)^2$ & has $4$ $1d$ subspaces preserving & $c=\frac{95}{8}$ \\
&$N=6$	&  &  & $U(1)^2\times G_1 \times G_2$ where &  \\
&	&  &  & $G_1, G_2 = USp(8)\times SU(2)$ &  \\
&	&  &  & or $USp(6)\times USp(4)$ &  \\
 \hline
$6$&$N_{AS}=N_{\overline{AS}}=2,$ & $29$ & $U(1)^3\times SU(2)^2$ & $\emptyset$  & $a=\frac{173}{24},$ \\
&	$N_{F}=N_{\overline{F}}=9,$ & & $\times SU(9)^2$ & has $4$ $1d$ subspaces preserving & $c=\frac{101}{12}$ \\
&$N=5$	&  &  & $U(1)^4\times G_1 \times G_2$ where &  \\
&	&  &  & $G_1, G_2 = USp(6)\times SU(2)$ &  \\
&	&  &  & or $USp(4)^2$ &  \\
 \hline
$7$ &$N_{AS}=4,$ & $82$ & $U(1)^2\times SU(4)$ & $U(1)$  & $a=\frac{223}{48},$ \\
&	$N_{F}=N_{\overline{F}}=8,$ & & $\times SU(8)^2$ & has $1d$ subspaces preserving & $c=\frac{133}{24}$ \\
&$N=4$	&  &  & $U(1)^4\times USp(4)^4$ &  \\
 \hline
$8$ &$N_{AS}=3,N_{\overline{AS}}=1,$ & $56$ & $U(1)^3\times SU(3)$ & $U(1)\times USp(8)$  & $a=\frac{165}{16},$ \\
&	$N_{F}=8,N_{\overline{F}}=12,$ & & $\times SU(8)$ & has $1d$ subspace preserving & $c=\frac{95}{8}$ \\
&$N=6$	&  & $\times SU(12)$ & $U(1)^3\times USp(8)^2 \times USp(4)$ &  \\
 \hline
$9$ &$N_{AS}=3,N_{\overline{AS}}=1,$ & $39$ & $U(1)^3\times SU(3)$ & $U(1)$  & $a=\frac{173}{24},$ \\
&	$N_{F}=8,N_{\overline{F}}=10,$ & & $\times SU(8)$ & has $1d$ subspace preserving & $c=\frac{101}{12}$ \\
&$N=5$	&  & $\times SU(10)$ & $U(1)^4\times USp(6)^2 \times USp(4)$ &  \\
 \hline
$10$&$N_{F}=N_{\overline{F}}=9,$ & $7$ & $U(1)\times SU(9)^2$ & $\emptyset$  & $a=\frac{21}{8},$ \\
&	$N=3$ & & & has $1d$ subspace preserving & $c=\frac{13}{4}$ \\
&	&  &  & $SU(3)^6$ &  \\
 \hline
                        
\end{tabular}
\end{adjustwidth}
 \end{center}
\caption{} 
\label{suNASF1}
\end{table}
 
\begin{table}[h!]
\begin{center}
\begin{adjustwidth}{-0.3cm}{}
\begin{tabular}{|c|c|c|c|c|c|}
  \hline 
   &  Matter & $dim\, {\cal M}$ & $G_F^{free}$ & $G^{gen}_F$  & $a$, $c$  \\
    \hline
$11$&$N_{AS}=N_{\overline{AS}}=4,$ & $15$ & $U(1)^3\times SU(2)^2$ & $SU(2)^2$ & $a=\frac{153}{16},$ \\
&	$N_{F}=N_{\overline{F}}=2,$ & & $\times SU(4)^2$ & has $1d$ subspace preserving & $c=\frac{83}{8}$ \\
&$N=6$	&  &  & $U(1)^4\times SU(2)^2$ &  \\
 \hline
$12$&$N_{AS}=5, N_{\overline{AS}}=3,$ & $26$ & $U(1)^2\times SU(3)$ & $\emptyset$ & $a=\frac{153}{16},$ \\
&	$N_{\overline{F}}=4,$ & & $\times SU(4)$ & has $1d$ subspace preserving & $c=\frac{83}{8}$ \\
&$N=6$	&  & $\times SU(5)$ & $U(1)^5\times USp(4)$ &  \\
 \hline
$13$&$N_{AS}=4, N_{\overline{AS}}=3,$ & $21$ & $U(1)^3\times SU(3)^2$ & $\emptyset$ & $a=\frac{39}{4},$ \\
&	$N_{F}=3, N_{\overline{F}}=5,$ & & $\times SU(4)$ & has $1d$ subspace preserving & $c=\frac{43}{4}$ \\
&$N=6$	&  & $\times SU(5)$ & $U(1)^4 \times SU(2)$ &  \\
 \hline
$14$&$N_{AS}=6, N_{\overline{F}}=12,$ & $273$ & $U(1)\times SU(6)$ & $\emptyset$ & $a=\frac{159}{16},$ \\
&	$N=6,$ & & $\times SU(12)$ & has $1d$ subspace preserving & $c=\frac{89}{8}$ \\
&	 & & & $U(1)^{4}\times USp(8)\times USp(4)$ & \\
 \hline
$15$&$N_{AS}=5, N_{\overline{AS}}=1,$ & $136$ & $U(1)^3\times SU(2)$ & $SU(2)$ & $a=\frac{159}{16},$ \\
&	$N_{F}=2, N_{\overline{F}}=10,$ & & $\times SU(5)$ & has $1d$ subspaces preserving & $c=\frac{89}{8}$ \\
&$N=6$	&  & $\times SU(10)$ & $U(1)^{3}\times SU(2)^2\times USp(8)$, &  \\
&	&  &  & $U(1)^{3}\times SU(2)\times USp(4)$ &  \\
&	&  &  & $\times USp(6),\ SU(2)^3$ &  \\
 \hline
$16$&$N_{AS}=4, N_{\overline{AS}}=2,$ & $55$ & $U(1)^3\times SU(2)$ & $SU(2)^{2}$ & $a=\frac{159}{16},$ \\
&	$N_{F}=4, N_{\overline{F}}=8,$ &  & $\times SU(4)^2$ & has $1d$ subspace preserving & $c=\frac{89}{8}$ \\
& $N=6$	&  & $\times SU(8)$ & $U(1)^4\times USp(4)\times USp(8)$ &  \\
 \hline
$17$&$N_{AS}=4, N_{\overline{AS}}=2,$ & $20$ & $U(1)^3\times SU(2)$ & $U(1)^3\times SU(2)$ & $a=7,$ \\
&	$N_{F}=5, N_{\overline{F}}=7,$ & & $\times SU(4)\times $ & has $1d$ subspace preserving & $c=8$ \\
&$N=5$	&  & $SU(5)\times SU(7)$ & $U(1)^4\times SU(2)^2 \times USp(4)$ &  \\
 \hline
$18$&$N_{AS}=5,$ & $152$ & $U(1)^2\times SU(5)^2$ & $U(1)$ & $a=\frac{341}{48},$ \\
&	$N_{F}=5, N_{\overline{F}}=10,$ & & $\times SU(10)$ & has $1d$ subspace preserving & $c=\frac{197}{24}$ \\
&$N=5$	&  &  & $U(1)^5 \times SU(2)^5$ &  \\
 \hline
$19$&$N_{AS}=4, N_{\overline{AS}}=1,$ & $72$ & $U(1)^3\times SU(4)$ & $U(1)$ & $a=\frac{341}{48},$ \\
&	$N_{F}=6, N_{\overline{F}}=9,$ & & $\times SU(6)$ & has $1d$ subspace preserving & $c=\frac{197}{24}$ \\
&$N=5$	&  & $\times SU(9)$ & $U(1)^3 \times SU(2)^2 \times USp(8)$ &  \\
 \hline                         
\end{tabular}
\end{adjustwidth}
 \end{center}
\caption{} 
\label{suNASF2}
\end{table}

 Let us make several specific comments about the various cases.

\begin{itemize}
\item
In case 1 the generically preserved $U(1)$ groups are part of the Cartan of each $USp(6)\subset SU(6)$ group. The $U(1)^2$ that is preserved on the subspace are different and are combinations of the Cartans of each $SU(3)$ group and the intrinsic $U(1)$ groups.
\item  In case 2 the $SU(2)$ is embedded as $SO(3)\subset SU(3)$ where it is in the diagonal of $SU(3)^2$ and $SU(6)^2$, where for the latter we use the embedding $\bold{6}_{SU(6)}\rightarrow \overline{\bold{6}}_{SU(3)}$.
\item  In case 3 one breaks $SU(9)\to SU(3)^3\times U(1)^2 \to SO(3)^3\times U(1)^2$, and then take the diagonal $SU(2)$ of the three $SO(3)$ groups. The other two $SU(2)$ groups are embedded as $SU(2)^2\times U(1)^2 \subset SU(7)$, and these are the ones generally preserved. The rest of the symmetries are broken completely.
\item  In case 4 the $USp(4)$ is embedded as $USp(4)\times U(1)^3\subset SU(7)$, and the $SU(2)^3$ is embedded as $SU(2)^3\times U(1)^4\subset SU(8)$. The $U(1)^5$ are combinations of the intrinsic $U(1)^3$, the Cartans of $SU(2)$ and $SU(3)$, and the $U(1)$ commutants in the non-abelian groups.
\item In case 5 the maximal breaking is $SU(10)\rightarrow SU(2)^5$ for both $SU(10)$ groups. The minimal breaking involves either $SU(10)\rightarrow U(1)\times SU(2)\times USp(8)$ or $SU(10)\rightarrow U(1)\times USp(4)\times USp(6)$ for each $SU(10)$ group. There are then $4$ distinct subspaces, one for each choice. The $U(1)^2$ are then the combinations of the commutant and intrinsic $U(1)$ groups. \item Similarly, in case 6 the minimal breaking involves either $SU(9)\rightarrow U(1)^2\times SU(2)\times USp(6)$ or $SU(9)\rightarrow U(1)^2\times USp(4)^2$ for each $SU(9)$ group. There are then $4$ distinct subspaces, one for each choice. The $U(1)^4$ are then the combinations of the commutant and intrinsic $U(1)$ groups. 
\item In case 7 the minimal breaking involves $SU(8)\rightarrow U(1)\times USp(4)^2$ for each $SU(8)$ group.  The $SU(4)$ is broken to its Cartan. The $U(1)^4$ are then three combinations of the commutant and intrinsic $U(1)$ groups, and one of the intrinsic $U(1)$ groups which is never broken on the conformal manifold.
\item In case 8 the $USp(4)$ and one of the $USp(8)$ are embedded inside $USp(4)\times USp(8)\times U(1)\subset SU(12)$, and the second $USp(8)$ which is also the one generally preserved is embedded inside $USp(8)\subset SU(8)$. The $U(1)^3$ are combinations of the intrinsic $U(1)^3$, and the $U(1)$ commutant in the non-abelian group. One of the intrinsic $U(1)$'s is the one preserved generally.
\item In case 9 the $USp(4)$ and one of the $USp(6)$ are embedded inside $USp(4)\times USp(6)\times U(1)\subset SU(10)$, and the second $USp(6)$ is embedded inside $USp(6)\times U(1)^2\subset SU(8)$. The $U(1)^4$ are combinations of the intrinsic $U(1)^3$, and the $U(1)$ commutants in the non-abelian groups. One of the intrinsic $U(1)$'s is the one preserved generally.
\item In case 10 the minimal breaking is $SU(9)\rightarrow U(1)^2\times SU(3)^3$ for both $SU(9)$ groups.
\item  In case 11 the $SU(2)$ groups rotating the fundamentals and their conjugates cannot be broken. It is possible to preserve some of the Cartan of the $SU(4)$ groups on subspaces of the conformal manifold. 
\item In case 12 the breaking is $SU(4)\rightarrow USp(4)$ and the other non-abelian groups are broken down to the Cartan. 
\item In case 13 the $SU(2)$ is embedded as $SU(2)\times  U(1)^3\subset SU(5)$. The $U(1)^4$ are combinations of the Cartans of $SU(3)^2$ and $SU(4)$, and the $U(1)$ commutants in $SU(5)$.
\item In case 14 the $USp(8)$ and the $USp(4)$ are embedded as $USp(8)\times USp(4)\times U(1)\subset SU(12)$. The $U(1)^4$ are combinations of the intrinsic $U(1)$, the Cartan of $SU(6)$, and the $U(1)$ commutant in the non-abelian group.
\item In case 15 one breaks $SU(10)\to SU(2)\times USp(8)\times U(1)$ for the first $1d$ subspace, $SU(10)\to USp(4)\times USp(6)\times U(1)$ for the second $1d$ subspace. For the third $1d$ subspace one breaks $SU(10)\to SU(5)\times SU(2)$ such that $\textbf{10}\to (\textbf{5},\textbf{2})$, then taking the diagonal of the remaining $SU(5)$ and of $SU(5)_F$ breaking it as $SU(5)\to SU(2)$ such that $\textbf{5}_{SU(5)}\to \textbf{5}_{SU(2)}$. The additional $SU(2)$ for all of these cases is the intrinsic one and is also preserved generally. The $U(1)^3$ in the first two $1d$ subspaces are combinations of the intrinsic $U(1)^3$, the Cartan of $SU(5)$, and the $U(1)$ commutant in $SU(10)$.
\item In case 16 the $USp(4)$ is embedded as $USp(4)\subset SU(4)_F$, and the $USp(8)$ is embedded as $USp(8)\subset SU(8)$. The $U(1)^4$ is a combination of the $SU(4)_{AS}$ and $SU(2)$ Cartans and the intrinsic $U(1)$ groups. The $SU(2)^2$ generally preserved is embedded as $SU(2)^2\subset USp(4)$.
\item In case 17 one breaks $SU(7)\to USp(4)\times SU(3)\times U(1) \to USp(4)\times SO(3)\times U(1)$ and taking the diagonal $SU(2)$ of this $SO(3)$ and $SU(2)_{\overline{AS}}$ to get the generally preserved  $SU(2)$. In addition one breaks $SU(5)\to SU(3)\times U(1)^2\to SO(3)\times U(1)^2$ and $SU(4)\to SU(2)\times U(1)^2$ and takes the diagonal $SU(2)$ of these $SO(3)$ and $SU(2)$. The $U(1)^4$ is a combination of the intrinsic $U(1)$ groups and the $U(1)$ commutants in the non-abelian groups. The generally preserved $U(1)^3$ are a combination of the $U(1)$ coming from the breaking $USp(4)\to SU(2)^2 \to SU(2) \to U(1)$ and the $U(1)^4$ preserved on the $1d$ subspace.
\item In case 18 one of the intrinsic $U(1)$ groups cannot be broken on the conformal manifold. On the special subspace, the preserved symmetry includes $U(1)^4\times SU(2)^5$, where the $SU(2)^5$ is embedded as $SU(2)^5\times U(1)^4\subset SU(10)$, and the $U(1)$ groups are a combination of the intrinsic $U(1)$'s, the Cartan of the two broken $SU(5)$ groups and the $U(1)$ commutants in the non-abelian groups.
\item In case 19 the $USp(8)$ is embedded as $USp(8)\times U(1)\subset SU(9)$. In addition one breaks $SU(6)\to SU(3)\times SU(2)\times U(1)^2 \to SO(3)\times SU(2)\times U(1)^2$ and $SU(4)\to SU(2)\times U(1)^2$ and take the diagonal $SU(2)$ of $SO(3)$ and the $SU(2)\subset SU(4)$.  The $U(1)^3$ groups are a combination of the intrinsic $U(1)$'s, and the $U(1)$ commutants in the non-abelian groups. The generally preserved $U(1)$ is a combination of the above $1d$ subspace $U(1)^3$.
\end{itemize}

\subsection{Exotic cases}

Finally we consider representations which are only possible for low values of $N$. The full list of possibilities appear in table \ref{SUExRep}. As can be seen, there are only a handful of cases extending up to $SU(12)$. Most of these cases are the rank $3$ antisymmetric, and a few rank $4$ antisymmetric as well. The groups $SU(4)$ and $SU(5)$ have a few more exotic choices. 

\begin{table}[h!]
\begin{center}
\begin{tabular}{|c|c|c|c|c|c|}
  \hline 
   Group & Label & Dimension  & Dynkin  & \mancube\;   Casimir& Reality  \\
    \hline
 $SU(4)$  & $(1,1,0), (0,1,1)$ & $\bold{20}, \overline{\bold{20}}$  & $\frac{13}{2}$ & $7, -7$ & Complex \\
\hline
$SU(4)$  & $(0,2,0)$ & $\bold{20'}$  & $8$ & $0$ & Real \\
\hline
\hline
$SU(5)$  & $(1,0,1,0), (0,1,0,1)$ & $\bold{45}, \overline{\bold{45}}$  & $12$ & $6, -6$ & Complex \\
\hline
\hline
$SU(6)$  & $(0,0,1,0,0)$ & $\bold{20}$  & $3$ & $0$ & Pseudo-real \\
\hline
\hline
$SU(7)$  & $(0,0,1,0,0,0), (0,0,0,1,0,0)$ & $\bold{35}, \overline{\bold{35}}$  & $5$ & $2, -2$ & Complex \\
\hline
\hline
$SU(8)$  & $(0,0,1,0,0,0,0),$ & $\bold{56},$  & $\frac{15}{2}$ & $5,$ & Complex \\
$SU(8)$  & $(0,0,0,0,1,0,0)$ & $\overline{\bold{56}}$  &  & $-5$ &  \\
\hline
$SU(8)$  & $(0,0,0,1,0,0,0)$ & $\bold{70}$  & $10$ & $0$ & Real \\
\hline
\hline
$SU(9)$  & $(0,0,1,0,0,0,0,0),$ & $\bold{84}, $  & $\frac{21}{2}$ & $9,$ & Complex \\
 & $(0,0,0,0,0,1,0,0)$ & $\overline{\bold{84}}$  &  & $-9$ & \\
\hline
$SU(9)$  & $(0,0,0,1,0,0,0,0),$ & $\bold{126},$  & $\frac{35}{2}$ & $5,$ & Complex \\
 & $(0,0,0,0,1,0,0,0)$ & $\overline{\bold{126}}$  &  & $-5$ & \\
\hline
\hline
$SU(10)$  & $(0,0,1,0,0,0,0,0,0),$ & $\bold{120},$  & $14$ & $14,$ & Complex \\
 & $(0,0,0,0,0,0,1,0,0)$ & $ \overline{\bold{120}}$  &  & $-14$ &  \\
\hline
\hline
$SU(11)$  & $(0,0,1,0,0,0,0,0,0,0),$ & $\bold{165},$  & $18$ & $20,$ & Complex \\
 & $(0,0,0,0,0,0,0,1,0,0)$ & $ \overline{\bold{165}}$  &  & $-20$ &  \\
\hline
\hline
$SU(12)$  & $(0,0,1,0,0,0,0,0,0,0,0),$ & $\bold{220},$  & $\frac{45}{2}$ & $27,$ & Complex \\
  & $(0,0,0,0,0,0,0,0,1,0,0)$ & $\overline{\bold{220}}$  &  & $-27$ & \\
\hline
 \end{tabular}
 \end{center}
\caption{Various group theory data for $SU(N)$ representations associated with tensors with more than two indices. Here the group entry stands for the group in question, and the label entry stands for the standard Dynkin label of the representation. The remaining entries list the dimension, Dynkin index (AKA quadratic Casimir), cubic Casimir and the reality properties of each representation.} 
\label{SUExRep}
\end{table}

We first review the list of possible superpotentials. We begin with the case of $SU(4)$. Here we have two types of twenty dimensional representations. The $\bold{20}$ and $\overline{\bold{20}}$ are the product of the fundamental and the antisymmetric. The antisymmetric product of either the $\bold{20}$ or the $\overline{\bold{20}}$ can be coupled to the antisymmetric and the symmetric product of either can be coupled to either the symmetric or its conjugate. The $\bold{20}$ can also couple to an anti-fundamental and an antisymmetric or conjugate symmetric, and likewise for the conjugate. As usual, the product of the $\bold{20}$ and $\overline{\bold{20}}$ can couple to the adjoint, and also to the $\bold{20'}$. Finally the $\bold{20}$ can also linearly couple to an adjoint or $\bold{20'}$ and a fundamental, and likewise for the conjugate. 

The $\bold{20'}$ is the symmetric traceless product of two antisymmetrics. It has a cubic symmetric coupling, and being a real representation, its anti-symmetric square can couple to the adjoint. It can also linearly couple to the symmetric square of the antisymmetric, adjoint, symmetric and conjugate symmetric representations. 

For $SU(5)$ we can have the $\bold{45}$ which appears only in one combination with anti-fundamentals. There are no cubic superpotentials that involve only these two representations.

For $N=6-12$, we can also have the three index antisymmetric representation. This representation can couple to the product of an anti-fundamental and conjugate antisymmetric, and likewise for the complex conjugate. Like all other representations, the product with its conjugate can be coupled to the adjoint. There are a few special cases. For $N=6$, the representation is pseudo-real so it can couple linearly to the fundamental and antisymmetric and also to the anti-fundamental and conjugate antisymmetric. Its symmetric product can be coupled to the adjoint representation. For $N=7$ and $N=8$, we can couple its antisymmetric square to the fundamental and antisymmetric representations, respectively, and likewise for the conjugate. Finally, for $N=9$ there is an antisymmetric cubic coupling, though this won't play a role here. 

The only other possible representation is the four index antisymmetric representation, which is possible for $N=8,9$. Its generic coupling is to the symmetric square of the conjugate antisymmetric and to the product of the conjugate three index antisymmetric and an anti-fundamental. For $N=8$, it is a real representation, so all these couplings are also possible for the non-conjugate representation. Also the antisymmetric square can be coupled to the adjoint, though this won't play a role here. For $N=9$, we can also couple its symmetric square to the fundamental. In that case it can also be coupled to the product of the two and three index antisymmetric representations, though this also won't play a role here.

We list the many theories which are anomaly free and have vanishing one loop beta function in Appendix \ref{appendix:exotics}.
All that is now left is to find out which cases support a conformal manifold. We list the cases where this is so, together with some of their properties, in tables \ref{SUEx1} and \ref{SUEx2}.

\begin{table}[h!]
\begin{center}
\begin{adjustwidth}{-0.1cm}{}
\begin{tabular}{|c|c|c|c|c|c|}
  \hline 
   &  Matter & $dim\, {\cal M}$ & $G_F^{free}$ & $G^{gen}_F$  & $a$, $c$  \\
 \hline
$1$&  $G=SU(4),$ & $1$ & $U(1)^4\times SU(2)$ & $U(1)^2\times SU(2)$  & $a=\frac{61}{16},$ \\
&	$N_{\bold{20}}=1, N_{\overline{S}}=1,$ &  &  & & $c=\frac{31}{8}$ \\
& $ N_{AS}=1,$ & & & & \\
&	$N_{\overline{F}}=1, N_{F}=2$ &  &  & & \\
 \hline
 $2$&$G=SU(4),$ & $2$ & $U(1)$ & $\emptyset$  & $a=\frac{85}{24},$ \\
&	$N_{\bold{20'}}=1, N_{Ad}=1$ &  &  & & $c=\frac{10}{3}$ \\
 \hline
$3$&$G=SU(4),$ & $1$ & $U(1)\times SU(4)$ & $SU(2)^2$  & $a=\frac{179}{48},$ \\
&	$N_{\bold{20'}}=1, N_{AS}=4$ &  &  & & $c=\frac{89}{24}$ \\
 \hline
$4$&$G=SU(4),$ & $2$ & $U(1)^3\times SU(2)^3$ & $SU(2)^2$  & $a=\frac{61}{16},$ \\
&	$N_{\bold{20'}}=1, N_{AS}=2,$ &  & & has a $1d$ subspace & $c=\frac{31}{8}$ \\
&	$N_{F}=N_{\overline{F}}=2$ &  &  & preserving $U(1)\times SU(2)^2$ & \\
 \hline
$5$&$G=SU(6),$ & $1$ & $U(1)^3\times SU(2)$ & $SU(2)^2$  & $a=\frac{437}{48},$ \\
&	$N_{\bold{20}}=3,N_{Ad}=1, $ & & $\times SU(3)$ & & $c=\frac{227}{24}$ \\
&	$N_{AS}=1, N_{\overline{F}}=2$ &  &  &  & \\
 \hline
$6$&$G=SU(6),$ & $3$ & $U(1)\times SU(2)^2$ & $\emptyset$  & $a=\frac{425}{48},$ \\
&	$N_{\bold{20}}=2, N_{Ad}=2$ & & & has $1d$ subspace & $c=\frac{215}{24}$ \\
&	 &  &  & preserving $U(1)^2$ & \\
 \hline
$7$&$G=SU(6),$ & $3$ & $U(1)^3\times SU(2)^2$ & $U(1)\times SU(2)^2$  & $a=\frac{37}{4},$ \\
&	$N_{\bold{20}}=2, N_{Ad}=1,$ & & $\times SU(4)$ & has a $1d$ subspace & $c=\frac{39}{4}$ \\
&	$N_{AS}=2, N_{\overline{F}}=4$ &  &  & preserving $U(1)^2\times USp(4)$ & \\
 \hline
$8$&$G=SU(6),$ & $2$ & $U(1)^4\times SU(2)\times$ & $U(1)^2\times SU(2)$  & $a=\frac{151}{16},$ \\
&	$ N_{Ad}=1, N_{\bold{20}}=2,$ & & $SU(3)\times SU(5)$ & has $1d$ subspace & $c=\frac{81}{8}$ \\
&	$N_{AS}=1,$ &  &  & preserving $U(1)^3\times USp(4)$ & \\
& $N_{F}=3, N_{\overline{F}}=5$ &  &  & also has $1d$ subspace & \\
& &  &  & preserving $U(1)^{2}$ & \\
& &  &  & $\times SU(2)\times SU(3)$ & \\
 \hline
\end{tabular}
\end{adjustwidth}
 \end{center}
\caption{} 
\label{SUEx1}
\end{table}

\begin{table}[h!]
\begin{center}
\begin{adjustwidth}{-0.7cm}{}
\begin{tabular}{|c|c|c|c|c|c|}
  \hline 
   &  Matter & $dim\, {\cal M}$ & $G_F^{free}$ & $G^{gen}_F$  & $a$, $c$  \\
 \hline
$9$&$G=SU(6),$ & $2$ & $U(1)^3\times SU(2)^2$ & $SU(2)$  & $a=9,$ \\
&	$N_{\bold{20}}=1, N_{Ad}=2,$ & & & & $c=\frac{37}{4}$ \\
&	$N_{AS}=1, N_{\overline{F}}=2$ &  &  & & \\
 \hline
$10$&$G=SU(6)$ & $5$ & $U(1)^3\times SU(2)$ & $U(1)^3$ & $a=\frac{147}{16},$ \\
&	$N_{\bold{20}}=1, N_{Ad}=2$ & & $\times SU(3)^2$ & has $1d$ subspace preserving & $c=\frac{77}{8}$ \\
&	$N_{F}=N_{\overline{F}}=3$ &  &  & $U(1)^2\times SU(3)$ & \\
 \hline
$11$& $G=SU(6)$ & $12$ & $U(1)^3\times SU(3)$ & $U(1)$  & $a=\frac{451}{48},$ \\
&	$N_{\bold{20}}=1, N_{Ad}=1$ & & $\times SU(6)$ & has $1d$ subspace preserving & $c=\frac{241}{24}$ \\
&	$N_{AS}=3, N_{\overline{F}}=6$ &  &  & $U(1)\times USp(6)$ & \\
 \hline
$12$ & $G=SU(6), N_{Ad}=1,$ & $1$ & $U(1)^4\times SU(6)$ & $U(1)\times SU(2)$  & $a=\frac{469}{48},$ \\
&	$N_{\bold{20}}=1, N_{AS}=1,$ & & $\times SU(8)$ & $\times SU(6)$ & $c=\frac{259}{24}$ \\
&	$N_{F}=6, N_{\overline{F}}=8$ &  &  & & \\
 \hline
$13$&$G=SU(6),$ & $12$ & $U(1)^5\times SU(2)^2$ & $\emptyset$  & $a=\frac{451}{48},$ \\
&	$N_{Ad}=1, N_{\bold{20}}=1,$ & & $\times SU(4)$ & has $1d$ subspace preserving & $c=\frac{241}{24}$ \\
&	$N_{AS}=1, N_{\overline{AS}}=2,$ &  &  & $U(1)\times SU(2)\times USp(4)$ & \\
& $N_{F}=4, N_{\overline{F}}=2$ &  &  &  & \\
 \hline
$14$&$G=SU(6), N_{Ad}=1,$ & $7$ & $U(1)^4\times SU(2)\times$ & $U(1)$  & $a=\frac{115}{12},$ \\
&	$N_{\bold{20}}=1, N_{\overline{AS}}=2,$ & & $SU(3)\times SU(7)$ & has $1d$ subspace preserving & $c=\frac{125}{12}$ \\
&	$N_{F}=7, N_{\overline{F}}=3$ &  &  & $U(1)^2\times USp(6)$ & \\
&  &  &  & also has $1d$ subspace & \\
&  &  &  & preserving $U(1)^{2}$ & \\
&  &  &  & $\times SU(3)\times USp(4)$ & \\
 \hline
$15$&$G=SU(6), N_{\bold{20}}=1,$ & $3$ & $U(1)^4\times SU(3)^4$ & $U(1)$ & $a=\frac{461}{48},$ \\
&	$N_{AS}=N_{\overline{AS}}=3,$ & & & has $1d$ subspace preserving & $c=\frac{251}{24}$ \\
&	$N_{F}=N_{\overline{F}}=3$ &  &  & $U(1)^5$ & \\
 \hline  
\end{tabular}
\end{adjustwidth}
 \end{center}
\caption{} 
\label{SUEx2}
\end{table}

Let us make several comments about the cases in Tables \ref{SUEx1} and \ref{SUEx2}.

\begin{itemize}
\item 
In case 1 only two of the intrinsic $U(1)$ groups are broken. 
\item In case 3 the unbroken symmetry is embedded as $SO(4)\subset SU(4)$.
\item  In case 4 the unbroken $SU(2)$ symmetries are the ones rotating the fundamentals and antifundamentals. On a $1d$ subspace, a combination of the Cartan of $SU(2)_{AS}$ and one of the intrinsic $U(1)$ groups also remains unbroken.
\item  In case 5 one of the preserved $SU(2)$ groups is $SU(2)_{\overline{F}}$ while the other is $SO(3)\subset SU(3)$.
\item  In case 6 one of the preserved $U(1)$ groups is the Cartan of $SU(2)_{\bold{20}}$ while the other is a combination of the intrinsic $U(1)$ and the Cartan of the other $SU(2)$. 
\item In case 7 the preserved symmetries are embedded as: $SO(4)\subset SO(5)\subset SO(6)=SU(4)$. The always preserved $U(1)$ is the Cartan of $SU(2)_{\bold{20}}$, while the one preserved on a subspace is a combination of the Cartan of $SU(2)_{AS}$ and one of the intrinsic $U(1)$ groups.
\item  In case 8 the $USp(4)$ symmetry in the first $1d$ subspace is embedded as $U(1)\times USp(4)\subset SU(5)$. The $SU(2)$ symmetry in the second $1d$ subspace is embedded as $SU(3)\times SU(2) \times U(1)\subset SU(5)$, and the $SU(3)$ is the diagonal of the intrinsic $SU(3)$ and the one embedded in $SU(5)$.
\item In case 9 the preserved $SU(2)$ is $SU(2)_{\overline{F}}$.
\item In case 10 the preserved $SU(3)$ is the diagonal one.
\item In case 11 the preserved $USp(6)$ is embedded as $USp(6)\subset SU(6)$.
\item In case 12 one breaks $SU(8)\rightarrow U(1)\times SU(2)\times SU(6)$ where the preserved $SU(6)$ is the diagonal one.
\item In case 13 the $SU(2)$ groups is $SU(2)_{\overline{F}}$ while the $USp(4)$ group is embedded as $USp(4)\subset SU(4)$. This case also has a $1d$ subspace where one preserves $U(1)^2\times SU(2)$, where now the $SU(2)$ is one of the two $SU(2)$ groups in $SU(4)$ under the same embedding.
\item In case 14 the $USp(6)$ in the first $1d$ subspace is embedded as $USp(6)\times U(1)\subset SU(7)$. In the second $1d$ subspace the $USp(4)$ groups is embedded as $USp(4)\times SU(3)\times U(1)\subset SU(7)$ where the preserved $SU(3)$ is the diagonal one.
\item In case 15 the $SU(3)_F$ and $SU(3)_{AS}$ are reduced to the diagonal $SU(3)$ and the same for the conjugate $SU(3)$'s. The $U(1)$'s contained in the $1d$ subspace are the Cartans of the two diagonal $SU(3)$'s and one of the intrinsic ones. The same intrinsic $U(1)$ is the one generally preserved.
\end{itemize}

\section{$G=USp(2N)$}\label{sec:symplectic}

We shall next consider the case of group $G=USp(2N)$.

\subsection{Generic cases}

Like for $SU$ type groups, for generic $N$, the condition $\ref{BetaZero}$ can only be satisfied by representations containing at most two indices. These are the symmetric (adjoint), antisymmetric and fundamental. We have summarized the relevant group theory data on these in table \ref{USpGenRep}. All of these representations are real or pseudo-real and so there is no cubic anomaly constraint. For sufficiently low $N$, some other representations are possible. We shall refer to the cases using them as exotic cases, and discuss them in the next section. Cases made from the above mentioned representations will be dubbed generic cases, and considered in this section, even if some of them are only possible for small $N$.

\begin{table}[h!]
\begin{center}
\begin{tabular}{|c|c|c|c|c|}
  \hline 
   Symbol & Label & Dimension& Dynkin index  & Reality  \\
 \hline
 $F$  & $(1,0,0,...,0)$ & $\bold{2N}$  & $\frac{1}{2}$ & Pseudo-real \\
\hline
 $AS$ & $(0,1,0,...,0)$  & $\bold{N(2N-1)-1}$ & $N-1$ & Real \\
\hline
 $S$  & $(2,0,0,...,0)$ & $\bold{N(2N+1)}$ & $N+1$ & Real \\
\hline
 \end{tabular}
 \end{center}
\caption{Various group theory data for $USp(2N)$ representations associated with tensors with at most two indices. Here the label entry stands for the standard Dynkin label of the representation, and the symbol entry provides the shorthand symbol that will be employed for that representation throughout this article. Additionally, the dimension, Dynkin index and the reality properties of the representation are listed.} 
\label{USpGenRep}
\end{table}

We next wish to discuss the possible superpotential terms one can add. Groups of type $USp$ do not have a cubic Casimir, and so there is no cubic symmetric superpotential term for adjoints. There are, however, cubic superpotentials coupling the symmetric square of the adjoints to the antisymmetric, and the anti-symmetric square of the antisymmetrics to the adjoint. Also, with the exception of $USp(4)$, antisymmetrics do have a cubic symmetric superpotential, and as such all conformal cases involve antisymmetrics. Additionally, the adjoint and antisymmetric representations can couple to two fundamentals in a symmetric or antisymmetric manner, respectively. 

The only exceptional case is $N=2$, where there is no cubic symmetric superpotential term for the antisymmetric representation. As a result these cases are usually not conformal. 

We next examine the possible cases. We begin by listing the possible solutions to conditions (\ref{BetaZero}) for generic $N$:

\begin{enumerate}

\item - $N_{S}=2, \quad N_F=2N+2$

\item - $N_{S}=2, \quad N_{AS}=1, \quad N_F=4$

\item - $N_{S}=1, \quad N_{AS}=1, \quad N_F=2N+6$

\item - $N_{AS}=3, \quad N_F=12$

\item - $N_{AS}=2, \quad N_F=2N+10$ 

\item - $N_{AS}=1, \quad N_F=4N+8$

\item - $N_F=6N+6$  

\end{enumerate}

Besides these, there are several solutions that exist only for small $N$:

\begin{enumerate}

\item - $N_{S}=2, \quad N_{AS}=3, \quad N=2$

\item - $N_{S}=2, \quad N_{AS}=2, \quad N_F=2(3-N), \quad N=2,3$

\item - $N_{S}=1, \quad N_{AS}=5, \quad N_F=2, \quad N=2$

\item - $N_{S}=1, \quad N_{AS}=3, \quad N_F=2(5-N), \quad N=2,3,4,5$

\item - $N_{AS}=9, \quad N=2$

\item - $N_{AS}=8, \quad N_F=2, \quad N=2$

\item - $N_{AS}=7, \quad N_F=4, \quad N=2$

\item - $N_{AS}=6, \quad N_F=6(3-N), \quad N=2,3$

\item - $N_{AS}=5, \quad N_F=4(4-N), \quad N=2,3,4$

\item - $N_{AS}=4, \quad N_F=2(7-N), \quad N=2,3,4,5,6,7$

\end{enumerate}

\begin{table}[h]
\begin{center}
\begin{adjustwidth}{-0.5cm}{}
\begin{tabular}{|c|c|c|c|c|c|}
  \hline 
 &  Matter & $dim\, {\cal M}$ & $G_F^{free}$ & $G^{gen}_F$  & $a$, $c$  \\
 \hline
$1$& $N_{S}=1, N_{AS}=1,$  & $N+4$ & $U(1)^2\times  $ & $U(1)^{N+3}$  & $a=\frac{26N^2+21N-1}{48},$ \\
& $N_F=2N+6$ & ($5$ for & $SU(2N+6)$ & for $N>2$ has a $1d$ & $c=\frac{14N^2+15N-1}{24}$ \\
&	&  $N=2$) &  & subspace preserving &  \\
&	&  &  & $SO(2N+6-2x)$ &  \\
&	&  &  & $\times USp(2x)$ for $x<N-1$ &  \\
&		&  &  & for $N\geq 2$ has a $1d$ &  \\
&	&  &  & subspace preserving $U(1)$ &  \\
&	&  &  & $ \times USp(2N-2) \times SO(8)$ &  \\
 \hline
$2$&$N_{S}=2, N_{AS}=3,$  & $1$ & $U(1)\times SU(2) $ & $U(1)\times SU(2)$  & $a=\frac{125}{48},$ \\
 &$N=2$ & & $\times SU(3)$ &  & $c=\frac{65}{24}$ \\
 \hline
$3$&$N_{S}=2, N_{AS}=2,$  & $1$ & $U(1)\times SU(2)^2 $ & $U(1)^2$  & $a=\frac{259}{48},$ \\
 &$N=3$ & & &  & $c=\frac{133}{24}$ \\
 \hline
$4$&$N_{S}=1, N_{AS}=5,$  & $1$ & $U(1)^2\times SU(2) $ & $U(1)\times SU(2)$  & $a=\frac{133}{48},$ \\
 &$N_{F}=2, N=2$ & & $\times SU(5)$ & $\times USp(4)$ & $c=\frac{73}{24}$ \\
 \hline
$5$&$N_{S}=1, N_{AS}=3,$  & $7$ & $U(1)\times SU(3) $ & $\emptyset$  & $a=\frac{341}{24},$ \\
 &$N=5$ & & & has a $1d$ subspace & $c=\frac{44}{3}$ \\
&&  &  & preserving $SU(2)$ &  \\
 \hline
$6$&$N_{S}=1, N_{AS}=3,$  & $10$ & $U(1)^2\times SU(2) $ & $U(1)$  & $a=\frac{457}{48},$ \\
 &$N_{F}=2, N=4$ & & $\times SU(3)$ & has a $1d$ subspace & $c=\frac{241}{24}$ \\
&&  &  & preserving $SU(2)^2$ &  \\
 \hline
$7$&$N_{S}=1, N_{AS}=3,$  & $19$ & $U(1)^2\times SU(3) $ & $\emptyset$  & $a=\frac{23}{4},$ \\
 &$N_{F}=4, N=3$ & & $\times SU(4)$ & has a $1d$ subspace & $c=\frac{25}{4}$ \\
&&  &  & preserving $U(1)\times SU(2)$ &  \\
&&  &  & $\times USp(4)$ &  \\
\hline
$8$&$N_{S}=1, N_{AS}=3,$  & $27$ & $U(1)^2\times SU(3) $ & $\emptyset$  & $a=\frac{139}{48},$ \\
 &$N_{F}=6, N=2$ & & $\times SU(6)$ & has a $1d$ subspace & $c=\frac{79}{24}$ \\
&&  &  & preserving $U(1)\times SU(2)^4$ &  \\
\hline
\end{tabular}
\end{adjustwidth}
 \end{center}
\caption{} 
\label{USpGen1}
\end{table}

Finally, we need to consider the possible superpotentials and perform the K$\ddot{a}$hler quotient. There are several different cases where this is non-trivial so we shall split them to two groups. The cases involving symmetric chirals appear in table \ref{USpGen1}, while those without appear in tables \ref{USpGen2} and \ref{USpGen3}.

Let us make several comments about the cases in Table \ref{USpGen1}.

\begin{itemize}

\item Case 1 is pretty involved. Besides the $1d$ subspace preserving $SO(2N+6)$, for $N\neq 2$, there is also a $1d$ subspace preserving $USp(2x)\times SO(2N+6-2x)$ for $x\leq N-1$, where when the inequality is saturated, one preserves an extra $U(1)$ on this $1d$ subspace. For $N=2$, we are missing one operator and so only the case where the inequality is saturated is possible. The symmetries then are embedded as $USp(2x)\times SO(2N+6-2x) \subset SU(2x)\times SU(2N+6-2x)\times U(1) \subset SU(10)$. The $U(1)$ that is preserved on the $x=N-1$ subspace is a combination of the commutant and intrinsic $U(1)$ groups. The Cartans of $USp(2x)\times SO(2N+6-2x)$ cannot be broken on the conformal manifold. 

\item In case 2 the preserved $SU(2)$ is the diagonal of $SU(2)_S$ and $SO(3)\subset SU(3)_{AS}$. The $U(1)$ cannot be broken. 

\item In case 3 the $U(1)$ and a combination of the Cartans of the $SU(2)$ groups is preserved. 
\item In case 4 the breaking goes as $USp(4)\subset U(1)\times SU(4)\subset SU(5)$. The $USp(4)$, $SU(2)$ and some combination of the commutant and intrinsic $U(1)$ groups are preserved.
\item  In case 5 the embedding is $SU(2)\subset SU(2)\times U(1) \subset SU(3)$.
\item  In case 6 the preserved $SU(2)$ groups are the one rotating the fundamentals and the $SU(2)\subset SU(2)\times U(1) \subset SU(3)$. The Cartan of the $SU(2)$ rotating the fundamentals cannot be broken. 
\item In case 7 the symmetries are embedded as follows $USp(4)\subset SU(4)$, $SU(2)\subset SU(2)\times U(1) \subset SU(3)$. There is also a $1d$ subspace preserving the $SO(4)\subset SU(4)$, $SU(2)\subset SU(2)\times U(1) \subset SU(3)$ symmetry. 
\item In case 8 the symmetries are embedded as $USp(2)\times SO(4)\subset SU(6)$ and $SU(2)\subset SU(2)\times U(1) \subset SU(3)$. The preserved $U(1)$ is a combination of the commutant and intrinsic $U(1)$ groups.
\end{itemize}

\begin{table}[h]
\begin{center}
\begin{adjustwidth}{-0.8cm}{}
\begin{tabular}{|c|c|c|c|c|c|}
  \hline 
 &  Matter & $dim\, {\cal M}$ & $G_F^{free}$ & $G^{gen}_F$  & $a$, $c$  \\
 \hline
$9$& $N_{AS}=3,$  & $56$ & $U(1)\times SU(3) $ & $\emptyset$ ($U(1)$ for $N=2$) & $a=\frac{8N^2+10N-1}{16},$ \\
& $N_F=12$ & ($47$ for & $\times SU(12) $ & for $N>4$ has a $4d$ subspace  & $c=\frac{4N^2+8N-1}{8}$ \\
&	&  $N=2$) &  & preserving $USp(12)$ &  \\
&	&  &  & for $N=4$ has a $1d$ subspace &  \\
&	&  &  & preserving $U(1)\times USp(12)$ &  \\
&	&  &  & for $N\geq 3$ has a $1d$ subspace &  \\
&	&  &  & preserving $U(1)\times USp(6)^2$ &  \\
&	&  &  & for $N= 3$ has a $1d$ &  \\
&	&  &  & subspace preserving $U(1)^2$ &  \\
&	&  &  & $\times USp(4)\times USp(8)$ &  \\
&		&  &  & for $N\geq 2$ has a $1d$ subspace &  \\
&	&  &  & preserving $USp(4)^3\times U(1)^2$ &  \\
&	&  &  & ($USp(4)^3\times U(1)^3$ for $N=2$) &  \\
 \hline
$10$& $N_{AS}=2, $ & $N+6$ & $U(1)\times SU(2) $ & $SU(2)^{N+5}$  & $a=\frac{26N^2+27N-2}{48},$ \\
 &$N_F=2N+10,$ & ($6$ for & $\times SU(2N+10) $ & ($U(1)\times SU(2)^{8}$ for $N=3$)  & $c=\frac{14N^2+21N-2}{24}$ \\
 &$N\neq 2$ &  $N=3$) &  & for $N>7$ has a $2d$ subspace &  \\
&	&  &  & preserving $USp(2N+10)$ &  \\
&	&  &  & for $N=7$ has a $1d$ &  \\
&	&  &  & subspace preserving &  \\
&	&  &  & $U(1)\times USp(2N+10)$ &  \\
&	&  &  & for $N=4,5,6$ has a $3d$ &  \\
&	&  &  & subspace preserving &  \\
&	&  &  & two $USp$ subgroups &  \\
&	&  &  & for $N>3$ has a $1d$ &  \\
&	&  &  & subspace preserving $U(1)$ &  \\
&	&  &  & $\times USp(8)\times USp(2N+2)$ &  \\
&	&  &  & for $N=3$ has a $1d$ subspace &  \\
&	&  &  & preserving $U(1)^2\times USp(8)^2$ &  \\
	\hline
\end{tabular}
\end{adjustwidth}
 \end{center}
\caption{} 
\label{USpGen2}
\end{table}

Let us make several comments about the cases in Tables \ref{USpGen2}.

\begin{itemize}

\item In case 9 for $N\geq 4$ you can preserve the $USp(12)\subset SU(12)$, where when the inequality is saturated an additional combination of the intrinsic $U(1)$ and the Cartans of $SU(3)$ can be preserved. For $N\geq 3$ you can preserve the $USp(6)^2\subset SU(12)$ and a $U(1)$ that is a combination of the commutant and intrinsic $U(1)$ groups. For $N\geq 2$ you can preserve the $USp(4)^3\subset SU(12)$ and a $U(1)^2$ that are combinations of the commutant and intrinsic $U(1)$ groups. For $N=3$ one can preserve $ USp(8) \times USp(4)\subset \SU(12)$ and a $U(1)^2$ that are combinations of the commutant and intrinsic $U(1)$ groups. The $N=2$ case is special in that there are less superpotentials, but they are all uncharged under the intrinsic $U(1)$. As a result the conformal manifold is smaller in this case, and an additional $U(1)$ is always preserved with respect to the $N>2$ cases. The $N=2$ case also has a $1d$ subspace preserving $U(1)\times SU(3)\times SU(2)^2$ with the symmetry embedded as $SU(3)\times SO(4) \subset SU(3)\times SU(4) \subset SU(12)$ such that $\bold{12}\rightarrow (\bold{3},\bold{2},\bold{2})$, with the preserved $SU(3)$ the diagonal of this and the one acting on the antisymmetrics. The $N=2$ case also has a $1d$ subspace preserving $U(1)\times SU(2)\times USp(6)$ with the symmetry embedded as $SU(2)\times USp(6) \subset SO(12) \subset SU(12)$ such that $\bold{12}\rightarrow (\bold{2},\bold{6})$, with the preserved $SU(2)$ being the diagonal of this and the one coming from the breaking $SU(3) \to SO(3)$ of the $SU(3)$ acting on the antisymmetrics.

\item Case 10 behaves similarly to case 9. For $N\geq 7$ one can always preserve the $USp(2N+10)\subset SU(2N+10)$, where when the inequality is saturated, one can also preserve an additional $U(1)$ which is a combination of the Cartan of the $SU(2)$ and the intrinsic $U(1)$. For $N>2$ one can preserve two $USp$ subgroups inside $SU(2N+10)$. In any case the $SU(2)^{N+5}\subset USp(2N+10)\subset SU(2N+10)$ is always preserved. For $N>3$ one can preserve $USp(8)\times USp(2N+2)\subset SU(2N+10)$ and a $U(1)$ that is a combination of the commutant and intrinsic $U(1)$ groups. For $N=3$ the superpotential coupling the antisymmetrics and two flavors is not charged under the $U(1)$. As a result, the cubic one coupling the antisymmetrics is marginally irrelevant and there are less superpotentials, and the $U(1)$ is never broken on the conformal manifold. Here there is a $1d$ subspace preserving the $USp(8)^2\subset U(1)\times SU(8)^2 \subset SU(2N+10)$, and an additional $U(1)$ which is a combination of the Cartan of the $SU(2)$ and the commutant $U(1)$.

\end{itemize}

\begin{table}[h]
\begin{center}
\begin{tabular}{|c|c|c|c|c|c|}
  \hline 
 &  Matter & $dim\, {\cal M}$ & $G_F^{free}$ & $G^{gen}_F$  & $a$, $c$  \\
 \hline
$11$&	$N_{AS}=6,$  & $21$ & $SU(6) $ & $\emptyset$  & $a=\frac{91}{16},$ \\
&	$N=3$  & &  & has $1d$ subspace & $c=\frac{49}{8}$ \\
&  & &  & preserving $SU(3)$ &  \\
	\hline
$12$&	$N_{AS}=5,$  & $11$ & $SU(5) $ & $\emptyset$  & $a=\frac{153}{16},$ \\
&	$N=4$  & &  & has $1d$ subspace & $c=\frac{81}{8}$ \\
&  & &  & preserving $SU(2)$ &  \\
	\hline
$13$&	$N_{AS}=5,$  & $25$ & $U(1)\times SU(4)$ & $\emptyset$  & $a=\frac{283}{48},$ \\
&	$N_{F}=4,$  & & $\times SU(5)$ & has a $1d$ subspace  & $c=\frac{157}{24}$ \\
&$N=3$	&  &  & preserving $U(1)^3 \times USp(4)$ &  \\
	\hline
$14$&	$N_{AS}=4,$  & $5$ & $SU(4)$ & $\emptyset$  & $a=\frac{435}{16},$ \\
&	$N=7$  & & & has a $1d$ subspace & $c=\frac{225}{8}$ \\
&  &  &  & preserving $U(1)^{2}$ &  \\
	\hline
$15$&	$N_{AS}=4,$  & $8$ & $U(1)\times SU(2)$ & $SU(2)$  & $a=\frac{493}{24},$ \\
&	$N_F=2, \; N=6$  & & $\times SU(4)$ & has a $1d$ subspace & $c=\frac{259}{12}$ \\
&  &  &  & preserving $U(1)^{2}\times SU(2)$ &  \\
	\hline
$16$&	$N_{AS}=4,$  & $14$ & $U(1)\times SU(4)^2$ & $U(1)$  & $a=\frac{237}{16},$ \\
&	$N_F=4$,  & & & has a $1d$ subspace & $c=\frac{127}{8}$ \\
&	$N=5$  & & & preserving $U(1)^{2}\times USp(4)$ & \\
	\hline
$17$&	$N_{AS}=4,$  & $29$ & $U(1)\times SU(4)$ & $\emptyset$  & $a=10,$ \\
&	$N_F=6, \; N=4$  & & $\times SU(6)$ & has $1d$ subspace & $c=11$ \\
&  & &  & preserving $U(1)^{2}\times USp(6)$ &  \\
	\hline
$18$&	$N_{AS}=4,$  & $53$ & $U(1)\times SU(4)$ & $\emptyset$  & $a=\frac{293}{48},$ \\
&	$N_F=8, \; N=3$  & & $\times SU(8)$ & has $1d$ subspace  & $c=\frac{167}{24}$ \\
&	  & & & preserving $U(1)^3\times USp(8)$ & \\
	\hline
\end{tabular}
 \end{center}
\caption{} 
\label{USpGen3}
\end{table}

Let us make several comments about the cases in Tables \ref{USpGen3}.

\begin{itemize}
\item In case 11 the embedding is $SU(3)\subset SU(6)$ such that $\bold{6}\rightarrow \bold{6}$.
\item  In case 12 the embedding is $SU(2)\subset SU(5)$ such that $\bold{5}\rightarrow \bold{5}$. 
\item In case 13 the symmetry is embedded as $USp(4)\subset SU(4)$. The $U(1)^3$ are a combination of the $SU(5)$ Cartans and the intrinsic $U(1)$.
\item In case 14 the $SU(4)$ is broken to three $U(1)$'s and the preserved $U(1)^2$ are two of these. 
\item In case 15 the $SU(2)$ is the one rotating the fundamentals. The $SU(4)$ is broken to three $U(1)$'s and the preserved $U(1)^2$ are two of these.
\item In case 16 the symmetry is embedded as $SO(2)\subset SO(3)\subset SO(4)\subset SO(5)=USp(4)\subset SU(4)$. The $SU(4)$ is broken to three $U(1)$'s and the preserved $U(1)^2$ are two of these.
\item In case 17 the symmetry is embedded as $USp(6)\subset SU(6)$. The $SU(4)$ is broken to three $U(1)$'s and the preserved $U(1)^2$ are two of these.
\item  In case 18 the symmetry breaking pattern is as follows: $USp(8)\subset SU(8)$, $U(1)^2 \subset SU(3)\subset U(1)\times SU(3)\subset SU(4)$ plus an additional $U(1)$ that is a combination of the commutant and intrinsic ones.
\end{itemize}

\subsection{Exotic cases}

Finally we consider representations which are only possible for low values of $N$. The full list of possibilities appear in table \ref{USpExRep}. As can be seen, there are only a handful of cases extending up to $USp(10)$. 

\begin{table}[h!]
\begin{center}
\begin{tabular}{|c|c|c|c|c|}
  \hline 
  Group & Label & Dimension & Dynkin index  & Reality  \\
 \hline
$USp(4)$ & $(0,2)$  & $\bold{14}$  & $7$ & Real \\
\hline
$USp(4)$ & $(1,1)$  & $\bold{16}$  & $6$ & Pseudo-real \\
\hline
\hline
 $USp(6)$ & $(0,0,1)$  & $\bold{14'}$  & $\frac{5}{2}$ & Pseudo-real \\
\hline
\hline
 $USp(8)$ & $(0,0,1,0)$  & $\bold{48}$ & $7$ & Pseudo-real \\
\hline
$USp(8)$ & $(0,0,0,1)$  & $\bold{42}$ & $7$ & Real \\
\hline
\hline
$USp(10)$ & $(0,0,1,0,0)$  & $\bold{110}$ & $\frac{27}{2}$ & Pseudo-real \\
\hline
 \end{tabular}
 \end{center}
\caption{Various group theory data for $USp(2N)$ representations associated with tensors with more than two indices. Here the group entry stands for the group in question, and the label entry stands for the standard Dynkin label of the representation. The remaining entries list the dimension, Dynkin index (AKA quadratic Casimir) and the reality properties of each representation.} 
\label{USpExRep}
\end{table}

We first review the list of possible superpotentials. We begin with the case of $USp(4)$. Here the two additional representations have no analogues for higher $N$. The $\bold{14}$ has a cubic symmetric invariant, which proves useful for getting conformal theories. Additionally, it by definition can be coupled to the symmetric product of two antisymmetrics, but also can couple to the symmetric product of two symmetrics. Finally, the antisymmetric product can couple to the symmetric as it is a real representation.   

The $\bold{16}$, however, has no cubic symmetric invariant. By definition, it can be coupled to a fundamental and an antisymmetric, and its symmetric square can be coupled to a symmetric as it is a pseudo-real representation. Additionally, it can also be coupled to a fundamental and a symmetric, and its anti-symmetric square can be coupled to an antisymmetric.

For $USp(6)$, $USp(8)$ and $USp(10)$ we have the thee-index traceless anti-symmetric representation. This representation has only a limited number of possible cubic superpotentials. The basic ones are the ones coupling it to a a fundamental and an antisymmetric, and the one coupling its symmetric square to a symmetric, owning to its pseudo-reality. For rank $4$ and above, its antisymmetric square can be coupled to an antisymmetric.

The only remaining representation is the four-index traceless anti-symmetric representation of $USp(8)$. By definition, it can couple to the symmetric square of the antisymmetric and to a fundamental and a thee-index anti-symmetric. As it is a real representation, its anti-symmetric square can be coupled to a symmetric, though that will not be of use here.

We list the exotic cases solving the condition \eqref{BetaZero} in Appendix \ref{appendix:spexotics}.
All that is now left is to find out which cases support a conformal manifold. We list the cases where this is so, together with some of their properties, in table \ref{USpEx}.

\begin{table}[h!]
\begin{center}
\begin{tabular}{|c|c|c|c|c|c|}
  \hline 
 &  Matter & $dim\, {\cal M}$ & $G_F^{free}$ & $G^{gen}_F$  & $a$, $c$  \\
 \hline
$1$&  $G=USp(4)$ & $1$ & $U(1)\times SU(2) $ & $U(1)$  & $a=\frac{19}{8},$ \\
&	$N_{\bold{14}}=1, \; N_{AS}=2$ &  &  & & $c=\frac{9}{4}$ \\
 \hline
 $2$&$G=USp(4)$ & $1$ & $U(1)^2\times SU(2) $ & $SU(2)$  & $a=\frac{39}{16},$ \\
&	$N_{\bold{14}}=1, \; N_{AS}=1,$ &  &  & & $c=\frac{19}{8}$ \\
&	$N_{F} = 2$ &  &  & &  \\
 \hline
$3$&$G=USp(6)$ & $4$ & $U(1)^3\times SU(2)^2 $ & $\emptyset$  & $a=\frac{11}{2},$ \\
&	$N_{\bold{14'}}=2, \; N_{S}=1, $ &  &  & has a $1d$ subspace & $c=\frac{23}{4}$ \\
&$ N_{AS}=1, \; N_{F} = 2$	&  &  & preserving $U(1)\times SU(2)$ &  \\
 \hline
$4$&$G=USp(6)$ & $8$ & $U(1)^3\times SU(7) $ & $\emptyset$  & $a=\frac{35}{6},$ \\
&	$N_{\bold{14'}}=1, \; N_{S}=1, $ &  &  & has a $1d$ subspace & $c=\frac{77}{12}$ \\
&$ N_{AS}=1, \; N_{F} = 7$	&  &  & preserving $SO(7)$ &  \\
 \hline
$5$&$G=USp(8)$ & $1$ & $U(1)^2$ & $\emptyset$  & $a=\frac{145}{16},$ \\
&	$N_{\bold{48}}=1, \; N_{S}=1, $ &  &  & & $c=\frac{73}{8}$ \\
&$ N_{AS}=1$	&  &  &  &  \\
 \hline
\end{tabular}
 \end{center}
\caption{} 
\label{USpEx}
\end{table}

Let us make some comments about Table \ref{USpEx}.

\begin{itemize}
\item For case 3 the $SU(2)$ is the one rotating the fundamentals. 
\item For case 4 there is also a $2d$ subspace preserving $USp(6)$.
\end{itemize}

\section{$G=SO(N)$}\label{sec:orthogonal}

We shall next consider the case of group $G=SO(N)$.

\subsection{Generic cases}

Like the previous cases, for generic $N$, the condition $\ref{BetaZero}$ can only be satisfied by representations containing at most two indices. These are the symmetric, antisymmetric (adjoint) and fundamental. We have summarized the relevant group theory data on these in table \ref{SOGenRep}. All of these representations are real, so there is no cubic anomaly constraint. For sufficiently low $N$, some other representations are possible. We shall refer to the cases using them as exotic cases, and discuss them in the next section. Here we only consider cases with $N>6$. Cases with lower $N$, are locally identical to same of the previously discussed groups and so were already considered there, notably, $SO(5) = USp(4)$ and $SO(6) = SU(4)$.

\begin{table}[h!]
\begin{center}
\begin{tabular}{|c|c|c|c|c|}
  \hline 
   Symbol & Label & Dimension & Dynkin index  & Reality  \\
 \hline
 $V$  & $(1,0,0,...,0)$ & $\bold{N}$  & $1$ & Real \\
\hline
 $AS$ & $(0,1,0,...,0)$  & $\bold{\frac{N(N-1)}{2}}$ & $N-2$ & Real \\
\hline
 $S$  & $(2,0,0,...,0)$ & $\bold{\frac{N(N+1)}{2}-1}$ & $N+2$ & Real \\
\hline
 \end{tabular}
 \end{center}
\caption{Various group theory data for $SO(N)$ representations associated with tensors with at most two indices. Here the label entry stands for the standard Dynkin label of the representation, and the symbol entry provides the shorthand symbol that will be employed for that representation throughout this article. Additionally, the dimension, Dynkin index and the reality properties of the representation are listed.} 
\label{SOGenRep}
\end{table}

We next wish to discuss the possible superpotential terms one can add. Groups of type $SO$ do not have a cubic Casimir, with the exception of $N=6$, and so there is no cubic symmetric superpotential term for adjoints. There are, however, cubic superpotentials coupling the symmetric square of the adjoints to the symmetric, and the anti-symmetric square of the symmetrics to the adjoint. Also the symmetrics do have a cubic symmetric superpotential, and as such all conformal cases involve symmetrics. Additionally, the adjoint and symmetric representations can couple to two fundamentals in a antisymmetric or symmetric manner, respectively. 

We next examine the possible cases. We begin by listing the possible solutions to conditions (\ref{BetaZero}) for generic $N$:

\begin{enumerate}

\item - $N_{S}=2, \quad N_V=N-10$

\item - $N_{S}=1, \quad N_{AS}=1, \quad N_V=N-6$

\item - $N_{S}=1, \quad N_V=2N-8$

\item - $N_{AS}=2, \quad N_V=N-2$

\item - $N_V=3N-6$  

\end{enumerate}

\begin{table}[h]
\begin{center}
\begin{adjustwidth}{-.3cm}{}
\begin{tabular}{|c|c|c|c|c|c|}
  \hline 
 &  Matter & $dim\, {\cal M}$ & $G_F^{free}$ & $G^{gen}_F$  & $a$, $c$  \\
 \hline
$1$ & $N_{S}=2,$  & $N-9$ & $U(1)\times SU(2)$ & $\emptyset$  & $a=\frac{13N^2-27N-4}{96},$ \\
& $N_V=N-10$ & & $\times SU(N-10)$ & for $N>10$ has a $2d$ & $c=\frac{7N^2-21N-4}{48}$ \\
&	 &  &  & subspace preserving &  \\
&	 &  &  & $SO(N-10)$ &  \\
 \hline
 $2$ & $N_{S}=1, N_{AS}=1,$  & $\lfloor \frac{N-4}{2} \rfloor$ & $U(1)^2\times$ & $U(1)^{\lfloor \frac{N-6}{2} \rfloor}$  & $a=\frac{13N^2-21N-2}{96},$ \\
& $N_V=N-6$ & & $SU(N-6)$ & has a $1d$ subspace & $c=\frac{7N^2-15N-2}{48}$ \\
&	&  &  & preserving $USp(2M)$ &  \\
&	&  &  & $\times SO(N-6-2M)$ &  \\
&	&  &  & for any $0\leq M \leq \lfloor \frac{N-6}{2} \rfloor$ &  \\
 \hline
 $3$ & $N_{S}=1,$  & $1$ & $U(1)\times$ & $SO(2N-8)$  & $a=\frac{7N^2-12N-1}{48},$ \\
&  $N_V=2N-8$ & & $SU(2N-8)$ & & $c=\frac{4N^2-9N-1}{24}$ \\
 \hline
\end{tabular}
\end{adjustwidth}
 \end{center}
\caption{} 
\label{SOGen}
\end{table}

Finally, we need to consider the possible superpotentials and perform the K$\ddot{a}$hler quotient. The results are summarized in table \ref{SOGen}. 

Let us make some comments.

\begin{itemize}
\item In case 2 the $USp(2M)\times SO(N-6-2M)$ preserved symmetries are embedded in $SU(N-6)$, and there is a $1d$ subspace with this breaking for every $0\leq M \leq \lfloor \frac{N-6}{2} \rfloor$. On a generic point, these symmetries are further broken to their Cartan subalgebra.
\end{itemize}

\subsection{Exotic cases}

Finally we consider representations which are only possible for low values of $N$. The full list of possibilities appear in table \ref{SOExRep}. Most of these are spinors which can be added up to $N=18$. For $N<10$, there are also several other representations like three index antisymmetric ones.   

\begin{table}[h!]
\begin{center}
\begin{tabular}{|c|c|c|c|c|}
  \hline 
  Group & Label & Dimension  & Dynkin  & Reality  \\
 \hline
$SO(7)$ & $(0,0,1)$  & $\bold{8}$  & $1$ & Real \\
\hline
$SO(7)$ & $(0,0,2)$  & $\bold{35}$  & $10$ & Real \\
\hline
$SO(7)$ & $(1,0,1)$  & $\bold{48}$  & $14$ & Real \\
\hline
\hline
 $SO(8)$ & $(0,0,1,0), \; (0,0,0,1)$  & $\bold{8}_S, \; \bold{8}_C$  & $1$ & Real \\
\hline
$SO(8)$ & $(0,0,2,0), \; (0,0,0,2)$  & $\bold{35}_S, \; \bold{35}_C$  & $10$ & Real \\
\hline
$SO(8)$ & $(0,0,1,1), \; (1,0,0,1), \; (1,0,1,0)$  & $\bold{56}_V, \; \bold{56}_S, \; \bold{56}_C$  & $15$ & Real \\
\hline
\hline
 $SO(9)$ & $(0,0,0,1)$  & $\bold{16}$ & $2$ & Real \\
\hline
$SO(9)$ & $(0,0,1,0)$  & $\bold{84}$ & $21$ & Real \\
\hline
\hline
$SO(10)$ & $(0,0,0,1,0), \; (0,0,0,0,1)$  & $\bold{16}, \; \bold{\overline{16}}$ & $2$ & Complex \\
\hline
\hline
$SO(11)$ & $(0,0,0,0,1)$  & $\bold{32}$ & $4$ & Pseudo-real \\
\hline
\hline
$SO(12)$ & $(0,0,0,0,1,0), \; (0,0,0,0,0,1)$  & $\bold{32}, \; \bold{32'}$ & $4$ & Pseudo-real \\
\hline
\hline
$SO(13)$ & $(0,0,0,0,0,1)$  & $\bold{64}$ & $8$ & Pseudo-real \\
\hline
\hline
$SO(14)$ & $(0,0,0,0,0,1,0), \; (0,0,0,0,0,0,1)$  & $\bold{64}, \; \bold{\overline{64}}$ & $8$ & Complex \\
\hline
\hline
$SO(15)$ & $(0,0,0,0,0,0,1)$  & $\bold{128}$ & $16$ & Real \\
\hline
\hline
$SO(16)$ & $(0,0,0,0,0,0,1,0), \; (0,0,0,0,0,0,0,1)$  & $\bold{128}, \; \bold{128'}$ & $16$ & Real \\
\hline
\hline
$SO(17)$ & $(0,0,0,0,0,0,0,1)$  & $\bold{256}$ & $32$ & Real \\
\hline
\hline
$SO(18)$ & $(0,0,0,0,0,0,0,1,0),$  & $\bold{256},$ & $32$ & Complex \\
 & $(0,0,0,0,0,0,0,0,1)$  & $\bold{\overline{256}}$ & &  \\
\hline
 \end{tabular}
 \end{center}
\caption{Various group theory date for $SO(N)$ representations associated with spinors and tensors with more than two indices. Here the group entry stands for the group in question, and the label entry stands for the standard Dynkin label of the representation. Additionally, the dimension, Dynkin index and the reality properties of the representation are listed.} 
\label{SOExRep}
\end{table}

We next review the list of possible superpotentials. First, we consider the spinors. The behavior of these states vary depending on $N$ with a periodicity of $8$, and so the discussion on these will be separated to various cases. Nevertheless, there are a few properties that are shared by all cases. First the only possible superpotentials must involve an even number of spinors, and so we are limited to terms quadratic in the spinors. These can then interact with either the vector or antisymmetric representation, but not with the symmetric. The exact manner of interaction depends on $N$ mode $8$, which we next explore. 

We first consider the case of $N$ odd, where there is a single spinor representation. There are then superpotentials coupling two spinors to the adjoint and vector representations, where the symmetry properties of the product varies with $N$. The spinor representation is real for $N=7, 9$ and pseudo-real for $N=11, 13$, where here and in what follows all numbers are mode $8$. As a result, for $N=7, 9$ the antisymmetric spinor product can be coupled to the adjoint while for $N=11, 13$, the symmetric spinor product can. Similarly, the vector coupling is symmetric for $N=9, 11$ and antisymmetric for $N=7, 13$.

Next we turn to the case of even $N$, where there are two spinor representations. If $N=4n+2$, then these are complex representations, and the product of the spinor with the conjugate spinor couples with the adjoint. The square of either of them couples to the vector, where it is symmetric for $N=10$ and antisymmetric for $N=14$. For $N=4n$ there are two self-conjugate spinor representations where the product of one with the other couples to the vector. The square of either of them couples to the adjoint, where it is symmetric for $N=12$ and antisymmetric for $N=8$, as these have different reality conditions.  

For $SO(7)$, $SO(8)$ and $SO(9)$ we also have the three-index anti-symmetric representation, and for $SO(8)$ also the representations related to it via triality. For $N=8,9$ it does not have a symmetric cubic invariant, which combined with its high contribution to the Beta function, impedes a K$\ddot{a}$hler quotient so we need not consider them any further. However, for $SO(7)$, there is a symmetric cubic invariant, making this representation viable. Additionally, it can couple to the symmetric square of both the adjoint and spinor representations, and its antisymmetric square can couple to the adjoint representation.

For $SO(7)$ we also have the representation of dimension $\bold{48}$. However, it does not posses a cubic symmetric invariant, which together with its high Dynkin index, is sufficient to rule it out for our porpuses, redeeming us from the need to consider it further. The only other representation is the self-dual and anti self-dual four-index anti-symmetric representations of $SO(8)$. These are related by triality to the symmetric of $SO(8)$, the $\bold{35}_v$, and so the possible superpotentials can be generated by applying the triality transformation on these cases.

We list all the cases solving the condition \eqref{BetaZero} in Appendix \ref{appendix:soexotics}.
All that is now left is to find out which cases support a conformal manifold. As there are a lot of cases, we have split the results across multiple tables. Specifically, table \ref{SOEx1} shows the cases with $G=SO(7), SO(8)$, table \ref{SOEx2} shows the cases with $G=SO(9)$ and some of the $G=SO(10)$ cases, and table \ref{SOEx3} shows the remaining cases with $G=SO(10)$.

\begin{table}[h!]
\begin{center}
\begin{adjustwidth}{-.3cm}{}
\begin{tabular}{|c|c|c|c|c|c|}
  \hline 
 &  Matter & $dim\, {\cal M}$ & $G_F^{free}$ & $G^{gen}_F$  & $a$, $c$  \\
 \hline
$1$ &  $G=SO(7)$ & $102$ & $U(1)\times SU(5)$ & $U(1)$  & $a=\frac{19}{3},$ \\
&	$N_{\bold{8}}=10, \; N_{V}=5$ &  & $\times SU(10)$ & has a $1d$ subspace & $c=\frac{89}{12}$ \\
&	&  &  & preserving $U(1)^5\times SU(2)^5$ &  \\
 \hline
 $2$ & $G=SO(7), \; N_{S}=1$ & $1$ & $U(1)^2\times SU(2)$ & $U(1)\times SU(2)^2$  & $a=\frac{65}{12},$ \\
&	$N_{\bold{8}}=2, \; N_{V}=4$ &  & $\times SU(4)$ &  & $c=\frac{67}{12}$ \\
 \hline
 $3$ & $G=SO(7), \; N_{S}=1$ & $1$ & $U(1)^2\times SU(3)^2$ & $SU(2)$  & $a=\frac{87}{16},$ \\
&	$N_{\bold{8}}=3, \; N_{V}=3$ &  & & & $c=\frac{45}{8}$ \\
 \hline
$4$ &$G=SO(7), \; N_{S}=1$ & $1$ & $U(1)^2\times SU(2)$ & $U(1)^2\times SU(2)^2$  & $a=\frac{131}{24},$ \\
&	$N_{\bold{8}}=4, \; N_{V}=2$ &  & $\times SU(4)$ & & $c=\frac{17}{3}$ \\
 \hline
$5$ &$G=SO(7), \; N_{AS}=1$ & $1$ & $U(1)$ & $\emptyset$  & $a=\frac{245}{48},$ \\
&	$N_{\bold{35}}=1$ &  & & & $c=\frac{119}{24}$ \\
 \hline
$6$ &$G=SO(7), \; N_{\bold{35}}=1$ & $2$ & $U(1)^2\times SU(4)$ & $U(1)^2$  & $a=\frac{263}{48},$ \\
&	$N_{\bold{8}}=4, \; N_{V}=1$ &  & & has a $1d$ subspace & $c=\frac{137}{24}$ \\
&	&  &  & preserving $U(1)^2\times SU(2)$ &  \\
 \hline
$7$ &$G=SO(7), \; N_{\bold{35}}=1$ & $1$ & $U(1)\times SU(5)$ & $USp(4)$  & $a=\frac{11}{2},$ \\
&	$N_{\bold{8}}=5$ &  & & & $c=\frac{23}{4}$ \\
 \hline
$8$ &$G=SO(8), \; N_{\bold{8}_S}=6$ & $111$ & $U(1)^2\times SU(6)^3$ & $U(1)^2$  & $a=\frac{33}{4},$ \\
&	$N_{\bold{8}_C}=6, \; N_{V}=6$ &  & & has a $1d$ subspace & $c=\frac{19}{2}$ \\
&	&  &  & preserving $U(1)^4\times SU(3)^2$ &  \\
 \hline
$9$ &$G=SO(8),$ & $1$ & $U(1)^2\times SU(2)$ & $SU(2)$  & $a=\frac{331}{48},$ \\
&	$N_{S}=1, \; N_{AS}=1,$ &  & & & $c=\frac{163}{24}$ \\
&	$N_{\bold{8}_S}=2$ &  & & & \\
 \hline
$10$ &$G=SO(8), \; N_{S}=1,$ & $4$ & $U(1)^3\times SU(2)^2$ & $U(1)$  & $a=\frac{117}{16},$ \\
&	$N_{\bold{8}_S}=2, \; N_{\bold{8}_C}=2,$ &  & $\times SU(4)$ & has a $1d$ subspace & $c=\frac{61}{8}$ \\
&	$N_{V}=4$ &  & & preserving $U(1)\times SU(2)^2$ & \\
 \hline
$11$ &$G=SO(8), \; N_{S}=1,$ & $1$ & $U(1)^3\times SU(6)$ & $U(1)^2\times USp(4)$  & $a=\frac{117}{16},$ \\
&	$N_{\bold{8}_S}=1, \; N_{\bold{8}_C}=1,$ &  & & & $c=\frac{61}{8}$ \\
&	$N_{V}=6$ &  & & & \\
 \hline
\end{tabular}
\end{adjustwidth}
 \end{center}
\caption{} 
\label{SOEx1}
\end{table}

Let us make some comments about the cases in Table \ref{SOEx1}.

\begin{itemize}
\item In case 1 the $SU(2)^5$ is embedded in $SU(10)$. Its commutant is $U(1)^4$, which combines with the Cartan of $SU(5)$ to give four of the preserved $U(1)$ groups. The final one is the intrinsic $U(1)$ in the theory, which is also the one preserved generically.
\item  In case 2 one of the $SU(2)$ is the one rotating the spinors, while the other is embedded in $SU(4)$ as $SO(3)\times U(1) \subset SU(4)$. The $U(1)$ is a combination of the intrinsic $U(1)$ groups and the commutant of the $SU(2)$ in $SU(4)$.
\item  In case 3 the $SU(2)$ is embedded as $SO(3)\subset SU(3)$, where the $SU(3)$ is the diagonal one.
\item  In case 4 the $SU(2)^2$ is embedded in $SU(4)$, where its commutant is $U(1)$. This $U(1)$ combines with the Cartan of the $SU(2)$ to give one of the preserved $U(1)$ groups, while the other is one of the intrinsic $U(1)$ groups in the theory.
\item  In case 6 the preserved symmetry is embedded as $SO(2)\times USp(2) \subset SU(4)$, while the remaining $U(1)$ is a combination of the commutant and one of the intrinsic $U(1)$ groups. This combination can be further broken while also breaking $USp(2)$ to its Cartan.
\item  In case 7 the $USp(4)$ is embedded as $SO(5)\subset SU(5)$.
\item  In case 8 the $SU(3)^2$ symmetry is embedded in the diagonal $SU(6)$ group of the $SU(6)^3$ part. Its commutant is $U(1)$ for each $SU(6)$ of which the diagonal combination is broken. Additionally, the two intrinsic $U(1)$ groups are preserved on the entire conformal manifold.
\item In case 10 the $SU(2)^2$ is the diagonal of the intrinsic $SU(2)^2 = SO(4)$ and $SO(4)\subset SU(4)$. 
\item In case 11 the $USp(4)$ is embedded in $SU(6)$ as $SO(5)\subset SU(5)\subset SU(6)$. The additional preserved $U(1)$ group is the one acting on the two spinors with opposite charges, and some combination of the intrinsic $U(1)$ groups and the commutant of $SO(5)$ in $SU(6)$.
\end{itemize}

\begin{table}[h!]
\begin{center}
\begin{adjustwidth}{-.4cm}{}
\begin{tabular}{|c|c|c|c|c|c|}
  \hline 
 &  Matter & $dim\, {\cal M}$ & $G_F^{free}$ & $G^{gen}_F$  & $a$, $c$  \\
 \hline
$12$ &$G=SO(9),$ & $100$ & $U(1)\times SU(7)^2$ & $U(1)$  & $a=\frac{499}{48},$ \\
&	$N_{\bold{16}}=7, \; N_{V}=7$ &  & & has a $1d$ subspace & $c=\frac{283}{24}$ \\
&		&  &  & preserving $U(1)^2\times SU(3)$ &  \\
 \hline
$13$ & $G=SO(9),$ & $1$ & $U(1)^3$ & $U(1)$  & $a=\frac{143}{16},$ \\
&	$N_{S}=1,N_{AS}=1,$ &  & & & $c=\frac{71}{8}$ \\
&	$N_{\bold{16}}=1, N_{V}=1$ &  & & &  \\
 \hline
$14$ &$G=SO(9), N_{S}=1,$ & $10$ & $U(1)^2\times SU(3)$ & $\emptyset$  & $a=\frac{113}{12},$ \\
&	$N_{\bold{16}}=3, N_{V}=4$ &  & $\times SU(4)$ & has a $1d$ subspace & $c=\frac{59}{6}$ \\
&		&  &  & preserving $SU(2)^2$ &  \\
 \hline
$15$ &$G=SO(9), N_{S}=1,$ & $3$ & $U(1)^2\times SU(2)$ & $SU(2)$  & $a=\frac{227}{24},$ \\
&	$N_{\bold{16}}=2, N_{V}=6$ &  & $\times SU(6)$ & has a $1d$ subspace & $c=\frac{119}{12}$ \\
&		&  &  & preserving $SU(2)^2$ &  \\
 \hline
$16$ &$G=SO(9), N_{S}=1,$ & $1$ & $U(1)^2\times SU(8)$ & $U(1)\times SO(7)$  & $a=\frac{19}{2},$ \\
&	$N_{\bold{16}}=1, N_{V}=8$ &  & & & $c=10$ \\
 \hline
$17$ &$G=SO(10),$ & $162$ & $U(1)\times SU(8)$ & $U(1)$  & $a=\frac{613}{48},$ \\
&	$N_{\bold{16}}=8, N_{V}=8$ &  & $\times SU(8)$ & has a $1d$ subspace & $c=\frac{343}{24}$ \\
&	&  &  & preserving $U(1)\times SU(3)$ &  \\
 \hline
$18$ &$G=SO(10), N_{\bold{16}}=7,$ & $100$ & $U(1)^2\times SU(7)$ & $U(1)^2$  & $a=\frac{613}{48},$ \\
&	$N_{\overline{\bold{16}}}=1, N_{V}=8$ &  & $\times SU(8)$ & has a $1d$ subspace & $c=\frac{343}{24}$ \\
&	&  &  & preserving $U(1)^3\times SU(3)$ &  \\
 \hline
$19$ &$G=SO(10), N_{\bold{16}}=6,$ & $90$ & $U(1)^2\times SU(2)$ & $U(1)$  & $a=\frac{613}{48},$ \\
&	$N_{\overline{\bold{16}}}=2, N_{V}=8$ &  & $\times SU(6) $ & has a $1d$ subspace & $c=\frac{343}{24}$ \\
&	&  & $\times SU(8)$ & preserving $U(1)^3\times SU(3)$ &  \\
 \hline
$20$ &$G=SO(10), N_{\bold{16}}=5,$ & $72$ & $U(1)^2\times SU(3)$ & $U(1)$  & $a=\frac{613}{48},$ \\
&	$N_{\overline{\bold{16}}}=3, N_{V}=8$ &  & $\times SU(5) $ & has a $1d$ subspace & $c=\frac{343}{24}$ \\
&	&  & $\times SU(8)$ & preserving $U(1)^8$ &  \\
 \hline
$21$ &$G=SO(10), N_{\bold{16}}=4,$ & $66$ & $U(1)^2\times SU(4)^2$ & $U(1)$  & $a=\frac{613}{48},$ \\
&	$N_{\overline{\bold{16}}}=4, N_{V}=8$ &  & $\times SU(8)$ & has a $1d$ subspace & $c=\frac{343}{24}$ \\
&	&  &  & preserving $U(1)^4\times SU(2)^4$ &  \\
 \hline
$22$ &$G=SO(10),$ & $1$ & $U(1)^3$ & $U(1)$  & $a=\frac{67}{6},$ \\
&	$ N_{S}=1,N_{AS}=1,$ &  & & & $c=\frac{133}{12}$ \\
&	$N_{\bold{16}}=1, N_{\overline{\bold{16}}}=1$ &  & & &  \\
 \hline 
$23$ &$G=SO(10),$ & $1$ & $U(1)^3\times SU(2)$ & $U(1)$  & $a=\frac{45}{4},$ \\
&	$N_{S}=1, N_{AS}=1,$ &  & & & $c=\frac{45}{4}$ \\
&	$N_{\bold{16}}=1, N_{V}=2$ &  & & & \\
 \hline
\end{tabular}
\end{adjustwidth}
 \end{center}
\caption{} 
\label{SOEx2}
\end{table}

Let us make some comments about the cases in Table \ref{SOEx2}.

\begin{itemize}
\item In case 12 the $U(1)$ part of the global symmetry at the free point is always preserved on the conformal manifold. In the $1d$ subspace preserving an additional $U(1)\times SU(3)$ the symmetry is embedded as follows. Consider the $U(1)\times SU(6)$ subgroup of both $SU(7)$ groups. Then the $SU(3)$ is embedded in both $SU(6)$ groups such that $\bold{6}_{SU(6)_1} \rightarrow \bold{6}_{SU(6)_2}\rightarrow \bold{6}_{SU(3)}$, and the $U(1)$ is a combination of the two commutant $U(1)$ groups in the $SU(7)$groups. 

\item In case 13 the $U(1)$ acting only on the vector and spinor is preserved. 
\item In case 14 one of the $SU(2)$ groups is the diagonal combination of $SO(4)\subset SU(4)$ and the other is embedded as $SO(3)\subset SU(3)$.
\item  In case 15 the symmetry is embedded as $SO(3)\times SO(3)\subset SO(6)\subset SU(6)$ with one also being locked on the $SO(3)\subset SU(3)$. The $SO(3)$ not locked on the subgroup of the $SU(3)$ group cannot be broken on the conformal manifold. 
\item In case 16 the $SO(7)$ is a subgroup of $SU(8)$ such that $\bold{8}\rightarrow \bold{7} + \bold{1}$, and the $U(1)$ is a combination of the intrinsic and commutant $U(1)$ groups.
\item   In case 17 the $SU(3)$ is the diagonal one in the embedding $SU(3)\subset SU(8)$ such that $\bold{8}\rightarrow \bold{8}$. There is also a $1d$ subspace preserving $U(1)^8$, where the $U(1)^7$ factor is some diagonal combination of the Cartans of the two $SU(8)$ groups.

\item  In case 18 the symmetry is embedded as follows. Consider the $U(1)\times SU(6)$ and $U(1)\times SU(2)\times SU(6)$ subgroups of $SU(7)$ and $SU(8)$ respectively. Then the $SU(3)$ is embedded in the $SU(6)$ groups as $\bold{6}_{SU(6)}\rightarrow \bold{6}_{SU(3)}$ for both groups, and the $U(1)^3$ is some combination of the abelian part of the free point global symmetry, the Cartan of the $SU(2)$ in $SU(8)$ and the $U(1)$ commutants in the non-abelian symmetries. There are no superpotentials charged under the $U(1)$ group acting on both spinors with the same charge, and it is never broken on the conformal manifold. Another $U(1)$ in the $U(1)^3$ part has $21$ superpotential terms charged under it all with the same sign, and as a result, it is also never broken and the charged superpotential terms are marginally irrelevant. There is also a $1d$ subspace preserving $U(1)^8$, where the $U(1)^7$ factor is some combination of the Cartans of the $SU(8)$ and $SU(7)$, as well as one of the $U(1)$ groups.

\item  Case 19 behaves similarly to case 18. The $SU(3)$ is again embedded inside the $SU(6)$ and the $SU(6)\subset SU(8)$ such that $\bold{6}_{SU(6)}\rightarrow \bold{6}_{SU(3)}$, and the $U(1)^3$ is some combination of the abelian part of the free point global symmetry, the Cartan of the commutant and intrinsic $SU(2)$ groups and the $U(1)$ commutant in the $SU(8)$. Here only the symmetry acting on both spinors with the same charge is preserved. There is also a $1d$ subspace preserving $U(1)^8$, where the $U(1)^7$ factor is some combination of the Cartans of the $SU(8)$, $SU(6)$ and $SU(2)$, as well as one of the $U(1)$ groups.

\item  For case 20, there is a $1d$ subspace preserving $U(1)^8$, where the $U(1)^7$ factor is some combination of the Cartans of the $SU(8)$, $SU(5)$, $SU(3)$, and one of the $U(1)$ groups. Again, the symmetry acting on both spinors with the same charge cannot be broken.

\item  In case 21 the breaking is as follows. Both $SU(4)$ groups are broken as $SU(4)\rightarrow U(1)\times SU(2)^2$, and the $SU(8)$ group is broken as $SU(8)\rightarrow U(1)\times SU(4)^2\rightarrow U(1)\times SO(4)^2$. The $1d$ subspace is then given by locking the $SU(2)^4 \subset SU(8)$ on the $SU(2)^4 \subset SU(4)^2$ and also locking the $U(1)$ commutant in $SU(8)$ with the $U(1)$ acting on the spinors with opposite charges. Like in the other cases, the symmetry acting on both spinors with the same charge cannot be broken. There is also a $1d$ subspace preserving $U(1)^8$, where the $U(1)^7$ factor is some combination of the Cartans of the $SU(8)$, the two $SU(4)$ groups and one of the $U(1)$ groups.

\item In case 22 the symmetry acting on the spinors with opposite charges cannot be broken.

\item In case 23 the preserved $U(1)$ is the diagonal combination of the Cartan of the $SU(2)$ and the $U(1)$ acting only on the spinor and vectors.

\end{itemize}

\begin{table}[h!]
\begin{center}
\begin{adjustwidth}{-.2cm}{}
\begin{tabular}{|c|c|c|c|c|c|}
  \hline 
 &  Matter & $dim\, {\cal M}$ & $G_F^{free}$ & $G^{gen}_F$  & $a$, $c$  \\
 \hline
$24$ &$G=SO(10),$ & $10$ & $U(1)^2\times SU(4)^2$ & $U(1)$  & $a=\frac{563}{48},$ \\
&	$N_{S}=1,$ &  & & has a $1d$ subspace & $c=\frac{293}{24}$ \\
&	$N_{\bold{16}}=4, \; N_{V}=4$ &  & & preserving $U(1)^2\times SU(2)^2$ & \\
 \hline
$25$ &$G=SO(10),$ & $2$ & $U(1)^3\times SU(3)$ & $U(1)^2$  & $a=\frac{563}{48},$ \\
&	$N_{S}=1, \; N_{\bold{16}}=3,$ &  & $\times SU(4)$ & has a $1d$ subspace & $c=\frac{293}{24}$ \\
&	$N_{\overline{\bold{16}}}=1, \; N_{V}=4$ &  & & preserving $U(1)^4$ & \\
 \hline
$26$ &$G=SO(10),$ & $2$ & $U(1)^3\times SU(2)^2$ & $U(1)$  & $a=\frac{563}{48},$ \\
&	$N_{S}=1, \; N_{\bold{16}}=2,$ &  & $\times SU(4)$ & has a $1d$ subspace & $c=\frac{293}{24}$ \\
&	$N_{\overline{\bold{16}}}=2, \; N_{V}=4$ &  & & preserving $U(1)^4$ & \\
 \hline
$27$ &$G=SO(10),$ & $13$ & $U(1)^2\times SU(3)$ & $\emptyset$  & $a=\frac{189}{16},$ \\
&	$N_{S}=1,$ &  & $\times SU(6)$ & has a $1d$ subspace & $c=\frac{99}{8}$ \\
&	$N_{\bold{16}}=3, \; N_{V}=6$ &  & & preserving $U(1)\times SU(2)$ & \\
&  &  &  & also has a $2d$ subspace &  \\
&  &  & & preserving a & \\
&  &  & & different $SU(2)$ & \\
 \hline
$28$ &$G=SO(10),$ & $6$ & $U(1)^3\times SU(2)$ & $U(1)$  & $a=\frac{189}{16},$ \\
&	$N_{S}=1, \; N_{\bold{16}}=2,$ &  & $\times SU(6)$ & has a $1d$ subspace & $c=\frac{99}{8}$ \\
&	$N_{\overline{\bold{16}}}=1, \; N_{V}=6$ &  & & preserving $U(1)\times SU(2)$ & \\
 \hline
$29$ &$G=SO(10),$ & $3$ & $U(1)^2\times SU(2)$ & $USp(4)$  & $a=\frac{571}{48},$ \\
&	$N_{S}=1,$ &  & $\times SU(8)$ & has a $1d$ subspace & $c=\frac{301}{24}$ \\
&	$N_{\bold{16}}=2, \; N_{V}=8$ &  & & preserving $SU(2)\times USp(4)$ & \\
 \hline
$30$ &$G=SO(10),$ & $1$ & $U(1)^3\times SU(8)$ & $U(1)^2\times SU(4)$  & $a=\frac{571}{48},$ \\
&	$N_{S}=1, \; N_{\bold{16}}=1,$ &  & & & $c=\frac{301}{24}$ \\
&	$N_{\overline{\bold{16}}}=1, \; N_{V}=8$ &  & & & \\
 \hline
$31$ &$G=SO(10), $ & $1$ & $U(1)^2\times SU(10)$ & $U(1)\times SO(9)$  & $a=\frac{575}{48},$ \\
&	$N_{S}=1,N_{\bold{16}}=1,$ &  & & & $c=\frac{305}{24}$ \\
&	$N_{V}=10$ &  & & &  \\
 \hline
$32$ &$G=SO(11),$ & $2$ & $U(1)^3$ & $\emptyset$  & $a=\frac{329}{24},$ \\
&	$ N_{S}=1,N_{AS}=1,$ &  & & & $c=\frac{41}{3}$ \\
&	$N_{\bold{32}}=1, N_{V}=1$ &  & & &  \\
 \hline
$33$ &$G=SO(12),$ & $2$ & $U(1)^3\times SU(2)$ & $U(1)$  & $a=\frac{793}{48},$ \\
&	$N_{S}=1, \; N_{AS}=1,$ &  & & has a $1d$ subspace & $c=\frac{397}{24}$ \\
&	$N_{\bold{32}}=1, \; N_{V}=2$ &  & & preserving $SU(2)$ &  \\
 \hline
\end{tabular}
\end{adjustwidth}
 \end{center}
\caption{} 
\label{SOEx3}
\end{table}

Let us make some comments about the cases in Table \ref{SOEx3}.

\begin{itemize}
\item  In case 24 the breaking is $SU(4)_{Spinor}\rightarrow U(1)\times SU(2)^2$, $SU(4)_V\rightarrow SO(4)$ with the $SU(2)^2\subset SU(4)_{Spinor}$ locked on the $SO(4)\subset SU(4)_V$. The $U(1)$ acting on the spinor and vector only cannot be broken on the conformal manifold. There is also a $1d$ subspace preserving $U(1)^4$, where the $U(1)^3$ factor is some combination of the Cartans of the two $SU(4)$ groups.
\item  In case 25 there is a $1d$ subspace preserving $U(1)^4$, where the $U(1)^3$ factor is some combination of the Cartans of the $SU(4)$, $SU(3)$ and one of the $U(1)$ groups. The $U(1)$ acting on the spinor and vector only, as well as a combination of the $U(1)$ acting on the spinors and a Cartan of the $SU(4)$ cannot be broken on the conformal manifold.
\item  In case 26 the $U(1)^4$ consists of the $U(1)$ acting on the spinor and vector only, which is never broken, and the $U(1)^3$ factor, which is some combination of the Cartans of the $SU(4)$, the two $SU(2)$ groups, and the other $U(1)$. 
\item In case 27 the symmetries are embedded as follows: first we reduce the $SU(3)$ to $SU(2)\times U(1)$ for the $1d$ subspace and to $SO(3)$ for the $2d$ subspace. The $SU(6)$ is then broken to $U(1)^2\times SU(2)$ for the $1d$ subspace, embedded such that $\bold{6}\rightarrow \bold{3}+\bold{2}+\bold{1}$, and $U(1)\times SU(2)$ for the $2d$ subspace, embedded such that $\bold{6}\rightarrow \bold{5}+\bold{1}$. We then lock the two $SU(2)$ groups to the diagonal. The preserved abelian symmetry on the $1d$ subspace is a combination of the commutant and intrinsic $U(1)$ groups.

\item  In case 28 the breaking is as follows: first we take $SU(6)\rightarrow SO(6)\rightarrow SO(3)\times SO(3)$ then we further break one $SO(3)$ to its $SO(2)$ subgroup, which is the $U(1)$ that is preserved generically, and the other $SO(3)$ is broken to the diagonal group with the intrinsic $SU(2)$, becoming the $SU(2)$ that remains at a special point on the conformal manifold.

\item  In case 29 the breaking is as follows: we take $SU(8)\rightarrow SO(8)\rightarrow SO(5)\times SO(3)$, where the $SO(5)$ is the group preserved generically. The special preserved $SU(2)$ is the diagonal of the $SO(3)$ and the intrinsic $SU(2)$.

\item  In case 30 the breaking is as follows: $SU(8)\rightarrow SU(6)\times SU(2)\times U(1) \rightarrow SO(6)\times SO(2)\times U(1)$, where the $SU(4)$ is the $SO(6)$ and the two $U(1)$ groups are combinations of the commutant and intrinsic $U(1)$ groups.

\item  In case 31 the breaking is as follows: $SU(10)\rightarrow SO(10)\rightarrow SO(9)$. The $U(1)$ is a combination of the commutant of $SO(9)$ in $SU(10)$ and one of the intrinsic $U(1)$ groups. 

\item In case 33 the generically preserved $U(1)$ is the Cartan of the $SU(2)$ group.

\end{itemize}

\section{$G=E_6,\,E_7,\,E_8,\, F_4,\, G_2$}\label{sec:exceptional}

Finally we consider the case of exceptional groups. The list of possible representations consistent with condition (\ref{BetaZero}) appears in table \ref{ExpRep}. All of these groups have no cubic Casimir, and as a result there is no cubic symmetric invariant using adjoints. The only cases then with a conformal manifold when adjoint matter is involved are the $\mathcal{N}=2$ ones, so we only consider here cases without adjoints. The properties of the conformal theories are summarized in table \ref{Excep}.

\begin{table}[h!]
\begin{center}
\begin{tabular}{|c|c|c|c|c|}
  \hline 
  Group & Label & Dimension  & Dynkin index  & Reality  \\
 \hline
$G_2$ & $(1,0)$  & $\bold{7}$  & $1$ & Real \\
\hline
$G_2$ & $(0,1)$  & $\bold{14}$  & $4$ & Real \\
\hline
$G_2$ & $(2,0)$  & $\bold{27}$  & $9$ & Real \\
\hline
\hline
 $F_4$ & $(0,0,0,1)$  & $\bold{26}$  & $3$ & Real \\
\hline
$F_4$ & $(1,0,0,0)$  & $\bold{52}$  & $9$ & Real \\
\hline
\hline
$E_6$ & $(1,0,0,0,0,0) , \; (0,0,0,0,0,1)$  & $\bold{27}, \; \bold{\overline{27}}$ & $3$ & Complex \\
\hline
$E_6$ & $(0,1,0,0,0,0)$  & $\bold{78}$ & $12$ & Real \\
\hline
\hline
$E_7$ & $(0,0,0,0,0,0,1)$  & $\bold{56}$ & $6$ & Pseudo-real \\
\hline
$E_7$ & $(1,0,0,0,0,0,0)$  & $\bold{133}$ & $18$ & Real \\
\hline
\hline
$E_8$ & $(0,0,0,0,0,0,0,1)$  & $\bold{248}$ & $30$ & Real \\
\hline
 \end{tabular}
 \end{center}
\caption{Various group theory data for representations of the exceptional groups. Here the group entry stands for the group in question, and the label entry stands for the standard Dynkin label of the representation. Additionally, the dimension, Dynkin index and the reality properties of the representation are listed.} 
\label{ExpRep}
\end{table}

\subsection*{$E_8$}

Here the only available representation is the adjoint and so the only case consistent with condition (\ref{BetaZero}) is the $\mathcal{N}=4$ one.   

\subsection*{$E_7$}

Here the only other representation is the fundamental, which do not posses a cubic invariant. As a result the only possible conformal cases are the $\mathcal{N}=2$ and the $\mathcal{N}=4$ ones. For convenience we shall write the possible purely $\mathcal{N}=1$ solutions to (\ref{BetaZero}):

\begin{enumerate}

\item - $N_{\bold{133}}=2, \quad N_{\bold{56}}=3$

\item - $N_{\bold{56}}=9$

\end{enumerate}

\subsection*{$E_6$}

Here the only other representation is the fundamental and its complex conjugate. These posses a symmetric cubic invariant which allows for conformal theories made only from these representations. In fact, as there is no cubic Casimir, there is no constraint in having any combination of these. As a result, the possible cases are:

\begin{enumerate}

\item - $N_{\bold{27}}=N_{\overline{\bold{27}}}=6$

\item - $N_{\bold{27}}=7, \quad N_{\overline{\bold{27}}}=5$

\item - $N_{\bold{27}}=8, \quad N_{\overline{\bold{27}}}=4$

\item - $N_{\bold{27}}=9, \quad N_{\overline{\bold{27}}}=3$

\item - $N_{\bold{27}}=10, \quad N_{\overline{\bold{27}}}=2$

\item - $N_{\bold{27}}=11, \quad N_{\overline{\bold{27}}}=1$

\item - $N_{\bold{27}}=12$

\end{enumerate}

For convenience we shall also write the possible purely $\mathcal{N}=1$ solutions to (\ref{BetaZero}) which are not conformal:

\begin{enumerate}

\item - $N_{\bold{78}}=2, \quad N_{\bold{27}}=N_{\overline{\bold{27}}}=2$

\item - $N_{\bold{78}}=2, \quad N_{\bold{27}}=3 , \quad N_{\overline{\bold{27}}}=1$

\item - $N_{\bold{78}}=2, \quad N_{\bold{27}}=4$

\item - $N_{\bold{78}}=1, \quad N_{\bold{27}}=5 , \quad N_{\overline{\bold{27}}}=3$

\item - $N_{\bold{78}}=1, \quad N_{\bold{27}}=6 , \quad N_{\overline{\bold{27}}}=2$

\item - $N_{\bold{78}}=1, \quad N_{\bold{27}}=7 , \quad N_{\overline{\bold{27}}}=1$

\item - $N_{\bold{78}}=1, \quad N_{\bold{27}}=8$

\end{enumerate}

\subsection*{$F_4$}

This case is similar to the $E_6$ one in that the only other representation is the fundamental and it possess a symmetric cubic invariant. Unlike the previous case, though, it is real. This allows for conformal theories made only from the fundamental representation. The possible purely $\mathcal{N}=1$ solutions to (\ref{BetaZero}) are:

\begin{enumerate}

\item - $N_{\bold{52}}=2, \quad N_{\bold{26}}=3$

\item - $N_{\bold{26}}=9$

\end{enumerate}

From these, only the last is conformal.

\subsection*{$G_2$}

In this case we have two possible representations besides the adjoints, these are the $\bold{7}$ dimensional fundamental representation and the $\bold{27}$ dimensional symmetric representation. The former has a cubic antisymmetric invariant, while the latter has two cubic symmetric invariants. These allows us to build conformal theories when only these representations are used. The possible purely $\mathcal{N}=1$ solutions to (\ref{BetaZero}) are:

\begin{enumerate}

\item - $N_{\bold{14}}=2, \quad N_{\bold{7}}=4$

\item - $N_{\bold{7}}=12$

\item - $N_{\bold{27}}=1, \quad N_{\bold{7}}=3$

\end{enumerate}

From these, only the last two are conformal.

\begin{table}[h!]
\begin{center}
\begin{tabular}{|c|c|c|c|c|c|}
  \hline 
 &  Matter & $dim\, {\cal M}$ & $G_F^{free}$ & $G^{gen}_F$  & $a$, $c$  \\
 \hline
$1$&  $G=E_6$ & $41$ & $U(1)\times SU(6)^2 $ & $\emptyset$  & $a=\frac{171}{8},$ \\
&	$N_{\bold{27}}=N_{\overline{\bold{27}}}=6$ &  &  & has a $1d$ subspace & $c=\frac{93}{4}$ \\
&	&  &  & preserving $SU(3)^2$ &  \\
 \hline
 $2$&$G=E_6, N_{\bold{27}}=7$ & $46$ & $U(1)\times SU(5) $ & $\emptyset$  & $a=\frac{171}{8},$ \\
&	$N_{\overline{\bold{27}}}=5$ &  & $\times SU(7)$ & has a $1d$ subspace & $c=\frac{93}{4}$ \\
&	&  &  & preserving $SU(2)\times SU(3)$ &  \\
 \hline
 $3$&$G=E_6, N_{\bold{27}}=8$ & $61$ & $U(1)\times SU(4) $ & $\emptyset$  & $a=\frac{171}{8},$ \\
&	$N_{\overline{\bold{27}}}=4$ &  & $\times SU(8)$ & has a $1d$ subspace & $c=\frac{93}{4}$ \\
&	&  &  & preserving $U(1)^2 \times SU(3)$ &  \\
 \hline
 $4$&$G=E_6, N_{\bold{27}}=9$ & $86$ & $U(1)\times SU(3) $ & $\emptyset$  & $a=\frac{171}{8},$ \\
&	$N_{\overline{\bold{27}}}=3$ &  & $\times SU(9)$ & has a $1d$ subspace & $c=\frac{93}{4}$ \\
&	&  &  & preserving $U(1)^2 \times SU(3)^2$ &  \\
 \hline
$5$& $G=E_6, N_{\bold{27}}=10$ & $121$ & $U(1)\times SU(2) $ & $\emptyset$  & $a=\frac{171}{8},$ \\
&	$N_{\overline{\bold{27}}}=2$ &  & $\times SU(10)$ & has a $1d$ subspace & $c=\frac{93}{4}$ \\
&	&  &  & preserving $SU(3)^2$ &  \\
 \hline
$6$&$G=E_6, N_{\bold{27}}=11$ & $166$ & $U(1)\times SU(11) $ & $\emptyset$  & $a=\frac{171}{8},$ \\
&	$N_{\overline{\bold{27}}}=1$ &  & & has a $1d$ subspace & $c=\frac{93}{4}$ \\
&	&  &  & preserving $SU(3)^2$ &  \\
 \hline
$7$&$G=E_6, N_{\bold{27}}=12$ & $221$ & $SU(12) $ & $\emptyset$  & $a=\frac{171}{8},$ \\
&	 &  & & has a $1d$ subspace & $c=\frac{93}{4}$ \\
&	&  &  & preserving $U(1)^2\times SU(3)^2$ &  \\
 \hline
$8$&$G=F_4, N_{\bold{26}}=9$ & $85$ & $SU(9) $ & $\emptyset$  & $a=\frac{117}{8},$ \\
&	 &  & & has a $1d$ subspace & $c=\frac{65}{4}$ \\
&	&  &  & preserving $SU(3)^2$ &  \\
 \hline
$9$&$G=G_2, N_{\bold{7}}=12$ & $77$ & $SU(12) $ & $\emptyset$  & $a=\frac{35}{8},$ \\
&	 &  & & has a $1d$ subspace & $c=\frac{21}{4}$ \\
&	&  &  & preserving $SU(3)^4$ &  \\
 \hline
$10$&$G=G_2, N_{\bold{27}}=1$ & $3$ & $U(1)\times SU(3) $ & $SU(2)$  & $a=\frac{29}{8},$ \\
&	$N_{\bold{7}}=3$ &  & & has a $2d$ subspace & $c=\frac{15}{4}$ \\
&	&  &  & preserving $SU(3)$ &  \\
 \hline
\end{tabular}
 \end{center}
\caption{} 
\label{Excep}
\end{table}

Let us make several comments about specific cases,
\begin{itemize}
\item In all these cases the superpotentials are  in three index symmetric or antisymmetric representations. For $E_6$ and $F_4$ we use the symmetric cases.

\item  Here the basic preserved subgroups are embedded as: $U(1)^2\subset SU(3)$ as the Cartan, $SU(2)\subset SU(5)$ such that $\bold{5}\rightarrow \bold{5}$, $SU(3)\subset SU(6)$ such that $\bold{6}\rightarrow \bold{6}$, $SU(3)\subset SU(8)$ such that $\bold{8}\rightarrow \bold{8}$ and $SU(3)\times SU(3)\subset SU(9)$ such that $\bold{9}\rightarrow (\bold{3},\bold{3})$. In the other cases these are used as subgroups. For $G_2$ the $SU(3)^4$ are just the four independent $SU(3)$ groups in $SU(12)$ for the first case, and the $SU(2)$ in the last case is embedded in $SU(3)$ as $\bold{3}\rightarrow \bold{3}$.
\end{itemize}

\section{Extended supersymmetry}\label{sec:extended}

Let us now discuss here conformal manifolds with weakly coupled loci with extended supersymmetry. The possible conformal Lagrangians, even with non-simple gauge groups, with ${\cal N}>1$ supersymmetry in four dimensions were analyzed in \cite{Bhardwaj:2013qia} and here we add the analysis of the conformal manifold for the cases with a simple gauge group. All such models have a $1d$ subspace of the conformal manifold which preserves the full extended supersymmetry of the free point. Some of them have additional directions which preserve only ${\cal N}=1$ supersymmetry. For the ${\cal N}=4$ cases these are the well studied $\beta$ and $\gamma$ deformations, see {\it e.g.}  \cite{Parkes:1984dh,Leigh:1995ep,Lunin:2005jy}.

\subsection{Special unitary groups}

Let us first discuss ${\cal N}=4$ $SU(N)$ gauge theories. This case was widely studied in the past.
The theory has three fields in the adjoint representation. For $N>2$ we have marginal operators in the ${\bf 10}+{\bf 1}$ of the global $SU(3)$ symmetry group. For $N=2$ only the singlet is there. 
The dimension of the conformal manifold is three for $N>2$ and is one for $N=2$.  On the direction parameterized by the singlet, ${\cal N}=4$ supersymmetry is preserved and in particular the $SU(3)$ symmetry. Along another direction, the so called $\beta$ deformation, a $U(1)\times U(1)$ global symmetry can be preserved. On a generic point of the conformal manifold all the global symmetry is broken.

In tables \ref{UnitGenExt} and \ref{UnitGenExt2} we write all the purely $\mathcal{N}=2$ cases with a single $SU(N)$ gauge symmetry that solve \eqref{AnomalyC} and \eqref{BetaZero}. All cases have one adjoint chiral field coming from the vector multiplet and we do not write it in the table. In addition, all the theories have a $1d$ subspace preserving $\mathcal{N}=2$ supersymmetry, some cases have additional $1d$ subspaces preserving $\mathcal{N}=1$ supersymmetry and we note it explicitly.

\begin{table}[htbp]
\begin{center}
\begin{adjustwidth}{-.6cm}{}
\begin{tabular}{|c|c|c|c|c|c|}
  \hline 
 &  Matter & $dim\, {\cal M}$ & $G_F^{free}$ & $G^{gen}_F$  & $a$, $c$  \\
 \hline
$1$ &  $G=SU(2),$ & $1$ & $U(1) \times$ & $U(1)\times SO(8)$ & $a=\frac{23}{24},$ \\
&	$N_{F}=8$ & & $SU(8)$ & & $c=\frac{7}6$ \\
 \hline
$2$&  $G=SU(N),$ & $1$ & $U(1)^2 \times$ & $U(1)^2\times SU(2N)$ & $a=\frac{7N^2-5}{24},$ \\
&	$N_{F}=N_{{\overline F}}=2N,$ & ($7$ for & $SU(2N)^2$ & ($\emptyset$ for $N=3$) & $c=\frac{2N^2-1}6$ \\
&	$N>2$ &  $N=3$) &  & for $N=3$ has an $\mathcal{N}=1$ $1d$ & \\
&	 &  &  & subspace preserving $SU(3)^4$ & \\
 \hline
$3$&  $G=SU(N),$ & $\left\lfloor \frac{N+4}{2}\right\rfloor$ & $U(1)^{4}\times $ & $U(1)^{1+(N\text{mod}\,2)}$ & $a=\frac{13N^{2}+3N-10}{48},$ \\
&	$N_{AS}=N_{{\overline AS}}=1,$ & ($8,6$ for & $SU(N+2)^{2}$ & $\times SU(2)^{\left\lfloor \frac{N+2}{2}\right\rfloor }$ & $c=\frac{7N^{2}+3N-4}{24}$ \\
&	$N_{F}=N_{{\overline F}}=N+2$ & $N=5,6$) &  &  ($U(1),SU(2)^4$ & \\
&	$N>4$ &  &  &  for $N=5,6$) & \\
&	 &  &  & for even $N>6$ has an $\mathcal{N}=1$ & \\
&	 &  &  & $1d$ subspace preserving  & \\
&	 &  &  & $U(1)\times USp(N+2)^2$  & \\
&	 &  &  & for $N=6$ has an $\mathcal{N}=1$ $1d$ & \\
&	 &  &  & subspace preserving  & \\
&	 &  &  & $U(1)^2\times USp(8)^2$  & \\
 \hline
$4$&  $G=SU(4),$ & $22$ & $U(1)^{3}\times $ & $\emptyset$ & $a=\frac{35}{8},$ \\
&	$N_{AS}=2,$ &  & $SU(2)\times$ & & $c=5$ \\
&	$N_{F}=N_{{\overline F}}=6$ &  & $SU(6)^2$ & & \\
 \hline
$5$&  $G=SU(N),$ & $7$ & $U(1)^{4}\times  $ & $U(1)^{2}$ & $a=\frac{6N^{2}+3N-5}{24},$ \\
&	$N_{AS}=N_{{\overline AS}}=2,$ & ($29,14$ for & $SU(2)^{2}\times $ & ($\emptyset, U(1)$ for $N=5,6$) & $c=\frac{3N^{2}+3N-2}{12}$ \\
&	$N_{F}=N_{{\overline F}}=4$ & $N=5,6$) & $ SU(4)^{2}$ & has an $\mathcal{N}=1$ $1d$ & \\
&	$N>4$ &  &  & subspace preserving  & \\
&	 &  &  & $U(1)^{2}\times USp(4)^{2}$  & \\
&	 &  &  & for $N=5$ also has an $\mathcal{N}=1$ & \\
&	 &  &  & $1d$ subspace preserving  & \\
&	 &  &  & $U(1)^{3}\times SU(2)^{2}$  & \\
 \hline
$6$&  $G=SU(4),$ & $23$ & $U(1)^{3}\times  $ & $U(1)^{2}$ & $a=\frac{103}{24},$ \\
&	$N_{AS}=4,$ &  & $SU(4)^{3}$ & has an $\mathcal{N}=1$ $1d$ & $c=\frac{29}{6}$ \\
&	$N_{F}=N_{{\overline F}}=4$ & &  & subspace preserving & \\
&	 &  &  & $U(1)^{2}\times SU(2)\times USp(4)^{2}$  & \\
 \hline
$7$&  $G=SU(N),$ & $N-1$ & $U(1)^{4}\times  $ & $U(1)$ ($\emptyset$ for $N=3$) & $a=\frac{13N^{2}-3N-10}{48},$ \\
&	$N_{S}=N_{{\overline S}}=1,$ & ($3$ for & $SU(N-2)^{2}$ & has an $\mathcal{N}=1$ $1d$ & $c=\frac{7N^{2}-3N-4}{24}$ \\
&	$N_{F}=N_{{\overline F}}=N-2$ & $N=3$) &  & subspace preserving  & \\
&	 &  &  &  $U(1)\times SO(N-2)^{2}$ & \\
 \hline
 \end{tabular}
\end{adjustwidth}
 \end{center}
\caption{} 
\label{UnitGenExt}
\end{table}

\begin{table}[htbp]
\begin{center}
\begin{tabular}{|c|c|c|c|c|c|}
  \hline 
 &  Matter & $dim\, {\cal M}$ & $G_F^{free}$ & $G^{gen}_F$  & $a$, $c$  \\
\hline
$8$&  $G=SU(N),$ & $2$ & $U(1)^{4}$  & $U(1)^2$ ($\emptyset$ for $N=6$) & $a=\frac{6N^{2}-5}{24},$ \\
&	$N_{S}=N_{{\overline S}}=1,$ & ($3$ for & ($U(1)^3 \times$ & & $c=\frac{3N^{2}-2}{12}$ \\
&	$N_{AS}=N_{{\overline AS}}=1$ & $N=6$) & $SU(2)$ & & \\
&	 &  & for $N=4$) &  & \\
&	 &  &  & & \\
 \hline
$9$&  $G=SU(4),$ & $2$ & $U(1)^{3}\times$ & $U(1)\times USp(4)\times SU(2)$ & $a=\frac{101}{24},$ \\
&	$N_{AS}=6,$ &  & $SU(6)\times$ & has an $\mathcal{N}=1$ $1d$ & $c=\frac{14}{3}$ \\
&	$N_{F}=N_{{\overline F}}=2$ &  & $SU(2)^{2}$ & subspace preserving & \\
&	 &  &  & $U(1)\times USp(4)\times SU(2)^{2}$  & \\
 \hline
$10$&  $G=SU(4),$ & $1$ & $U(1)\times$ & $U(1)\times USp(8)$ & $a=\frac{33}{8},$ \\
&	$N_{AS}=8$ &  & $SU(8)$ &  & $c=\frac{9}{2}$ \\
 \hline
$11$&  $G=SU(5),$ & $4$ & $U(1)^{4}\times$ & $U(1)$ & $a=\frac{155}{24},$ \\
&	$N_{AS}=N_{{\overline AS}}=3,$ &  & $SU(3)^{2}$ & & $c=\frac{83}{12}$ \\
&	$N_{F}=N_{{\overline F}}=1$ &  &  &  & \\
 \hline
$12$&  $G=SU(6),$ & $12$ & $U(1)^{2}\times$ & $\emptyset$ & $a=\frac{55}{6},$ \\
&	$N_{AS}=N_{{\overline AS}}=3$ &  & $SU(3)^{2}$ & has an $\mathcal{N}=1$ $1d$ & $c=\frac{115}{12}$ \\
&	 &  &  & subspace preserving  & \\
&	 &  &  & $U(1)^{4}$  & \\
 \hline
$13$&  $G=SU(6),$ & $1$ & $U(1)^3 \times$ & $U(1)^2\times SU(9)$ & $a=\frac{239}{24},$ \\
&	$N_{\textbf{20}}=1,$ &  &  & $SU(9)^2$ & $c=\frac{67}{6}$ \\
&	$N_{F}=N_{{\overline F}}=9$ &  &  &  & \\
 \hline
$14$&  $G=SU(6),$ & $7$ & $U(1)^{5}\times$ & $\emptyset$ & $a=\frac{115}{12},$ \\
&	$N_{\textbf{20}}=1,$ &  & $SU(5)^{2}$ & has an $\mathcal{N}=1$ $1d$ subspace & $c=\frac{125}{12}$ \\
&	$N_{AS}=N_{{\overline AS}}=1,$ &  &  & preserving $U(1)^{2}\times USp(4)^{2}$ & \\
&	$N_{F}=N_{{\overline F}}=5$ &  &  & & \\
 \hline
$15$&  $G=SU(6),$ & $8$ & $U(1)^{5}\times$ & $\emptyset$ & $a=\frac{221}{24},$ \\
&	$N_{\textbf{20}}=1,$ &  & $SU(2)^{2}$ & & $c=\frac{29}{3}$ \\
&	$N_{AS}=N_{{\overline AS}}=2,$ &  &  & & \\
&	$N_{F}=N_{{\overline F}}=1,$ &  &  & & \\
 \hline
 \end{tabular}
 \end{center}
\caption{} 
\label{UnitGenExt2}
\end{table}

\begin{table}[htbp]
\begin{center}
\begin{tabular}{|c|c|c|c|c|c|}
  \hline 
 &  Matter & $dim\, {\cal M}$ & $G_F^{free}$ & $G^{gen}_F$  & $a$, $c$  \\
 \hline
$16$&  $G=SU(6),$ & $2$ & $U(1)^{5}$ & $U(1)$ & $a=\frac{53}{6},$ \\
&	$N_{\textbf{20}}=1,$ &  &  & & $c=\frac{107}{12}$ \\
&	$N_{S}=N_{{\overline S}}=1,$ &  &  & & \\
&	$N_{F}=N_{{\overline F}}=1$ &  &  &  & \\
 \hline
$17$&  $G=SU(6),$ & $1$ & $U(1)^{3}\times SU(2)$ & $U(1)^{3}\times SU(6)$ & $a=\frac{77}{8},$ \\
&	$N_{\textbf{20}}=2,$ &  & $\times SU(6)^{2}$ &  & $c=\frac{21}{2}$ \\
&	$N_{F}=N_{{\overline F}}=6$ &  &  &  & \\
 \hline
$18$&  $G=SU(6),$ & $7$ & $U(1)^{5}\times SU(2)^{3}$ & $\emptyset$ & $a=\frac{37}{4},$ \\
&	$N_{\textbf{20}}=2,$ &  &  & has an $\mathcal{N}=1$ $1d$ subspace & $c=\frac{39}{4}$ \\
&	$N_{AS}=N_{{\overline AS}}=1,$ &  &  & preserving $U(1)^{2}\times SU(2)^{2}$ & \\
&	$N_{F}=N_{{\overline F}}=2$ &  &  & & \\
 \hline
$19$&  $G=SU(6),$ & $1$ & $U(1)^{3}\times SU(3)^{3}$ & $U(1)^{2}\times SU(2)\times SU(3)$ & $a=\frac{223}{24},$ \\
&	$N_{\textbf{20}}=3,$ &  &  &  & $c=\frac{59}{6}$ \\
&	$N_{F}=N_{{\overline F}}=3$ &  &  &  & \\
 \hline
$20$&  $G=SU(6),$ & $1$ & $U(1)\times SU(4)$ & $U(1)\times SU(2)^{2}$ & $a=\frac{215}{24},$ \\
&	$N_{\textbf{20}}=4$ &  &  &  & $c=\frac{55}{6}$ \\
 \hline
$21$&  $G=SU(7),$ & $1$ & $U(1)^{4}\times SU(4)^{2}$ & $U(1)^{3}\times SU(4)$ & $a=\frac{101}{8},$ \\
&	$N_{\textbf{35}}=N_{\overline{\textbf{35}}}=1,$ &  &  &  & $c=\frac{53}{4}$ \\
&	$N_{F}=N_{{\overline F}}=4$ &  &  &  & \\
 \hline
$22$&  $G=SU(8),$ & $1$ & $U(1)^{4}$ & $U(1)^{3}$ & $a=\frac{379}{24},$ \\
&	$N_{\textbf{56}}=N_{\overline{\textbf{56}}}=1,$ &  &  &  & $c=\frac{95}{6}$ \\
&	$N_{F}=N_{{\overline F}}=1$ &  &  &  & \\
 \hline
 \end{tabular}
 \end{center}
\caption{} 
\label{UnitGenExt3}
\end{table}

Let us make several comments about specific cases,
\begin{itemize}
\item In case $2$ the generically preserved $SU(2N)$, for $N>3$, is the diagonal of the two intrinsic $SU(2N)$ groups. This space is also the $\mathcal{N}=2$ preserving one. This space also exists for $N=3$, but then there is also a $1d$ subspace preserving $\mathcal{N}=1$ and $SU(3)^4$ which is embedded as $SU(3)^2\times U(1) \subset SU(6)$ in both $SU(6)$ groups.

\item The conformal manifold in case $3$, being an $\mathcal{N}=2$ superconformal gauge theory with a simple gauge group, has an $\mathcal{N}=2$ preserving $1d$ subspace. The preserved symmetry on this subspace is $U(1)^3 \times SU(N+2)$, where the $SU(N+2)$ is the diagonal of the two intrinsic $SU(N+2)$ groups. This group is then broken as $SU(N+2)\to USp(N+2) \to SU(2)^{\frac{N+2}2}$ when $N$ is even, and as $SU(N+2)\to U(1)\times USp(N+1) \to U(1)\times SU(2)^{\frac{N+1}2}$ when $N$ is odd in general. For even $N>6$ we have an additional $1d$ subspace preserving $\mathcal{N}=1$ and $U(1)\times USp(N+2)^2$, where both $USp(N+2)$ groups are embedded as $USp(N+2) \subset SU(N+2)$ in both $SU(N+2)$ groups. For $N=6$ we find a similar $1d$ subspace preserving $U(1)^2\times USp(8)^2$ instead.

\item In case $4$ there is an $\mathcal{N}=2$ preserving $1d$ subspace, on which an $U(1)^{2}\times SU(2)\times SU(6)$ global symmetry is preserved. The $SU(6)$ here is the diagonal of the two intrinsic $SU(6)$ groups.

\item In case $5$ the $\mathcal{N}=2$ preserving subspace preserves a $U(1)^3\times SU(2)\times SU(4)$ global symmetry, where both $SU(4)$ and $SU(2)$ groups are the diagonal of the two intrinsic $SU(4)$ and $SU(2)$ groups, respectively. Has an additional $1d$ subspace preserving $\mathcal{N}=1$ and $U(1)^{2}\times USp(4)^{2}$, where both $USp(4)$ groups are embedded as $USp(4) \subset SU(4)$ in both $SU(4)$ groups. For $N=5$ we have another $\mathcal{N}=1$ $1d$ subspace preserving $U(1)^{3}\times SU(2)^{2}$, where we break both $SU(4)\to SO(3)\times U(1)$, and identify the diagonal $SU(2)$ of $SU(2)_{AS}$ and $SO(3)_F$ and the same for the conjugates.

\item In case $6$ the $1d$ subspace preserving $\mathcal{N}=2$ supersymmetry also preserves a $U(1)^2\times USp(4)\times SU(4)$ global symmetry, where the $SU(4)$ is the diagonal of $SU(4)_F$ and $SU(4)_{\overline{F}}$, while we break $SU(4)_{AS} \to USp(4)$. This case also has an additional $1d$ subspace preserving $\mathcal{N}=1$ and $U(1)^{2}\times SU(2)\times USp(4)^{2}$, where both $USp(4)$ groups are embedded as $USp(4) \subset SU(4)$ in $SU(4)_F$ and $SU(4)_{\overline{F}}$, while we break $SU(4)_{AS} \to SU(2)^2 \times U(1) \to SU(2)\times U(1)^2$.

\item In case $7$ the $\mathcal{N}=2$ preserving $1d$ subspace also preserves a $U(1)^{3}\times SU(N-2)$ global symmetry, where the $SU(N-2)$ is the diagonal of the two intrinsic $SU(N-2)$ groups. Also has an additional $1d$ subspace preserving $\mathcal{N}=1$ and $U(1)\times SO(N-2)^{2}$, where both $SO(N-2)$ groups are embedded as $SO(N-2) \subset SU(N-2)$ in both intrinsic $SU(N-2)$ groups.

\item In case $8$ the $\mathcal{N}=2$ preserving $1d$ subspace also preserves a $U(1)^{3}$ global symmetry for $N\neq 4$, which is enhanced to $U(1)^2\times SU(2)$ for $N=4$.

\item In case $9$ the $\mathcal{N}=2$ preserving $1d$ subspace also preserves a $U(1)^{2}\times USp(6)\times SU(2)$ global symmetry, where the $SU(2)$ is the diagonal of the two intrinsic $SU(2)$ groups while we break $SU(6)\to USp(6)$. The generally preserved $USp(4)$ group is embedded as $USp(4)\times U(1) \subset USp(6)$. In the $\mathcal{N}=1$ $1d$ subspace we break $SU(6)\to USp(4)\times U(1)$.

\item In case $10$ one breaks $SU(8)\to USp(8)$.

\item In case $11$ the $\mathcal{N}=2$ preserving $1d$ subspace also preserves a $U(1)^{3}\times SU(3)$ global symmetry, where the $SU(3)$ is the diagonal of the two intrinsic $SU(3)$ groups.

\item In case $12$ the $\mathcal{N}=2$ preserving $1d$ subspace also preserves a $U(1)^{2}\times SU(3)$ global symmetry, where the $SU(3)$ is the diagonal of the two intrinsic $SU(3)$ groups.

\item In case $13$ the $SU(9)$ is the diagonal of the two intrinsic $SU(9)$ groups.

\item In case $14$ the $\mathcal{N}=2$ preserving $1d$ subspace also preserves a $U(1)^{3}\times SU(5)$ global symmetry, where in the $SU(5)$ is the diagonal of the two intrinsic $SU(5)$ groups. In the $\mathcal{N}=1$ $1d$ subspace we break both $SU(5) \to USp(4)\times U(1)$.

\item In case $15$ the $\mathcal{N}=2$ preserving $1d$ subspace also preserves a $U(1)^{3}\times SU(2)$ global symmetry, where the $SU(2)$ is the diagonal of the two intrinsic $SU(2)$ groups.

\item In case $16$ the $\mathcal{N}=2$ preserving $1d$ subspace also preserves a $U(1)^{3}$ global symmetry.

\item In case $17$ the $SU(6)$ is the diagonal of the two intrinsic $SU(6)$ groups. In addition one breaks $SU(2)\to SO(2)$.

\item In case $18$ the $\mathcal{N}=2$ preserving $1d$ subspace also preserves a $U(1)^{4}\times SU(2)$ global symmetry, where the $SU(2)$ is the diagonal of the intrinsic $SU(2)_F$ and $SU(2)_{\overline{F}}$ groups. In addition one breaks $SU(2)_{\textbf{20}}\to SO(2)$ in both subspaces.

\item In case $19$ the $SU(3)$ is the diagonal of the intrinsic $SU(3)_F$ and $SU(3)_{\overline{F}}$ groups. In addition one breaks $SU(3)_{\textbf{20}}\to SO(3)$, where the $SU(2)$ is the $SO(3)$.

\item In case $20$ one breaks $SU(4)_{\textbf{20}}\to SO(4)$, where the $SU(2)^2$ is the $SO(4)$.

\item In case $21$ the $SU(4)$ is the diagonal of the two intrinsic $SU(4)$ groups.

\end{itemize}

\subsection{Symplectic groups}

Here we take the group to be $USp(2N)$. The case of $\mathcal{N}=4$ is trivial as there is no symmetric cubic invariant of the adjoint (symmetric) representation. Thus, there is only a $1d$ subspace preserving $\mathcal{N}=4$ and the full $SU(3)$ symmetry exchanging the three adjoints. All the purely $\mathcal{N}=2$ cases have one adjoint chiral field coming from the vector multiplet and we do not write it down in the table. All of the theories have a single ${\cal N}=2$ preserving direction. As in the special unitary cases some of the cases have additional ${\cal N}=1$ preserving $1d$ subspaces.

\begin{table}[htbp]
\begin{center}
\begin{adjustwidth}{-.2cm}{}
\begin{tabular}{|c|c|c|c|c|c|}
  \hline 
 &  Matter & $dim\, {\cal M}$ & $G_F^{free}$ & $G^{gen}_F$  & $a$, $c$  \\
 \hline
$1$&  $G=USp(2N),$ &  $1$ & $U(1)\times$ & $U(1)$ & $a=\frac{N\left(14N+9\right)}{24},$ \\
&	$N_{f}=4N+4$ &  & $SU(4N+4)$ & $\times SO(4N+4)$ & $c=\frac{N\left(4N+3\right)}{6}$ \\
\hline
$2$&  $G=USp(2N),$ & $31$ & $U(1)^{2}\times SU(2)$ & $\emptyset$ & $a=\frac{12N^2+12N-1}{24},$ \\
&	$N_{AS}=2,N_{f}=8$ & ($27$ for & $\times SU(8)$ & for $N>3$ has an $\mathcal{N}=1$ & $c=\frac{6N^2+9N-1}{12}$ \\
&	 & $N=2$) &  & $2d$ subspace preserving &  \\
&	 & &  & $U(1)\times USp(8)$ &  \\
&	 & &  & for $N=3$ has an $\mathcal{N}=1$ &  \\
&	 & &  & $1d$ subspace preserving &  \\
&	 & &  & $U(1)^2\times USp(8)$ &  \\
\hline
$3$&  $G=USp(4),$ & $3$ & $U(1)^{2}$ & $U(1)^{3}\times SU(2)$  & $a=\frac{17}{6},$ \\
&	$N_{AS}=4, N_{f}=4$ &  & $\times SU(4)^{2}$ & has an $\mathcal{N}=1$ $1d$ & $c=\frac{19}6$ \\
&	 & &  & subspace preserving &  \\
&	 & &  & $U(1)^{2}\times SU(2)\times USp(4)$ &  \\
\hline
$4$&  $G=USp(4),$ & $1$ & $U(1)\times SU(6)$ & $U(1)\times USp(6)$ & $a=\frac{65}{24},$ \\
&	$N_{AS}=6$ &  &  &  & $c=\frac{35}{12}$ \\
\hline
$5$&  $G=USp(4),$ & $2$ & $U(1)$ & $U(1)$  & $a=\frac{29}{12},$ \\
&	$N_{{\bf 16}}=1$ &  &  &  & $c=\frac{7}3$ \\
 \hline
 \end{tabular}
\end{adjustwidth}
 \end{center}
\caption{} 
\label{USpGenExt1}
\end{table}

\begin{table}[htbp]
\begin{center}
\begin{tabular}{|c|c|c|c|c|c|}
  \hline 
 &  Matter & $dim\, {\cal M}$ & $G_F^{free}$ & $G^{gen}_F$  & $a$, $c$  \\
\hline
$6$&  $G=USp(6),$ & $14$ & $U(1)\times SU(4)$  & $\emptyset$ & $a=\frac{133}{24},$ \\
&	$N_{AS}=4$ & &  & has an $\mathcal{N}=1$ $1d$ & $c=\frac{35}6$ \\
&	 & &  & subspace preserving &  \\
&	 & &  & $U(1)^3$ &  \\
\hline
$7$&  $G=USp(6),$ & $1$ & $U(1)^{2}$ & $U(1)\times SO(11)$ & $a=\frac{145}{24},$ \\
&	$N_{{\bf 14'}}=1,N_{f}=11$  &  & $\times SU(11)$ &  & $c=\frac{41}6$ \\
\hline
$8$&  $G=USp(6),$ & $12$ & $U(1)^{3}\times SU(2)$ & $\emptyset$ & $a=\frac{45}{8},$ \\
&	$N_{{\bf14'}}=1,$  &  & $\times SU(3)$ & & $c=6$ \\
&	$N_{AS}=2,N_{f}=3$  &  &  & & \\
\hline
$9$&  $G=USp(6),$ & $1$ & $U(1)^{2}\times SU(2)$ & $U(1)^{2}\times SU(4)$ & $a=\frac{137}{24},$ \\
&	$N_{{\bf 14'}}=2,N_{f}=6$  & & $\times SU(6)$ &  & $c=\frac{37}6$ \\
\hline
$10$&  $G=USp(6),$ & $1$ & $U(1)^{2}\times SU(3)$ & $U(1)\times SU(2)$ & $a=\frac{43}{8},$ \\
&	$N_{{\bf 14'}}=3,N_{f}=1$  &  &  & & $c=\frac{11}2$ \\
\hline
$11$&  $G=USp(8),$ & $1$ & $U(1)^{2}\times SU(6)$ & $U(1)\times SU(4)$ & $a=\frac{19}{2},$ \\
&	$N_{{\bf 48}}=1,N_{f}=6$  &  &  &  & $c=10$ \\
 \hline
 \end{tabular}
 \end{center}
\caption{} 
\label{USpGenExt2}
\end{table}

Let us make several comments about specific cases,
\begin{itemize}

\item In case $1$ one breaks $SU(4N+4)\to SO(4N+4)$.

\item In case $2$ the $\mathcal{N}=2$ preserving $1d$ subspace also preserves a $U(1)\times SU(2)\times SO(8)$ global symmetry, where one breaks $SU(8)\to SO(8)$. For $N\geq 3$, there is also an $\mathcal{N}=1$ $2d$ subspace along which one breaks $SU(8)\to USp(8)$ and $SU(2)\to U(1)$. For $N=3$, it also contains a $1d$ subspace where an additional $U(1)$ can be preserved. 

\item In case $3$ the $\mathcal{N}=2$ preserving $1d$ subspace also preserves a $U(1)\times SU(2)^{2}\times USp(4)$ global symmetry, where one breaks $SU(4)_{F}\to SO(4)$, and the $SU(2)^2$ is the $SO(4)$. In addition, one breaks $SU(4)_{AS}\to USp(4)$. One can further continue and break the $SO(4)$ to its Cartan, and $U(1)\times USp(4)\rightarrow U(1)\times SU(2)^2\rightarrow U(1)\times SU(2)$. For the $\mathcal{N}=1$ $1d$ subspace one instead breaks $SU(4)_{F}\to USp(4)$ and $SU(4)_{AS}\to SU(2)\times U(1)^2$. 

\item In case $4$ one breaks $SU(6)\to USp(6)$.

\item In case $6$ the $\mathcal{N}=2$ preserving $1d$ subspace also preserves a $U(1)\times USp(4)$ global symmetry, where one breaks $SU(4)\to USp(4)$. For the $\mathcal{N}=1$ $1d$ subspace one breaks $SU(4)\to U(1)^3$.

\item In case $7$ one breaks $SU(11)\to SO(11)$.

\item In case $8$ the $\mathcal{N}=2$ preserving $1d$ subspace also preserves a $U(1)\times SU(2)^{2}$ global symmetry, where one breaks $SU(3)\to SO(3)$, which is one of the $SU(2)$ groups, while the other is the intrinsic $SU(2)_{AS}$ group.

\item In case $9$ one breaks $SU(6)\to SO(6)$, where the $SU(4)$ is the $SO(6)$. In addition one breaks $SU(2)\to SO(2)$, where one of the $U(1)^2$ groups is the $SO(2)$ and the other is a combination of the intrinsic $U(1)$ groups.

\item In case $10$ one breaks $SU(3)\to SO(3)$, where the $SU(2)$ is the $SO(3)$.

\item In case $11$ one breaks $SU(6)\to SO(6)$, where the $SU(4)$ is the $SO(6)$.

\end{itemize}

\subsection{Orthogonal groups}

Here we take the group to be $SO(N)$. The case of $\mathcal{N}=4$ is trivial as there is no symmetric cubic invariant of the adjoint (antisymmetric) representation. Thus, there is only a $1d$ subspace preserving $\mathcal{N}=4$ and the full $SU(3)$ symmetry exchanging the three adjoints. All the $\mathcal{N}=2$ cases have one adjoint chiral field coming from the vector multiplet and we do not write it down in the table. All of the theories have only a single ${\cal N}=2$ preserving direction.

\begin{table}[htbp]
\begin{center}
\begin{adjustwidth}{-.2cm}{}
\begin{tabular}{|c|c|c|c|c|c|}
  \hline 
 &  Matter & $dim\, {\cal M}$ & $G_F^{free}$ & $G^{gen}_F$  & $a$, $c$  \\
 \hline
$1$&  $G=SO(N),$ &  $1$ & $U(1)$ & $U(1)$ & $a=\frac{N\left(7N-9\right)}{48},$ \\
&	$N_{v}=2N-4$ &  & $\times SU(2N-4)$ & $\times USp(2N-4)$ & $c=\frac{N\left(2N-3\right)}{12}$ \\
\hline
$2$&  $G=SO(7),$ & $1$ & $U(2x)$ & $U(1)\times USp(2x)$  & $a=\frac{140+x}{24},$ \\
&	$N_{\textbf{8}}=2x,$ &  & $\times U(10-2x)$ & $\times USp(10-2x)$ & $c=\frac{77+x}{12}$ \\
&	$N_{v}=10-2x$ &  &  &  &  \\
\hline
$3$&  $G=SO(8),$ & $1$ & $U(2x)\times U(2y)\times$ & $U(1)\times USp(2x)$  & $a=\frac{47}{6},$ \\
&	$N_{\textbf{8}_S}=2x,N_{\textbf{8}_C}=2y,$ &  & $ U(12-2x-2y)$ & $\times USp(2y)\times$ & $c=\frac{26}{3}$ \\
&	$N_{v}=12-2x-2y$ &  &  & $USp(12-2x-2y)$ &  \\
\hline
$4$&  $G=SO(9),$ & $1$ & $U(1)^{2}\times SU(2x)$ & $U(1)\times USp(2x)$  & $a=\frac{243-2x}{24},$ \\
&	$N_{\textbf{16}}=2x,$ &  & $\times SU(14-4x)$ & $\times USp(14-4x)$ & $c=\frac{135-2x}{12}$ \\
&	$N_{v}=14-4x$ &  &  &  &  \\
\hline
$5$&  $G=SO(10),$ & $1$ & $U(x)^{2}$ & $U(1)^{2}\times SU(x)$  & $a=\frac{305-4x}{24},$ \\
&	$N_{\textbf{16}}=N_{\overline{\textbf{16}}}=x,$ &  & $\times U(16-4x)$ & $\times USp(16-4x)$ & $c=\frac{85-2x}{6}$ \\
&	$N_{v}=16-4x$ &  &  &  &  \\
\hline
$6$&  $G=SO(11),$ & $1$ & $U(1)^{2}\times SU(x)$ & $U(1)\times SO(x)$  & $a=\frac{187-3x}{12},$ \\
&	$N_{\textbf{32}}=x,$ &  & $\times SU(18-4x)$ & $\times USp(18-4x)$ & $c=\frac{209-6x}{12}$ \\
&	$N_{v}=18-4x$ &  &  &  &  \\
\hline
$7$&  $G=SO(12),$ & $1$ & $U(x)\times U(y)\times$ & $U(1)\times SO(x)$  & $a=\frac{225-4x-4y}{12},$ \\
&	$N_{\textbf{32}}=x,N_{\textbf{32}'}=y,$ &  & $ U(20-4x-4y)$ & $\times SO(y)\times $ & $c=\frac{63-2x-2y}{3}$ \\
&	$N_{v}=20-4x-4y$ &  &  & $USp(20-4x-4y)$ &  \\
\hline
$8$&  $G=SO(13),$ & $1$ & $U(1)^{2}\times SU(x)$ & $U(1)\times SO(x)$  & $a=\frac{533-20x}{24},$ \\
&	$N_{\textbf{64}}=x,$ &  & $\times SU(22-8x)$ & $\times USp(22-8x)$ & $c=\frac{299-20x}{12}$ \\
&	$N_{v}=22-8x$ &  &  &  &  \\
\hline
$9$&  $G=SO(14),$ & $1$ & $U(1)^{3}\times SU(8)$ & $U(1)^{2}\times USp(8)$  & $a=\frac{575}{24},$ \\
&	$N_{\textbf{64}}=N_{\overline{\textbf{64}}}=1,$ &  &  & & $c=\frac{151}{6}$ \\
&	$N_{v}=8$ &  &  &  &  \\
\hline
 \end{tabular}
\end{adjustwidth}
 \end{center}
\caption{} 
\label{SOGenExt}
\end{table}

Let us make several comments about specific cases,
\begin{itemize}

\item In case $1$ one breaks $SU(2N-4)\to USp(2N-4)$.

\item In case $2$ one breaks $SU(2x)\to USp(2x)$, and $SU(10-2x)\to USp(10-2x)$ when $x<5$.

\item In case $3$ one breaks $SU(2x)\to USp(2x)$, $SU(2y)\to USp(2y)$, and $SU(12-2x-2y)\to USp(12-2x-2y)$ when the groups are non vanishing.

\item In case $4$ one breaks $SU(2x)\to USp(2x)$, and $SU(14-4x)\to USp(14-4x)$.

\item In case $5$ one takes the diagonal $SU(x)$ of $SU(x)_{\textbf{16}}$ and $SU(x)_{\overline{\textbf{16}}}$. In addition, one breaks $SU(16-4x)\to USp(16-4x)$ when $x<4$.

\item In case $6$ one breaks $SU(x)\to SO(x)$, and $SU(18-4x)\to USp(18-4x)$.

\item In case $7$ one breaks $SU(x)\to SO(x)$, $SU(y)\to SO(y)$, and $SU(20-4x-4y)\to USp(20-4x-4y)$ when the groups are non vanishing.

\item In case $8$ one breaks $SU(x)\to SO(x)$, and $SU(22-8x)\to USp(22-8x)$.

\item In case $9$ one breaks $SU(8)\to USp(8)$.

\end{itemize}

\subsection{Exceptional groups}

With exceptional gauge groups we can only have one dimensional conformal manifolds on which the supersymmetry is preserved. The reason is that none of these groups has a cubic symmetric invariant of the adjoint representation, or any other cubic invariant containing at least two adjoints, and thus the only exactly marginal superpotential is the ${\cal N}=2$ one.

\begin{table}[htbp]
\begin{center}
\begin{tabular}{|c|c|c|c|c|c|}
  \hline 
 &  Matter & $dim\, {\cal M}$ & $G_F^{free}$ & $G^{gen}_F$  & $a$, $c$  \\
 \hline
$1$&  $G=E_8,\; N_{{\bf 248}}=3$ & $1$ & $SU(3)$ & $SU(3)$  & $a=62,$ \\
&	 &  &  & ${\cal N}=4$ & $c=62$ \\
 \hline
 $2$&  $G=E_7,\; N_{{\bf 133}}=3$ & $1$ & $SU(3) $ & $SU(3)$  & $a=\frac{133}4,$ \\
&	 &  &  & ${\cal N}=4$ & $c=\frac{133}4$ \\
 \hline
  $3$&  $G=E_7,\; N_{{\bf 133}}=1$ & $1$ & $SU(6)\times U(1) $ & $SU(4)\times U(1)$  & $a=\frac{833}{24},$ \\
&	$N_{{\bf 56}}=6$ &  &  & ${\cal N}=2$ & $c=\frac{217}6$ \\
 \hline
  $4$&  $G=E_6,\; N_{{\bf 78}}=3$ & $1$ & $SU(3)$ & $SU(3)$  & $a=\frac{39}{2},$ \\
&	 &  &  & ${\cal N}=4$ & $c=\frac{39}2$ \\
 \hline
   $5$&  $G=E_6,\; N_{{\bf 78}}=1$ & $1$ & $SU(4)^2\times U(1)^2 $ & $SU(4)\times U(1)^2$  & $a=\frac{83}{4},$ \\
&	$N_{{\bf 27}}=N_{{\bf \overline{27}}}=4$ &  &  & ${\cal N}=2$ & $c=22$ \\
 \hline
    $6$&  $G=F_4,\; N_{{\bf 52}}=3$ & $1$ & $SU(3)$ & $SU(3)$  & $a=13,$ \\
&	 &  &  & ${\cal N}=4$ & $c=13$ \\
 \hline
    $7$&  $G=F_4,\; N_{{\bf 52}}=1$ & $1$ & $SU(6)\times U(1) $ & $USp(6)\times U(1)$  & $a=\frac{169}{16},$ \\
&	$N_{{\bf 26}}=6$ &  &  & ${\cal N}=2$ & $c=\frac{91}6$ \\
 \hline
    $8$&  $G=G_2,\; N_{{\bf 14}}=3$ & $1$ & $SU(3)$ & $SU(3)$  & $a=\frac{7}2,$ \\
&	 &  &  & ${\cal N}=4$ & $c=\frac{7}2$ \\
 \hline
    $9$&  $G=G_2,\; N_{{\bf 14}}=1$ & $1$ & $SU(8)\times U(1) $ & $USp(8)\times U(1)$  & $a=\frac{49}{12},$ \\
&	$N_{{\bf 7}}=8$ &  &  & ${\cal N}=2$ & $c=\frac{14}{3}$ \\
  \hline
\end{tabular}
 \end{center}
\caption{} 
\label{ExcExt}
\end{table}

Let us make several comments about specific cases,
\begin{itemize}
\item In case $3$ one breaks $SU(6)\to SO(6)$, where the $SU(4)$ is the $SO(6)$ group.

\item In case $5$ the $SU(4)$ is the diagonal of the two intrinsic $SU(4)$ groups.

\item In case $7$ one breaks $SU(6)\to USp(6)$.

\item In case $9$ one breaks $SU(8)\to USp(8)$.

\end{itemize}

\

\section{Epilogue}\label{sec:epilogue}

In this paper we have obtained a rich variety of conformal gauge theories with a simple gauge group.
Let us end the discussion with two simple examples of further possible developments of the results presented in the paper. First we will discuss an IR duality connecting one of the conformal theories we have obtained
to a somewhat simpler non-conformal gauge theory. Second,  we will discuss an ${\cal N}=1$ extension of the conformal manifolds of $A$ type class ${\cal S}$ theories by gauging the symmetry of a maximal puncture with an ${\cal N}=1$ vector multiplet and two adjoint chiral superfields.

\subsection{An IR duality}
\label{sec:irduality}
 
Some of the conformal theories we have classified might be related by IR dualities, of the sort first discussed in \cite{Seiberg:1994pq}, to other models. As an example let us consider the $USp(4)=Spin(5)$ theory with two fields, $q$ and $\widetilde q$, in the five dimensional two index anti-symmetric representation and one field, $\phi$, in the ${\bf 14}$.  We have obtained that this is a conformal theory with a one dimensional conformal manifold with a $U(1)$ symmetry preserved on a generic locus. The superpotential is $W=q\widetilde q \phi +\phi^3$ and under the preserved $U(1)_a$, $\phi$ is neutral while $q$ is charged $+1$ and $\widetilde q$ charged $-1$. The $\text{Tr} R U(1)_a^2$ anomaly is $-\frac{10}3$, the conformal anomalies are $a=\frac{19}8$ and $c=\frac94$, while the rest of the `t Hooft anomalies vanish. The free fixed point has an $SU(2)_a\times U(1)_b$ symmetry such that $U(1)_a$ is the Cartan of $SU(2)_a$ and under $U(1)_b$ the fields $q$ and $\widetilde q$ have charge $+1$ while $\phi$ has charge $-\frac27$. This means that the marginal operators $q^2\phi$, $\widetilde q^2 \phi$, and $q\widetilde q \phi$ are in the ${\bf 3}$ of $SU(2)_a$ and have charge $\frac{12}7$ under $U(1)_b$. The marginal operator $\phi^3$ is a singlet of $SU(2)_a$ and has charge $-\frac67$ under $U(1)_b$. There is a single independent invariant we can build out of ${\bf 3}_{\frac{12}7}$ and ${\bf 1}_{-\frac67}$, implying that the conformal manifold is one dimensional. The Cartan of $SU(2)_a$ is not broken on the conformal manifold.
 
Now, let us consider an $SU(4)$ gauge theory with two fields in the ten dimensional two index symmetric representation, field $Q$ in ${\bf 10}$ and field $\widetilde Q$ in $\overline{\bf 10}$. We also add a gauge singlet field $\Phi$ coupling through a superpotential $W=\Phi Q\widetilde Q$. The superconformal R-charges of the fields $Q$ and $\widetilde Q$ computed by $a$ maximization \cite{Intriligator:2003jj}  is $\frac13$ meaning that $Q\widetilde Q$ is a free field \cite{Kutasov:2003iy}. The superpotential removes this operator in the IR \cite{Barnes:2004jj} (see also {\it e.g.} \cite{Benvenuti:2017lle}). The superpotential has a non-anomalous $U(1)$ symmetry under which $Q$ is charged $+\frac12$ and $\widetilde Q$ is charged $-\frac12$. The $\text{Tr} R U(1)^2$ anomaly is $-\frac{10}3$, the conformal anomalies are $a=\frac{19}8$ and $c=\frac94$, while the rest of the `t Hooft anomalies vanish.
This matches perfectly the $USp(4)$ theory. Computing the superconformal index in both duality frames we obtain,

\be
&&1+\left(a+\frac1a\right)^2 x^4+\left(a+\frac1a\right)^2 \left(n+\frac1n\right) x^7+\left(a^4+\frac{1}{a^4}+a^2+\frac{1}{a^2}+a+\frac{1}{a}+4\right) x^8-\nonumber\\
&&\;\;\;\;\;\;\;(n+\frac1n)x^9+\left(1+(2+a^2+\frac{1}{a^2}+a+\frac{1}{a})(n^2+\frac1{n^2}+1)\right)x^{10}+\nonumber\\
&&\;\;\;\;\;\;\;\;\;\;\;\;\;\;\;\;(n+\frac1n)\left(a^4+\frac{1}{a^4}+3 a^2+\frac{3}{a^2}+a+\frac{1}{a}+5\right)x^{11}\\
&&+\left(a^6+\frac{1}{a^6}+a^4+\frac{1}{a^4}+a^3+\frac{1}{a^3}+2 a^2+\frac{2}{a^2}+2-(n^2+\frac1{n^2})(1+a+\frac1a)\right)x^{12}+\cdots\,,\nonumber
\ee  where we define $p/q =n^2$ and $p q= x^6$, while $a$ is the fugacity for $U(1)_a$. Note that at order $pq$, at which we should observe the contribution of marginal operators minus the conserved currents \cite{Bah:2012dg}, we have a zero. However, as we have the $U(1)_a$ symmetry conserved current, which contributes $-1$ at order $pq$,  there has to be a $+1$ contribution which comes from the exactly marginal deformation.

\subsection{Extending the conformal manifolds of class ${\cal S}$ with two adjoints}\label{sec:classs}

Let us consider general ${\cal N}=2$ class ${\cal S}$ \cite{Gaiotto:2009we} theory corresponding to compactification of the $6d$ $A_{N-1}$ type $(2,0)$ theory on a genus $g$ Riemann surface with $s$ 
maximal punctures. From ${\cal N}=1$ perspective these theories have a $U(1)_t$ global symmetry which is the Cartan of the $6d$ $SU(2)$ global symmetry, in addition to the $(1,0)$  R-symmetry, that the $(2,0)$ models have. Each puncture contains a moment map operator $M_i$ which is in the adjoint representation of the $SU(N)_i$ symmetry associated to the $i$'th puncture. The moment map operators have the superconformal R-charge assignment 
$4/3$ and are charged $+1$ under the $U(1)_t$ symmetry.  We also note that dimension $r$ Coulomb branch operators have R-charge assignment $\frac23\, r$ and have $U(1)_t$ charge $-r$. 
The  anomalies of the $SU(N)_i$ symmetry are those of $N$ free hypermultiplets in the fundamental representation of $SU(N)$ using this R-charge.
 Let us consider the case  of $N>2$  and gauge the $SU(N)$ symmetry associated to one of the punctures with an addition of two adjoint fields $X$ and $Y$.  The gauging is conformal for $N>2$ with the superconformal R-symmetry of the adjoint 
fields being $\frac23$.  The  $U(1)_t$ charge of the adjoint fields is $-\frac14$.  Note that we have an $SU(2)_a$ symmetry rotating the two adjoint fields. To explore  the conformal manifold we need to turn on some superpotentials which will break the $SU(2)_a\times U(1)_t$ symmetry, to which we turn next. 
The most general marginal superpotential built from operators charged under the gauged symmetry of the puncture is,

\be\label{marginalsup}
W=M\,(X+Y)+\left(X^3+Y^3+X^2Y+Y^2X\right)\,.
\ee Here $M$ is the moment map operator associated with the puncture.  Note that the model has also dimension $3$ Coulomb branch operators charged $-3$ under $U(1)_t$ which is marginal. The first term in the superpotential above is a doublet of $SU(2)_a$ and has $U(1)_t$ charge $\frac34$ while the second term is ${\bf 4}$ of $SU(2)_a$ with $U(1)_t$ charge $-\frac34$. We can easily compute the K\"ahler quotient and see that it has dimension one sub-locus which preserves a linear combination of $U(1)_a$ and $U(1)_t$. We can use the $SU(2)_a$ symmetry to choose this combination to be $U(1)_{t-\frac34 a}\equiv U(1)_{t'}$ and thus the superpotential we turn on is,\footnote{This  is reminiscent of superpotential deformations of SQCD with two adjoints \cite{Intriligator:2003mi}. Note that considering more general ${\cal N}=1$ class ${\cal S}$ theories of \cite{Benini:2009mz,Bah:2012dg} RG flows are involved and the superconformal R-symmetries might allow for more choices  of superpotentials from the $ADE$ classification of \cite{Intriligator:2003mi,Kutasov:1995np,Brodie:1996vx,Kutasov:2014yqa,Intriligator:2016sgx}. It would be interesting to explore these possibilities.}

\be\label{superpotential}
W=\lambda_1 M\, X+\lambda_2 Y^2X\,.
\ee Note that out of the $6$ superpotential terms in \eqref{marginalsup} the operators $M\, Y$ and $X^2Y$ recombine with the conserved currents of $SU(2)_a$, while  $Y^3$ and $X^3$ have charges $2$ and $-4$ respectively under $U(1)_{t'}$. The dimension three Coulomb branch operators also have $U(1)_{t'}$. We can in principle further  break $U(1)_{t'}$ but we will refrain from doing so.
Note also that out of the  two independent combinations of the form \eqref{superpotential} one recombined with one of the broken conserved current and one remains exactly marginal.
The gauging with the deformation preserving $U(1)_{t'}$ thus produces a theory with  $(3g-3+s)+g+1$ exactly marginal deformations, where the first $3g-3+s$ come from the complex structure moduli of  the original class ${\cal S}$ theory and $g$ from the flat connections of the $U(1)_t$ symmetry \cite{Bah:2012dg,Razamat:2016dpl}. Performing this procedure for  the other punctures we add one exactly marginal dimension per puncture and the preserved symmetry is $U(1)_{t-\frac34\sum_{i=1}^s a_i}\equiv U(1)_{t'}$. If we apply the procedure to all punctures, 
the theory will have $3g-3+2s+g$ exactly marginal deformations. Note that the original class ${\cal S}$ model has a duality group acting on the conformal manifold and as our procedure extends this manifold, we expect this duality also to act after the extension. It is tempting to entertain the possibility that applying the procedure outlined here one can interpret the resulting model as corresponding to genus $g$ surface but now with $2s$ punctures of some sort (say $s$ of one kind and $s$ of another) which would mean that the duality group would be the mapping class group of such a surface. It would be very interesting to understand the structure of the conformal manifold in more detail.

\section*{Acknowledgments}

We are grateful to Zohar Komargodski, Samson Shatashvili, Jaewon Song and Yuji Tachikawa for very useful discussions.
The research of SSR and ES is supported in part by Israel Science Foundation under grant no. 2289/18, by I-CORE  Program of the Planning and Budgeting Committee, and by BSF grant no. 2018204. GZ is supported in part by World Premier International Research Center Initiative (WPI), MEXT, Japan, by the ERC-STG grant 637844-HBQFTNCER and by the INFN. SSR would like to thank the Simons Center for Geometry and Physics, the organizers of the 2019 Summer Simons Workshop, and the organizers of the Pollica summer workshop for hospitality
during different stages of this project. The Pollica workshop was
funded in part by the Simons Foundation (Simons Collaboration on the Non-perturbative Bootstrap) and in
part by the INFN, and the authors are grateful for this support.

\appendix

\section{$\mathcal{N}=1$ superconformal index}\label{app:indexdef}
In this appendix we give some definitions on the $\mathcal{N}=1$ superconformal index \cite{Kinney:2005ej,Romelsberger:2005eg,Dolan:2008qi}, in addition to some related notations, and useful results.
In four dimensions the index is defined as a trace over the Hilbert space of the theory quantized on $\mathbb{S}^3$,
\be
\mathcal{I}\left(\mu_i\right)=Tr(-1)^F e^{-\beta \delta} e^{-\mu_i \mathcal{M}_i},
\ee
where $\delta\triangleq \half \left\{\mathcal{Q},\mathcal{Q}^{\dagger}\right\}$, with $\mathcal{Q}$ one of the Poincar\'e supercharges, and $\mathcal{Q}^{\dagger}=\mathcal{S}$ its conjugate conformal supercharge, $\mathcal{M}_i$ are $\mathcal{Q}$-closed conserved charges and $\mu_i$ their associated chemical potentials. The non-vanishing contributions come from states with $\delta=0$ making the index independent on $\beta$, since states with $\delta>0$ come in boson/fermion pairs.

The full contribution for a chiral superfield in the fundamental representation of $SU(N)$ or $USp(2N)$ with R-charge $r$ can be written in terms of elliptic gamma functions, as follows
\be
\mathcal{I}_{\chi}(SU(N)) & \equiv & \prod_{i=1}^{N} \Gamma_e \left((pq)^{\half r} z_i \right), \nonumber \\
\mathcal{I}_{\chi}(USp(N)) & \equiv & \prod_{i=1}^{N} \Gamma_e \left((pq)^{\half r} z_i^{\pm 1} \right), \nonumber \\
\Gamma_e(z)\triangleq\Gamma\left(z;p,q\right) & \equiv & \prod_{n,m=0}^{\infty} \frac{1-p^{n+1} q^{m+1}/z}{1-p^n q^m z},
\ee
where $\{z_i\}$ with $i=1,...,N$ are the fugacities parameterizing the Cartan subalgebra of $SU(N)$, with $\prod_{i=1}^{N} z_i = 1$ or $USp(2N)$. In addition, we will often use the shorten notation
\be
\Gamma_e \left(x y^{\pm n} \right)=\Gamma_e \left(x y^{n} \right)\Gamma_e \left(x y^{-n} \right).
\ee

In a similar manner we can write the full contribution of the vector multiplet in the adjoint of $SU(N)$ or $USp(2N)$, together with the matching Haar measure and projection to gauge singlets as
\be
\mathcal{I}_{V}(SU(N)) & = & \frac{\kappa^{N-1}}{N !} \oint_{\mathbb{T}^{N-1}} \prod_{i=1}^{N-1} \frac{dz_i}{2\pi i z_i} \prod_{k\ne \ell} \frac1{\Gamma_e(z_k/z_\ell)}\cdots,\nonumber\\
\mathcal{I}_{V}(USp(2N)) & = & \frac{\kappa^{N}}{2^N N!} \oint_{\mathbb{T}^{N}} \prod_{a=1}^{N} \frac{dz_a}{2\pi i z_a} \frac1{\Gamma_e\left(z_a^{\pm 2}\right)} \prod_{1<a<b<N} \frac1{\Gamma_e\left(z_a^{\pm 1} z_b^{\pm 1}\right)}\cdots,
\ee
where the dots denote that it will be used in addition to the full matter multiplets transforming in representations of the gauge group. The integration is a contour integration over the maximal torus of the gauge group. $\kappa$ is the index of $U(1)$ free vector multiplet defined as
\be
\kappa \triangleq (p;p)(q;q),
\ee
where
\be
(x;y) \triangleq \prod_{n=0}^\infty \left( 1-xy^n \right)
\ee
is the q-Pochhammer symbol.

\section{Algorithm for finding conformal manifolds and full calculation examples}\label{app:confmancalc}

In the bulk of the paper we have given the results for the analysis of the conformal manifold for a variety of cases, but have not provided details for how the calculation is done. In this appendix we will provide such details and illustrate them with a variety of examples. The general algorithm we will use is\footnote{Such algorithms can be applied more generally to approach the problem of analyzing a K\"ahler quotient. One notable example of physical use is the analysis of the Higgs branch of ${\cal N}=2$ theories, which can be described as a hyperk\"ahler quotient. The methods used here then, also have parallels in the analysis of such Higgs branches, see \cite{Bourget:2020sja}.}:
 
\begin{enumerate}

\item Pick a theory consistent with conditions \eqref{AnomalyC} and \eqref{BetaZero}.

\item List all the marginal operators invariant under the gauge symmetry, these operators must be cubic in our case\footnote{Here we ignore the marginal operator associated with the gauge coupling constant, and similarly we ignore the anomalous symmetry when listing the global symmetry of the theory. This comes about as the gauge coupling of the simple gauge groups discussed here becomes marginally irrelevant as it recombine with the broken current operator associated with the anomalous symmetry. However, it should be noted that in some quiver theories, the number of anomalous symmetries may be smaller than the number of coupling constants, in which case the analysis should be done more carefully.}.

\item The next stage is to find at least one exactly marginal operator. The problem of finding such operators can be mapped to finding operators composed of the above marginal operators that are invariant under the flavor symmetry of the theory. This is done in order to prove that the K\"ahler quotient isn't empty. It is convenient here to separate the global symmetry into its abelian and non-abelian parts, and consider each in turn:

	\begin{enumerate}
	
	\item First look for invariants under the abelian symmetries, this can be easily found after assigning non-anomalous charges to all the fields, by finding operators of vanishing charge. If no operators are found, then there is no conformal manifold.
	
	\item Next, one should look for powers of the operators invariant under the abelian symmetries that will give invariants under the non-abelian symmetries. When taking powers of the marginal operators one should only consider the symmetric product as these are bosonic operators. If no such operator can be found, then there is no conformal manifold.
	
	\end{enumerate}
	
\item With the exactly marginal operator at hand one needs to find the symmetry preserved by this operator. This can be done by considering what keeps a given representation fixed. Some examples are:

	\begin{enumerate}
	
	\item A fundamental breaks $SU(N)$ to $SU(N-1)$ which is the group fixing a chosen fundamental.
	
	\item The subgroup of $SU(N)$ fixing a chosen symmetric matrix is $SO(N)$ and an antisymmetric matrix is $USp(N)$ for $N$ even.
	
	\end{enumerate}
	
\item Some cases are more intricate then the ones above, and one needs to rely on group theory branching rules and verify the breaking using the next stage. Another possibility is taking a diagonal combination of two symmetries.
	
\item Next, we verify that the chosen subgroup of the flavor symmetry is the preserved one under the chosen exactly marginal operator. This is done by breaking the symmetries of all the currents and marginal operators under the alleged preserved symmetry, and checking that all the broken currents are eaten by the marginal operators. The remaining singlet operators give the dimension of the subspace on the conformal manifold preserving the chosen symmetry. In a more algebraic manner we break the operators $Ops \to Ops_{new}$ and break the currents that are in the adjoint representation of the flavor symmetry $G_f$ as $Adj_{G_f}\to Adj_{G_{new}} \oplus b.c.$ and need to verify that the broken currents $b.c.$ appear in $Ops_{new}$ and remove them.

\item Finally one should take the remaining operators and symmetry and go back to stage 3 of the algorithm. This recursive process stops when no additional exactly marginal operators can be found or when the flavor symmetry is broken completely.

\end{enumerate}

Next, we shall illustrate this algorithm using various examples, of cases both with and without a K\"ahler quotient. We will use this opportunity to further clarify various aspects of the calculation.

\subsection{Examples of cases with no K\"ahler quotient}

We shall first consider cases where there is no K\"ahler quotient. 

\subsubsection{Lack of a K\"ahler quotient due to no marginal operators}

The most trivial way for a K\"ahler quotient to be non-existent is that there are no marginal operators, that is the list of all marginal operators, which we determine in step 2, is empty. Examples of this include a $USp(2N)$ gauge theory with $6(N+1)$ fundamental chiral fields, an $SO(N)$ gauge theory with $3(N-2)$ vector chiral fields and an $SU(N)$ gauge theory with $3N$ chiral fields in the fundamental and anti-fundamental representations, for $N\neq 3$.  

\subsubsection{Lack of a K\"ahler quotient due to abelian symmetries}

A common reason for a K\"ahler quotient to be non-existent is due to the abelian symmetries. This happens when there is a non-anomalous $U(1)$ global symmetry under which the charges of all marginal operators have the same sign, and thus there is no K\"ahler quotient. As a simple example, consider the case of an $SU(5)$ gauge theory with six antisymmetric chiral fields, three conjugate antisymmetric chiral fields and three anti-fundamental chiral fields. We can define two non-anomalous $U(1)$ groups, say $U(1)_x \times U(1)_y$. A convenient choice is to take one, say $U(1)_x$, to act only on the antisymmetric and conjugate antisymmetric chiral fields, with charge $+1$ for the former and $-2$ for the latter, and take the other, $U(1)_y$, to act only on the conjugate antisymmetric and anti-fundamental chiral fields, with charge $+1$ for the former and $-3$ for the latter. The theory has two marginal operators, given by $AS\times \overline{F}^2$ and $\overline{AS}^2 \times \overline{F}$, both of which have negative charge under $U(1)_y$. As a result there is no quotient in this case.

As another simple example, we consider an $SU(N)$ gauge theory with three antisymmetric chiral fields, three conjugate antisymmetric chiral fields, six fundamental chiral fields and six anti-fundamental chiral fields. We can define a non-anomalous $U(1)$ group acting on the antisymmetric and conjugate antisymmetric chiral fields with charge $2$, and on the fundamental and anti-fundamental chiral fields with charge $-(N-2)$. For $N>6$, the only marginal operators are given by $AS \times \overline{F}^2$ and $\overline{AS} \times F^2$, and these both have negative charge under this $U(1)$. As a result, there is no K\"ahler quotient in these cases.  

While, in the two examples we have shown, it was readily apparent that there is no quotient under the abelian groups, in some cases, depending on the choice of $U(1)$ groups, this may be more obscure. As an example, consider an $SU(6)$ gauge theory with the following combination of chiral fields, and a specific choice of $U(1)$ groups:

\begin{table}[H]
\begin{center}
\begin{tabular}{|c|c|c|c|c|c|}
 \hline 
Representation & Number of fields & $U(1)_{x}$ & $U(1)_{y}$ & $U(1)_{b}$  & $U(1)_{a}$  \\
 \hline
 \hline
$\boldsymbol{20}$& $1$ &  $2$ &  $3$ & $0$ & $0$ \\
\hline
$AS$& $1$ &  $-3$ &  $-1$ & $0$ & $1$ \\
\hline
$\overline{AS}$& $1$ &  $0$ &  $2$ & $0$ & $-1$ \\
\hline
$F$& $11$ & $0$ &  $-1$ & $1$ & $0$ \\
\hline
$\overline{F}$& $11$ & $0$ &  $-1$ & $-1$ & $0$ \\
\hline
 \end{tabular}
 \end{center}
\caption{} 
\label{ExSU6Charges}
\end{table}

Here we list the representations of the chiral fields that participates in this case, the number of chiral fields charged under each representation, and their charges under a choice of four non-anomalous $U(1)$ groups. There are six different marginal operators that we can turn on: $AS^3$, $\overline{AS}^3$, $AS \times \overline{F}^2$, $\overline{AS} \times F^2$, $\boldsymbol{20}\times AS\times F$ and $\boldsymbol{20}\times \overline{AS}\times \overline{F}$. We can write down their charges under the four $U(1)$ symmetries. For this it is convenient to use fugacities, where we use the fugacity $i$ for $U(1)_i$. In this way, if the charge of the operator under $U(1)_i$ is $q_i$ then we write its charges under all symmetries concisely as $\prod_i i^{q_i}$. In this notation, the charges of the operators under the $U(1)$ symmetries are:

\be
\frac{a^3}{y^3 x^9}, \frac{y^6}{a^3}, \frac{b^2}{a}, \frac{a}{b^2 x^3 y^3}, \frac{a b y}{x}, \frac{x^2 y^4}{a b} . \nonumber
\ee  

Here it is not immediately apparent that there is no quotient, even though this is indeed the case. One way to see this is to redefine $U(1)_y \rightarrow U(1)_y' - U(1)_x$, which we can implement on fugcities by taking $y\rightarrow \frac{y'}{x}$. Then the first and last two operators have negative charge under $U(1)_x$, while the third and fourth one have zero charges, and as these two alone don't have a quotient under the remaining abelian symmetries, there is indeed no quotient. A more general approach is as follows. Say there is a quotient, then we can find a set of positive integers $n_i$ such that $\prod_i O^{n_i}_i = 1$, where $O_i$ are the charges of the operators in fugacities. We can then form a set of linear equations, number of which is the number of $U(1)$ groups, providing restrictions on the set of numbers $n_i$. Then, there is a quotient under the abelian symmetries if and only if there is a non-trivial solution for $n_i$ positive integers. In this case, it is possible to show with a little algebra that there is no such solution.

In many cases it is possible to know immediately if there is no quotient under the abelian symmetries thanks to the following observation. Say we have $n_i$ chiral fields charged under the gauge symmetry in the representation $\boldsymbol{R}_i$, with Dynkin index $T_{\boldsymbol{R}_i}$, then the contribution of these to the mixed gauge-$U(1)_R$ anomaly, which in turn is related to the one loop beta function, is $-\frac{n_i T_{\boldsymbol{R}_i}}{3}$. The contribution of all the other chiral fields, from the fact that $U(1)_R$ is anomaly free, is $-h^{\vee}_G + \frac{n_i T_{\boldsymbol{R}_i}}{3}$. We also note that if we define a $U(1)$ to act on all of the $n_i$ chiral fields with the same charge, $q_i$, then its contribution to the mixed anomaly between this global $U(1)$ and the gauge symmetry is $q_i n_i T_{\boldsymbol{R}_i}$, and in particular is proportional to its contribution to the mixed gauge-$U(1)_R$ anomaly. There is a similar relation also for the remaining chiral fields. We then note the following:  

\begin{enumerate}
	
	\item If we have a collection of chiral fields whose contribution to the mixed gauge-$U(1)_R$ anomaly (equivalently, to the beta function) is the same as two adjoint chiral fields, then we can define a $U(1)$ symmetry such that all chiral fields in this collection have charge $1$, while all other chiral fields have charge $-2$. It then also follows that if we have a collection of chiral fields whose contribution to the mixed gauge-$U(1)_R$ anomaly is greater than that of two adjoint chiral fields, then if we define a $U(1)$ symmetry such that all chiral fields in this collection have charge $1$, then for it to be anomaly free, all other chiral fields must have charge greater than $-2$. In this case, the only marginal operator with positive charge under this symmetry is the one when all three chiral fields that build this operator belong to this collection. All other operators have negative charge.
	
	\item If we have a collection of chiral fields whose contribution to the mixed gauge-$U(1)_R$ anomaly (equivalently, to the beta function) is the same as one adjoint chiral field, then we can define a $U(1)$ symmetry such that all chiral fields in this collection have charge $2$, while all other chiral fields have charge $-1$. It then also follows that if we have a collection of chiral fields whose contribution to the mixed gauge-$U(1)_R$ anomaly is greater than that of one adjoint chiral field, then if we define a $U(1)$ symmetry such that all chiral fields in this collection have charge $2$, then for it to be anomaly free, all other chiral fields must have charge greater than $-1$. In this case, the only marginal operator with positive charge under this symmetry is the one when at least two chiral fields that build this operator belong to this collection. All other operators have negative charge. 
	
	\end{enumerate}

With these observations in mind, it is straightforward to immediately show that in all three examples there is no quotient. In our first example, we had an $SU(5)$ gauge theory with six chiral fields in the antisymmetric representation and three chiral fields in the conjugate antisymmetric representation. On one hand, these together contribute more than two adjoints to the gauge-$U(1)_R$ anomaly, while on the other hand, there is no marginal superpotential made solely from these fields, and so there is no quotient. Similarly in our second example, we had an $SU(N)$ gauge theory with three chiral fields in the antisymmetric representation. When $N>6$, these together contribute more than one adjoint to the gauge-$U(1)_R$ anomaly. However, there is no marginal superpotential made from two or more of these fields, and so there is no quotient. Finally, we have the third example, with the matter content given in table \eqref{ExSU6Charges}. Here it is a bit trickier, but we can again notice that the contribution to the gauge-$U(1)_R$ anomaly of the chiral fields in the fundamental, the anti-fundamental and the three index antisymmetric is greater than two adjoints while there is no marginal superpotential made solely from these fields. Therefore, we can again conclude that there is no quotient. In this way it is possible to immediately rule out many cases.

\subsubsection{Lack of a K\"ahler quotient due to non-abelian symmetries}

There can be instances where it is possible to find a quotient under the abelian symmetries, but it is impossible to find one under the non-abelian symmetries. This usually happens due to one of the following two reasons:

\begin{enumerate}
	
	\item For an $SU(N)$ group with only $n$ operators in the fundamental representation, there is a quotient if and only if $n\geq N$.
	
	\item For an $SU(N)$ with only a single operator in the antisymmetric representation, there is a quotient if and only if $N$ is even. 
	
	\end{enumerate} 

As an example, we consider a $Spin(18)$ gauge theory with a single chiral field in the spinor representation and $16$ chiral fields in the vector representation. In this case there is a single marginal operator made from the product of the vector and the symmetric product of the spinor chiral fields. The global symmetry consists of an $SU(16)$ acting on the vector chiral fields and a $U(1)$, which can be chosen so as to act on the spinor chiral field with charge $1$ and on the vector chiral fields with charge $-2$. The marginal operator then is uncharged under the $U(1)$, but is in the fundamental representation of the $SU(16)$. In this case then, it is trivial that there is a quotient under the $U(1)$, but as there is only one operator in the fundamental of $SU(16)$, there is no quotient under the non-abelian symmetry. 

\subsubsection{Lack of a K\"ahler quotient due to a combination of abelian and non-abelian symmetries}

The last case where there is no quotient is when there is a quotient under both the abelian and non-abelian symmetries, but these are mutually exclusive. As a simple example of this, consider an $SU(6)$ gauge theory with four antisymmetric chiral fields, a single chiral field in the conjugate antisymmetric representation, five fundamental chiral fields and eleven anti-fundamental chiral fields. There are four marginal operators: $AS^3$, $\overline{AS}^3$, $AS\times \overline{F}^2$, $\overline{AS}\times F^2$. The free point global symmetry is $U(1)^3\times SU(4)\times SU(5)\times SU(11)$. It is straightforward to show that there is a quotient under the abelian symmetries, if it is allowed to use all four operators. However, the only operator charged under the $SU(5)$ is $\overline{AS} \times F^2$, which is in its antisymmetric representation, and as there is no quotient for a single antisymmetric of $SU(5)$, it is in fact marginally irrelevant. The remaining three have a quotient under the non-abelian symmetries, however, it is impossible to have a quotient under the abelian symmetries with only these three operators. Therefore, although there is a quotient under both the abelian and non-abelian symmetries, there is no quotient under both simultaneously.  

Finally, as a more complicated example we consider an $SU(6)$ gauge theory with a single chiral field in its $\boldsymbol{20}$ representation, four chiral fields in the antisymmetric representation, a single chiral field in the conjugate antisymmetric representation, two chiral fields in the fundamental representation and eight chiral fields in the anti-fundamental representation. The symmetry of the free point is $U(1)^{4}\times SU(4)\times SU(2)\times SU(8)$, and we assign charges under the non-anomalous $U(1)$ groups as follows. We assign charge $(1,0,0,0)$ to the chiral field in the $\boldsymbol{20}$, $(0,1,0,0)$ to the $AS$ chiral fields, $(0,0,1,0)$ to the $\overline{AS}$ chiral field, $(1,0,2,3)$ to the $F$ chiral fields and $(-1,-2,-1,-1)$ to the $\overline{F}$ chiral fields, under $U(1)_x \times U(1)_y \times U(1)_z \times U(1)_a$. The marginal operators and their representations under the flavor symmetries are
\be
A=AS\times\overline{F}^{2} & \sim & \left(\boldsymbol{4},\boldsymbol{1},\boldsymbol{28}\right)x^{-2}y^{-3}z^{-2}a^{-2}\nonumber\\
B=\overline{AS}\times F^{2} & \sim & \left(\boldsymbol{1},\boldsymbol{1},\boldsymbol{1}\right)x^{2}z^{5}a^{8}\nonumber\\
C=AS^{3} & \sim & \left(\boldsymbol{20}'',\boldsymbol{1},\boldsymbol{1}\right)y^{3}\nonumber\\
D=\overline{AS}^{3} & \sim & \left(\boldsymbol{1},\boldsymbol{1},\boldsymbol{1}\right)z^{3}\nonumber\\
E=\boldsymbol{20}\times\overline{AS}\times\overline{F} & \sim & \left(\boldsymbol{1},\boldsymbol{1},\boldsymbol{8}\right)y^{-2}a^{-1}\nonumber\\
G=\boldsymbol{20}\times AS\times F & \sim & \left(\boldsymbol{4},\boldsymbol{2},\boldsymbol{1}\right)x^{2}yz^{2}a^{4}\,,
\ee
where we again use fugacities to represent charges for abelian groups. Here the representations under the non-abelian groups are ordered in the same order as the symmetries were listed when we noted the free point symmetry. 

Like in the previous cases, we seek first an invariant under the abelian symmetries. We attempt to find one by writing down the constraints of zero charge under each $U(1)$ as linear equations in the number of times we use each operator, and look for positive integer solutions. Using this, we find that the only invariant operators under the abelian symmetries are powers of $AC^{2}E^{2}G$. Next we would like to build an invariant under the non-abelian symmetries. Focusing first on building an $SU(8)$ invariant, we find two marginal operators transforming under the $SU(8)$ symmetry $A$ and $E$. In the case of operator $E$, it transforms only under the $SU(8)$ out of the non-abelian symmetries forcing us to take its symmetric power only. Taking $n$ copies of the operator invariant under the abelian symmetries we will take $E^{2n\, sym}$, giving us an operator transforming under the $2n$ index symmetric representation of $SU(8)$ which is represented by a Young tableau with a single row of length $2n$. To get an $SU(8)$ invariant, we must contract this representation with the conjugate $2n$ index symmetric representation of $SU(8)$, represented by a Young tableau of seven rows of length $2n$. Generating such a representation requires at least $7n$ copies of the two index antisymmetric representation, while we only have $n$ such copies. We see then that while we can generate an invariant under the abelian part of the global symmetry, it is impossible to make an invariant under the full symmetry from these abelian invariants. Similarly, we note that while it is possible to generate invariants under the non-abelian part of the global symmetry, it is impossible to make an invariant under the full symmetry from these non-abelian invariants. Therefore, in this case it is impossible to generate an invariant under the full flavor symmetry, meaning there are no exactly marginal operators and the conformal manifold is empty.

\subsection{Examples of cases with a K\"ahler quotient}

After showing some examples without a K\"ahler quotient, we move on to show various examples of cases with a K\"ahler quotient. We shall start with a few standard examples, to illustrate the basic features, and then move on to show a few more examples with interesting features.

\subsubsection{Simple examples with a K\"ahler quotient}

In the first example we look at the first case of table \ref{suNtwoAd} ignoring the case of $N=3$ for simplicity. This is a theory of an $SU(N)$ gauge group with two adjoint chiral fields and $N_F=N_{\overline F}=N$. One can verify it obeys \eqref{AnomalyC} and \eqref{BetaZero}. The symmetry of the free point is $U(1)^{2}\times SU(2)\times SU(N)^{2}$, and we assign charges under the non-anomalous $U(1)$ groups as follows. We assign charge $(0,+1)$ to the adjoints and $(\pm1,-2)$ to the fundamentals and anti-fundamentals under $U(1)_x \times U(1)_y$. The marginal operators and their representations under the flavor symmetries are
\be
A=Ad^{3} & \sim &\left(\boldsymbol{4},\boldsymbol{1},\boldsymbol{1}\right)y^{3}\,, \nonumber\\
B=Ad\times F\times\overline{F} & \sim & \left(\boldsymbol{2},\boldsymbol{N},\overline{\boldsymbol{N}}\right)y^{-3}\, ,
\ee
where we again use fugacities to represent charges under $U(1)$ symmetries, and order the non-abelian symmetries in the same order in which the symmetries were presented when we introduced the symmetry at the free point.

An invariant under the abelian symmetries is the operator $AB$. An invariant under the non-abelian symmetries can be constructed by taking $N$ such copies, and contracting both the $SU(N)$ indices with epsilon tensors to create an invariant under both. As for the $SU(2)$, the $\textbf{2}$ and $\textbf{4}$ representations need to be taken to the $N$-th symmetric power and multiplied to generate an $SU(2)$ invariant.

The fact that the invariant under both $SU(N)$ groups is generated by a baryon of a bifundamental implies that the subspace preserves the diagonal $SU(N)$, while both the representations under $SU(2)$ imply that it is at least broken to $U(1)_a$. The marginal operators under this breaking
\be \label{Ex1mobr}
\left(\boldsymbol{4},\boldsymbol{1},\boldsymbol{1}\right)y^{3}\oplus\left(\boldsymbol{2},\boldsymbol{N},\overline{\boldsymbol{N}}\right)y^{-3} & \to & \left(\boldsymbol{1}\right)y^{3}\left(a^{3}+a+a^{-1}+a^{-3}\right)\nonumber\\
 & & \oplus\left(\boldsymbol{Ad}_{SU(N)}\oplus\boldsymbol{1}\right)y^{-3}\left(a+a^{-1}\right)\,.
\ee

When we insert operators into the superpotential, we break the symmetry to the subgroup such that the inserted operators are uncharged under the preserved subgroup. For the case at hand, to get a K\"ahler quotient we need to insert both the operators $A$ and $B$, and so the preserved symmetry must be such that at least one component of each is a singlet. This does not yet hold in \eqref{Ex1mobr}, implying that we need to break more symmetry. The minimal choice is to break one of the $U(1)$ groups by the identification
\be
y^{3}a^{-1}=1\Rightarrow a=y^{3}\,.
\ee
This implies that we are inserting into the superpotential the component with charge $y^{3}a^{-1}$ of $A$ and with charge $y^{-3} a$ of $B$. It is important to note that indeed $A B$ is a singlet as required. 

Breaking the symmetry further according to the identification above, the marginal operators are
\be \label{Ex1moab}
\to \left(\boldsymbol{1}\right)\left(y^{12}+y^{6}+1+y^{-6}\right)\oplus\left(\boldsymbol{Ad}_{SU(N)}\oplus\boldsymbol{1}\right)\left(1+y^{-6}\right)\,.
\ee
Next we write down the broken currents
\be
b.c.=\left(\boldsymbol{1}\right)\left(y^{6}+1+y^{-6}\right)\oplus\left(\boldsymbol{Ad}_{SU(N)}\right)\,.
\ee

In the first term there are the currents of the $SU(2)$, where we have used the relation $a=y^{3}$. The second term arises as we lose an $SU(N)$ group from the breaking of $SU(N)\times SU(N)$ to the diagonal $SU(N)$ group.  

The broken current can only become so by absorbing a marginally irrelevant operator. As a result some of the marginal operators in \eqref{Ex1moab} must in fact be marginally irrelevant and eaten by the broken currents. Indeed, we see that there are marginal operators with the correct charges to be eaten. This is an important consistency check that this breaking can exist.

The remaining marginal operators after being eaten by broken currents are
\be
\left(\boldsymbol{1}\right)\oplus\left(\boldsymbol{1}\right)\left(y^{12}+y^{-6}\right)\oplus\left(\boldsymbol{Ad}_{SU(N)}\right)y^{-6}.
\ee
We see that we are left with one singlet, meaning one exactly marginal operator, and thus there is a $1d$ subspace on the conformal manifold preserving $U(1)^2 \times SU(N)$ symmetry. This is again an important consistency check, as there must always be at least one singlet corresponding to the exactly marginal operator that we have turned on.  

We next consider turning on additional marginal operators. For that we look for a collection of operators, among the remaining marginal operators on this $1d$ subspace, with a K\"ahler quotient under the preserved symmetry on that subspace. Looking at the remaining operators one can easily see that we can generate another singlet by combining the two operators $\left(\boldsymbol{1}\right)y^{12}$ and $\left(\boldsymbol{1}\right)y^{-6}$ breaking $U(1)_y$. The remaining operators after the symmetry breaking and the reduction of the broken current are
\be
2\left(\boldsymbol{1}\right)\oplus\left(\boldsymbol{Ad}_{SU(N)}\right).
\ee
Thus, we have an additional direction preserving $U(1) \times SU(N)$ together with the one we already found preserving a higher amount of symmetry.

Next, we can break $SU(N)\to SU(N-1)\times U(1)$ to find
\be
3\left(\boldsymbol{1}\right)\oplus\left(\boldsymbol{Ad}_{SU(N-1)}\right),
\ee
with an additional direction preserving $U(1)^2 \times SU(N-1)$. We can keep breaking $SU(N-1)$ in the same manner serially until the $SU(N)$ has been completely broken down to its Cartan subalgebra. This will result in an $N+1$ dimensional conformal manifold preserving $U(1)^N$. In the end of this process we are left only with exactly marginal operators, and the full conformal manifold is found.

\subsubsection*{Another example}
Next, we wish to further illustrate our method with another example. In this example we look at case $27$ appearing in Table \ref{SOEx3}. This is a theory of an $SO(10)$ gauge group with one two index traceless symmetric chiral field, three spinor chiral fields and six vector chiral fields. One can verify it obeys \eqref{BetaZero}. The symmetry of the free point is $U(1)^{2}\times SU(3)\times SU(6)$, and we assign charges under the non-anomalous $U(1)$ groups as follows. We assign charge $(1,0)$ to the spinor chirals, $(0,1)$ to the symmetric chirals and $(-1,-2)$ to the vector chirals under $U(1)_x \times U(1)_y$. The marginal operators and their representations under the flavor symmetries are
\be
A=S\times V^{2}&\sim&\left(\boldsymbol{1},\overline{\boldsymbol{21}}\right)x^{-2}y^{-3}\nonumber\\
B=S^{3}&\sim&\left(\boldsymbol{1},\boldsymbol{1}\right)y^{3}\nonumber\\
C=\boldsymbol{16}^{2}_{SO(10)}\times V&\sim&\left(\boldsymbol{6},\overline{\boldsymbol{6}}\right)xy^{-2}\,.
\ee
An operator invariant under the abelian symmetries is given by $A^{3}B^{7}C^{6}$. The easiest way to form an invariant under the non-abelian symmetries is taking two copies of $A^{3}B^{7}C^{6}$, that is to consider instead $A^{6}B^{14}C^{12}$, and then each operator can be multiplied with itself to form a singlet. For $A$, we use the $SU(6)$ invariant given by the determinant of the symmetric matrix, which is a sixth order symmetric invariant, and in fact it is the only $SU(6)$ invariant that can be generated from a single symmetric matrix of $SU(6)$. For $B$, this is immediate as it is a singlet under the non-abelian symmetries. Finally for $C$, we take twice the sixth totally antisymmetric product for both the $SU(3)$ and $SU(6)$ indices, so that the total product is symmetric, which gives a singlet under both symmetries.

The antisymmetric contractions imply we should identify a diagonal $SU(3)$ symmetry. This requires first breaking $SU(6)\to SU(3)$ and then take the diagonal of this $SU(3)$ and the intrinsic one. This indeed gives a component in $C$ with a singlet under the non-abelian groups. We next need to have a singlet component under the non-abelian groups also in operator $A$. For this we note that the two index (conjugate) symmetric representation $\overline{\textbf{21}}$ of $SU(6)$ breaks to the symmetric $\textbf{6}$ and the $\textbf{15}'$ representations of $SU(3)$. As this does not give us a singlet we need to further break the $SU(3)$ to $SU(2)$, which can be accomplished by taking either $SU(3)\to SO(3)$ or $SU(3)\to SU(2)\times U(1)_z$. As we have taken the diagonal group with an $SU(3)$ subgroup of $SU(6)$, this implies that we should also break the $SU(6)$ accordingly. Here it is important to note that while the commutant of the chosen $SU(3)$ subgroup in $SU(6)$ is empty, and the commutant of $SU(2)$ in $SU(3)$ is either empty or $U(1)_z$, depending on the choice of embedding, the commutant of $SU(2)$ in $SU(6)$ is larger. Specifically, if we break $SU(3)\to SO(3)$, then we should break $SU(6)\rightarrow U(1)_a \times SU(2)$ such that $\textbf{6} \to a\textbf{5}_{SU(2)} + \frac{1}{a^5}$, while if we break $SU(3)\to SU(2)\times U(1)_z$, then we should break $SU(6)\rightarrow U(1)_a \times U(1)_b \times SU(2)$ such that $\textbf{6} \to a\textbf{3}_{SU(2)} + \frac{b}{a}\textbf{2}_{SU(2)}+\frac{1}{a b^2}$. The marginal operators under these breakings are
\be
&&\left(\boldsymbol{1},\overline{\boldsymbol{21}}\right)x^{-2}y^{-3}\oplus\left(\boldsymbol{1},\boldsymbol{1}\right)y^{3}\oplus\left(\boldsymbol{6},\overline{\boldsymbol{6}}\right)xy^{-2}\nonumber\\
1st&\to&\left(\boldsymbol{5}\right)x^{-2}y^{-3}a^{-2}\oplus\left(\boldsymbol{4}\right)x^{-2}y^{-3}b^{-1}\oplus\left(\boldsymbol{3}\right)x^{-2}y^{-3}(b^{2}+a^2b^{-2})\oplus\left(\boldsymbol{2}\right)x^{-2}y^{-3}(b^{-1}+a^2b)\nonumber\\
&&\oplus\underline{\left(\boldsymbol{1}\right)x^{-2}y^{-3}(a^{-2}+a^2b^{4})}\oplus\underline{\left(\boldsymbol{1}\right)y^{3}}\oplus\left(\boldsymbol{5}\right)xy^{-2}z^2a^{-1}\oplus\left(\boldsymbol{4}\right)xy^{-2}\left(z^{2}ab^{-1}+z^{-1}a^{-1}\right)\nonumber\\
&&\oplus\left(\boldsymbol{3}\right)xy^{-2}\left(z^{2}a^{-1}+ab^2z^2+ab^{-1}z^{-1}+a^{-1}z^{-4}\right)\nonumber\\
&&\oplus\left(\boldsymbol{2}\right)xy^{-2}\left(z^{2}ab^{-1}+ab^2z^{-1}+z^{-1}a^{-1}+ab^{-1}z^{-4}\right)\nonumber\\
&&\oplus\left(\boldsymbol{1}\right)xy^{-2}\left(\underline{z^{2}a^{-1}}+\underline{ab^{-1}z^{-1}}+ab^{2}z^{-4}\right)\,,\\
2nd&\to&\left(a^{-2}(\boldsymbol{9}\oplus\boldsymbol{5}\oplus\underline{\boldsymbol{1}})+a^4\boldsymbol{5}+a^{10}\boldsymbol{1}\right)x^{-2}y^{-3}\oplus\underline{\left(\boldsymbol{1}\right)y^{3}}\nonumber\\
&&\oplus\left(a^{-1}(\boldsymbol{9}\oplus\boldsymbol{7}\oplus\boldsymbol{5}\oplus\boldsymbol{3}\oplus\underline{\boldsymbol{1}})+\boldsymbol{5}(a^5+a^{-1})+\underline{a^5\boldsymbol{1}}\right)xy^{-2}\,,
\ee
where the first and second refer to the breaking $SU(3)\to SU(2)\times U(1)$ and $SU(3)\to SO(3)$, respectively. The underlined operators are non-abelian singlets that represent the operators to be added to the superpotential, and as such form a quotient as dictated by the found invariant operator. We next need to break the abelian symmetries such that all these operators are singlets under these symmetries as well. In the second case, this dictates that all the abelian symmetries are to be completely broken, while in the first case, we are to identify

\bea
&&x^{-2}y^{-3}a^{-2}=x^{-2}y^{-3}a^{2}b^4=y^{3}=xy^{-2}z^2a^{-1}=xy^{-2}a^{1}z^{-1}b^{-1}=1\nonumber\\
&&\Rightarrow y=1, z=a=b^{-1}=x^{-1}\,.
\eea
The resulting marginal operators are
\be
1st&\to&2\left(\boldsymbol{5}\right)\oplus\left(\boldsymbol{4}\right)\left(x^3+2x^{-3}\right)\oplus\left(\boldsymbol{3}\right)\left(4+x^6+x^{-6}\right)\oplus3\left(\boldsymbol{2}\right)\left(x^3+x^{-3}\right)\nonumber\\  && \oplus\left(\boldsymbol{1}\right)\left(5+x^{6}\right)\,,\\
2nd&\to&2\left(\boldsymbol{9}\right)\oplus\left(\boldsymbol{7}\right)\oplus5\left(\boldsymbol{5}\right)\oplus\left(\boldsymbol{3}\right)\oplus5\left(\boldsymbol{1}\right)\,.
\ee
Next we write down the broken currents
\be
b.c.^{1st}&\to&\left(\boldsymbol{5}\right)\oplus\left(\boldsymbol{4}\right)\left(x^3+x^{-3}\right)\oplus4\left(\boldsymbol{3}\right)\oplus3\left(\boldsymbol{2}\right)\left(x^3+x^{-3}\right)\oplus4\left(\boldsymbol{1}\right)\,,\\
b.c.^{2nd}&\to&\left(\boldsymbol{9}\right)\oplus\left(\boldsymbol{7}\right)\oplus4\left(\boldsymbol{5}\right)\oplus\left(\boldsymbol{3}\right)\oplus3\left(\boldsymbol{1}\right)\,.
\ee
The remaining marginal operators after being eaten by the broken currents are
\be
1st&\to&\left(\boldsymbol{5}\right)\oplus\left(\boldsymbol{4}\right)x^{-3}\oplus\left(\boldsymbol{3}\right)\left(x^3+x^{-3}\right)\oplus\left(\boldsymbol{1}\right)x^6\oplus\left(\boldsymbol{1}\right)\,,
\\2nd&\to&\left(\boldsymbol{9}\right)\oplus\left(\boldsymbol{5}\right)\oplus2\left(\boldsymbol{1}\right)\,.
\ee
We see that we are left with one singlet in the first case and two singlets in the second case, and thus there is a $1d$ subspace of the conformal manifold preserving $U(1)\times SU(2)$ symmetry and a $2d$ subspace preserving only a different $SU(2)$ symmetry. 

Looking at the remaining operators we see various representations under the remaining symmetries that break them completely, leaving us with a $16\ominus3=13$ dimensional conformal manifold on which no symmetry is generically preserving. In the end of this process we are left only with exactly marginal operators, and the full conformal manifold is found.

\subsubsection*{Yet another example}

In this example we look at the eighth case of Table \ref{USpGen1}. This is a theory of a $USp(4)$ gauge symmetry with one adjoint chiral field, three chirals in the two index traceless antisymmetric representation and six fundamental chirals. One can verify that it obeys \eqref{BetaZero}. The symmetry at the free point is $U(1)\times SU(3)\times SU(6)$, and we assign charges under the non-anomalous $U(1)$ groups as follows. We assign charge $(1,0)$ to the adjoint chiral, $(0,1)$ to the antisymmetric chirals and $(-1,-1)$ to the fundamental chirals under $U(1)_x \times U(1)_y$. The marginal operators and their representations under the flavor symmetries are
\be
A=S^{2}\times AS&\sim&\left(\boldsymbol{3},\boldsymbol{1}\right)x^{2}y\nonumber\\
B=S\times AS^{2}&\sim&\left(\overline{\boldsymbol{3}},\boldsymbol{1}\right)xy^{2}\nonumber\\
C=S\times F^{2}&\sim&\left(\boldsymbol{1},\boldsymbol{21}\right)x^{-1}y^{-2}\nonumber\\
D=AS\times F^{2}&\sim&\left(\boldsymbol{3},\boldsymbol{15}\right)x^{-2}y^{-1}.
\ee
There are two invariants under the abelian symmetries given by the operators $AD$ and $BC$. An invariant under the non-abelian symmetries can be constructed by taking two copies of $AD$ and four copies of $BC$ all with symmetric contractions. In total the invariant operator is $A^2B^4C^4D^2$.

We have both the symmetric and antisymmetric representation under the $SU(6)$ flavor symmetry group, and thus we expect the breaking of $SU(6)\to SU(n)\times SU(6-n) \times U(1)$ where one of the special unitary subgroups will be broken to a symplectic group by the antisymmetric representation and the other to an orthogonal group by the symmetric representation. The initial breaking is determined by the relation between the numbers of the two representations in the invariant operator. In our case it is $SU(6)\to SU(4)\times SU(2) \times U(1)_a$ where the $SU(4)$ is broken to $SO(4)$, which we take to be $SU(2)^2$, and the $SU(2)$ remains unbroken as it is the same as $USp(2)$. As for the $SU(3)$, it is broken by the fundamentals as $SU(3)\to SU(2)\times U(1)_z$. The marginal operators under this breaking are
\be
&&\left(\boldsymbol{3},\boldsymbol{1}\right)x^{2}y\oplus\left(\overline{\boldsymbol{3}},\boldsymbol{1}\right)xy^{2}\oplus\left(\boldsymbol{1},\boldsymbol{21}\right)x^{-1}y^{-2}\oplus\left(\boldsymbol{3},\boldsymbol{15}\right)x^{-2}y^{-1}\nonumber\\
&\to&\underline{\left(\boldsymbol{1},\boldsymbol{1},\boldsymbol{1},\boldsymbol{1}\right)x^{2}yz^{-2}}\oplus\left(\boldsymbol{2},\boldsymbol{1},\boldsymbol{1},\boldsymbol{1}\right)x^{2}yz\oplus\underline{\left(\boldsymbol{1},\boldsymbol{1},\boldsymbol{1},\boldsymbol{1}\right)xy^{2}z^{2}}\nonumber\\
&&\oplus\left(\boldsymbol{2},\boldsymbol{1},\boldsymbol{1},\boldsymbol{1}\right)xy^{2}z^{-1}\oplus\left(\boldsymbol{1},\boldsymbol{1},\boldsymbol{1},\boldsymbol{3}\right)x^{-1}y^{-2}a^{-4}\oplus\left(\boldsymbol{1},\boldsymbol{2},\boldsymbol{2},\boldsymbol{2}\right)x^{-1}y^{-2}a^{-1}\nonumber\\
&&\oplus\left(\boldsymbol{1},\boldsymbol{3},\boldsymbol{3},\boldsymbol{1}\right)x^{-1}y^{-2}a^{2}\oplus\underline{\left(\boldsymbol{1},\boldsymbol{1},\boldsymbol{1},\boldsymbol{1}\right)x^{-1}y^{-2}a^{2}}\nonumber\\
&&\oplus\underline{\left(\boldsymbol{1},\boldsymbol{1},\boldsymbol{1},\boldsymbol{1}\right)x^{-2}y^{-1}z^{-2}a^{-4}}\oplus\left(\boldsymbol{1},\boldsymbol{2},\boldsymbol{2},\boldsymbol{2}\right)x^{-2}y^{-1}z^{-2}a^{-1}\nonumber\\
&&\oplus\left(\boldsymbol{1},\boldsymbol{3},\boldsymbol{1},\boldsymbol{1}\right)x^{-2}y^{-1}z^{-2}a^{2}\oplus\left(\boldsymbol{1},\boldsymbol{1},\boldsymbol{3},\boldsymbol{1}\right)x^{-2}y^{-1}z^{-2}a^{2}\nonumber\\
&&\oplus\left(\boldsymbol{2},\boldsymbol{1},\boldsymbol{1},\boldsymbol{1}\right)x^{-2}y^{-1}za^{-4}\oplus\left(\boldsymbol{2},\boldsymbol{2},\boldsymbol{2},\boldsymbol{2}\right)x^{-2}y^{-1}za^{-1}\nonumber\\
&&\oplus\left(\boldsymbol{2},\boldsymbol{3},\boldsymbol{1},\boldsymbol{1}\right)x^{-2}y^{-1}za^{2}\oplus\left(\boldsymbol{2},\boldsymbol{1},\boldsymbol{3},\boldsymbol{1}\right)x^{-2}y^{-1}za^{2}.
\ee
Here the symmetries are broken as $\boldsymbol{3}_{SU(3)}\rightarrow z\left(\boldsymbol{2},\boldsymbol{1},\boldsymbol{1},\boldsymbol{1}\right) + \frac{1}{z^2}$, $\boldsymbol{6}_{SU(6)}\rightarrow a\left(\boldsymbol{1},\boldsymbol{2},\boldsymbol{2},\boldsymbol{1}\right) + \frac{1}{a^2}\left(\boldsymbol{1},\boldsymbol{1},\boldsymbol{1},\boldsymbol{2}\right)$. We have underlined four operators that are singlets under the non-abelian parts of the preserved global symmetry, each one a component of one of the four marginal operators $A$, $B$, $C$ and $D$. These should represent the components we expect to be turned on when going on the subspace corresponding to the invariant combination that we found, and indeed these have a quotient under the $U(1)$ symmetries. As the symmetry preserved by the subspace must be such that these operators transform as a singlet under it, we see that some of the $U(1)$ groups need to be identified with one another. In this case the required identifications are
\be
x^{2}yz^{-2}=xy^{2}z^{2}=x^{-1}y^{-2}a^{2}=x^{-2}y^{-1}z^{-2}a^{-4}=1\Rightarrow x=z^{-2},\ y=z^{2},\ a=z^{-1}\,.
\ee
Breaking the symmetry further according to the identification above, the marginal operators are
\be
&\to&4\left(\boldsymbol{1},\boldsymbol{1},\boldsymbol{1},\boldsymbol{1}\right)\oplus\left(\boldsymbol{2},\boldsymbol{1},\boldsymbol{1},\boldsymbol{1}\right)\left(2z^{3}+z^{-3}\right)\oplus\left(\boldsymbol{1},\boldsymbol{2},\boldsymbol{2},\boldsymbol{2}\right)\left(z^{3}+z^{-3}\right)\nonumber\\
&&\oplus\left(\boldsymbol{1},\boldsymbol{1},\boldsymbol{1},\boldsymbol{3}\right)z^{6}\oplus\left(\boldsymbol{1},\boldsymbol{3},\boldsymbol{3},\boldsymbol{1}\right)\oplus\left(\boldsymbol{1},\boldsymbol{3},\boldsymbol{1},\boldsymbol{1}\right)z^{-6}\oplus\left(\boldsymbol{1},\boldsymbol{1},\boldsymbol{3},\boldsymbol{1}\right)z^{-6}\nonumber\\
&&\oplus\left(\boldsymbol{2},\boldsymbol{2},\boldsymbol{2},\boldsymbol{2}\right)\oplus\left(\boldsymbol{2},\boldsymbol{3},\boldsymbol{1},\boldsymbol{1}\right)z^{-3}\oplus\left(\boldsymbol{2},\boldsymbol{1},\boldsymbol{3},\boldsymbol{1}\right)z^{-3}\,.
\ee
Next we write down the broken currents
\be
b.c.=3\left(\boldsymbol{1},\boldsymbol{1},\boldsymbol{1},\boldsymbol{1}\right)\oplus\left(\boldsymbol{2},\boldsymbol{1},\boldsymbol{1},\boldsymbol{1}\right)\left(z^{3}+z^{-3}\right)\oplus\left(\boldsymbol{1},\boldsymbol{2},\boldsymbol{2},\boldsymbol{2}\right)\left(z^{3}+z^{-3}\right)\oplus\left(\boldsymbol{1},\boldsymbol{3},\boldsymbol{3},\boldsymbol{1}\right)\,.
\ee
Here the first term comes from the fact we lost three abelian symmetries in the above identification, the second term comes from the breaking of $SU(3)\rightarrow SU(2)\times U(1)_z$ and the last two come from the breaking of $SU(6)\rightarrow SO(4)\times USp(2)\times U(1)_a$.

The remaining marginal operators after being eaten by the broken currents
\be
&\to&\underline{\left(\boldsymbol{1},\boldsymbol{1},\boldsymbol{1},\boldsymbol{1}\right)}\oplus\left(\boldsymbol{1},\boldsymbol{1},\boldsymbol{1},\boldsymbol{3}\right)z^{6}\oplus\left(\boldsymbol{1},\boldsymbol{3},\boldsymbol{1},\boldsymbol{1}\right)z^{-6}\oplus\left(\boldsymbol{1},\boldsymbol{1},\boldsymbol{3},\boldsymbol{1}\right)z^{-6}\nonumber\\
&&\oplus\left(\boldsymbol{2},\boldsymbol{1},\boldsymbol{1},\boldsymbol{1}\right)z^{3}\oplus\left(\boldsymbol{2},\boldsymbol{2},\boldsymbol{2},\boldsymbol{2}\right)\oplus\left(\boldsymbol{2},\boldsymbol{3},\boldsymbol{1},\boldsymbol{1}\right)z^{-3}\oplus\left(\boldsymbol{2},\boldsymbol{1},\boldsymbol{3},\boldsymbol{1}\right)z^{-3}\,.
\ee
We see that we are left with one singlet, meaning one exactly marginal operator, and thus there is a $1d$ subspace on the conformal manifold preserving $U(1)\times SU(2)^{4}$ global symmetry. 

Looking at the remaining operators one can continue and generate singlets breaking the symmetry completely as there are multiple representations under each $SU(2)$. The end result is thus
\be
1 \oplus 3 \oplus 3 \oplus 3 \oplus 2 \oplus 16 \oplus 6 \oplus 6 \ominus 4 \times 3 \ominus 1 = 27\, .
\ee
The result is a $27$ dimensional conformal manifold preserving no global symmetry on a generic point. In the end of this process we are left only with exactly marginal operators, and the full conformal manifold is found.

\subsubsection{Cases with marginally irrelevant operators due to unbreakable symmetries}

So far all the examples we discussed with a K\"ahler quotient had either exactly marginal operators or marginally irrelevant operators that are eaten by broken symmetries. However, there can be cases where there is a conformal manifold, but we have marginally irrelevant operators due to them being charged under a symmetry under which there is no quotient. These cases then look like a combination of the previous cases in this subsection, and the cases of the previous subsection. Like the previous cases in this subsection, they have a non-trivial quotient, and so a conformal manifold, but when going on it, the space of marginal operators that we can turn on eventually resembles that of the cases in the previous subsection. In these cases, the dimension of the conformal manifold is not equal to the number of marginal operators minus the number of broken currents. Like in the previous subsection, the eventual lack of a quotient can be due to the abelian symmetries, the non-abelian symmetries or both.

As a simple example of this case, we look at the fourth case of Table \ref{suNSym1}. The theory here is an $SU(N)$ gauge theory with a single symmetric chiral field, a single chiral field in the conjugate antisymmetric, $2N-4$ fundamental chiral fields and $2N+4$ anti-fundamental chiral fields. This theory has $U(1)^3\times SU(2N-4)\times SU(2N+4)$ global symmetry. We assign charges under the non-anomalous $U(1)$ groups as follows. We assign charges $(N-2,0,2)$ to the symmetric chiral, $(-N-2,0,2)$ to the conjugate antisymmetric chiral, $(0,N+2,-1)$ to the fundamental chirals and $(0,-N+2,-1)$ to the anti-fundamental chirals under $U(1)_a \times U(1)_b \times U(1)_c$. The marginal operators and their representations under the flavor symmetries are
\be
A=\overline{AS}\times F^{2}&\sim&\left(\boldsymbol{(N-2)(2N-5)},\boldsymbol{1}\right)b^{2(N+2)}a^{-(N+2)}\nonumber\\
B=S\times \overline{F}^{2}&\sim&\left(\boldsymbol{1},\boldsymbol{(N+2)(2N+5)}\right)b^{-2(N-2)}a^{N-2},
\ee
where we assume that $N>6$. We shall relax that assumption later. The interesting feature in this model is that the two marginal operators are charged only under a single combination of the $U(1)$ groups. We can form a non-abelian invariant from both operators by taking the determinant of the antisymmetric or symmetric matrix. As a result the operator $A^{2(N-2)}B^{2(N+2)}$ is invariant under the full global symmetry, and a quotient exists. Since we are inserting an operator in the symmetric representation of $SU(2N+4)$ and another in the antisymmetric representation of $SU(2N-4)$, the former global symmetry then is broken to $SO(2N+4)$, while the latter is broken to $USp(2N-4)$. Similarly, $U(1)_a$ and $U(1)_b$ are broken to the combination defined by $a=b^2$. We can next decompose the operators into representations of the preserved global symmetry, write down the broken currents and identify the marginally irrelevant operators, similarly to what was done in the previous examples. One then finds that there is only one exactly marginal operator, the rest eaten by the broken symmetries. As this is similar to the previous examples, we will not present the analysis in detail here.  

The different aspect enters when we consider cases with $N<7$. In these cases there are additional marginal operators so we might expect there to be a larger conformal manifold. However, with the exception of $N=3$, in all other cases these additional operators are actually marginally irrelevant as these are charged under the remaining $U(1)$ groups, such that there is no quotient. For instance, consider $N=6$. The additional marginal operator in this case is

\be
C=\overline{AS}^3&\sim&\left(\boldsymbol{1},\boldsymbol{1}\right)c^{6}a^{-24},
\ee 
and there is indeed no quotient.

In the example we have shown above, the unbreakable symmetries were one of the intrinsic $U(1)$ symmetries, and so the issue was readily apparent. However, there are cases where the unbreakable symmetry is some combination of the abelian and non-abelian symmetries, and its existence is not immediately apparent. As an example, we look at the sixth case of Table \ref{suNoneAd1}. This is a theory of an $SU(6)$ gauge theory with one adjoint chiral, two antisymmetric chirals, one conjugate antisymmetric chiral, five fundamental chirals and seven anti-fundamental chirals. One can verify it obeys \eqref{AnomalyC} and \eqref{BetaZero}. The symmetry of the free point is $U(1)^{4}\times SU(2)\times SU(5)\times SU(7)$, and we assign charges under the non-anomalous $U(1)$ groups as follows
\begin{table}[H]
\begin{center}
\begin{tabular}{|c||c|c|c|c|}
 \hline 
Field & $U(1)_{x}$ & $U(1)_{y}$ & $U(1)_{z}$  & $U(1)_{a}$  \\
 \hline
 \hline
$Ad$&  $0$ &  $0$ & $5$ & $5$ \\
\hline
$AS$&  $0$ &  $1$ & $0$ & $-3$ \\
\hline
$\overline{AS}$&  $0$ &  $-2$ & $0$ & $-4$ \\
\hline
$F$&  $7$ &  $0$ & $2$ & $-4$ \\
\hline
$\overline{F}$&  $-5$ &  $0$ & $-10$ & $0$ \\
\hline
 \end{tabular}
 \end{center}
\caption{} 
\label{Ex2Charges}
\end{table}

The marginal operators and their representations under the flavor symmetries are
\be
A=Ad^{3} & \sim & \left(\boldsymbol{1},\boldsymbol{1},\boldsymbol{1}\right)z^{15}a^{15}\nonumber\\
B=Ad\times F\times\overline{F}&\sim&\left(\boldsymbol{1},\boldsymbol{5},\overline{\boldsymbol{7}}\right)x^{2}z^{-3}a\nonumber\\
C=Ad\times AS\times\overline{AS}&\sim&\left(\boldsymbol{2},\boldsymbol{1},\boldsymbol{1}\right)y^{-1}z^{5}a^{-2}\nonumber\\
D=AS\times\overline{F}^{2}&\sim&\left(\boldsymbol{2},\boldsymbol{1},\overline{\boldsymbol{21}}\right)x^{-10}yz^{-20}a^{-3}\nonumber\\
E=\overline{AS}\times F^{2}&\sim&\left(\boldsymbol{1},\boldsymbol{10},\boldsymbol{1}\right)x^{14}y^{-2}z^{4}a^{-12}\nonumber\\
G=AS^{3}&\sim&\left(\boldsymbol{4},\boldsymbol{1},\boldsymbol{1}\right)y^{3}a^{-9}\nonumber\\
I=\overline{AS}^{3}&\sim&\left(\boldsymbol{1},\boldsymbol{1},\boldsymbol{1}\right)y^{-6}a^{-12}\, .
\ee
There are several invariants one can build under the abelian symmetries, and we will focus on the operators $AB^{5}C^{4}DG$ and $A^{3}BC^{2}D^{3}E^{2}G$. 

One invariant under the non-abelian symmetries can be constructed from one copy of $AB^{5}C^{4}DG$, by contracting the indices of both the $SU(5)$ and $SU(7)$ groups in $B^{5}$ with the respective epsilon tensors to create an invariant under the $SU(5)$ and the $\textbf{21}$ representation under $SU(7)$. The latter can then be contracted with the $\overline{\textbf{21}}$ of $SU(7)$ in $D$. Taking $C$ to the fourth symmetric power gives the $\textbf{5}$ representation of $SU(2)$ that can be contracted with the $\textbf{2}$ and $\textbf{4}$ representations to generate an $SU(2)$ invariant.

Another invariant under the non-abelian symmetries can be constructed from one copy of $A^{3}BC^{2}D^{3}E^{2}G$. This is again done by contracting the indices with the epsilon tensor, where for $SU(5)$ we use the two antisymmetric representations in $E^2$ and the single fundamental in $B$, which is also used for the $SU(7)$ with the three conjugate antisymmetric representations in $D^3$. Finally for the $SU(2)$, we can multiply the $\textbf{3}$ coming from $C^2$, with the two $\textbf{4}$ representations coming from $D^3$ and $G$ to form an $SU(2)$ singlet.

In the invariant operator $AB^{5}C^{4}DG$, we use an $SU(5)\times SU(7)$ baryon, implying that a diagonal $SU(5)$ is conserved out of the one coming from the breaking of $SU(7)\to SU(5)\times SU(2)\times U(1)_c$ and the intrinsic $SU(5)$. As for the intrinsic $SU(2)$, the presence of operators charged only under it imply that the $SU(2)$ is at least broken to $U(1)_b$.
In the invariant operator $A^{3}BC^{2}D^{3}E^{2}G$ we use only symmetric contractions, so we don't expect diagonal combinations of groups. The use of a single $SU(5)\times SU(7)$ bifundamental implies these break as $SU(5)\to SU(4)\times U(1)_c$ and $SU(7)\to SU(6)\times U(1)_d$. In addition the presence of operators in the antisymmetric representations under the $SU(5)$ and $SU(7)$ groups imply further breakings as $SU(6)\to USp(6)$ and $SU(4)\to USp(4)$. The intrinsic $SU(2)$ is at least broken to $U(1)_b$ for the same reasons mentioned before.
The marginal operators under the two breakings
\be
&& \left(\boldsymbol{1},\boldsymbol{1},\boldsymbol{1}\right)z^{15}a^{15}\oplus\left(\boldsymbol{1},\boldsymbol{5},\overline{\boldsymbol{7}}\right)x^{2}z^{-3}a\oplus\left(\boldsymbol{2},\boldsymbol{1},\boldsymbol{1}\right)y^{-1}z^{5}a^{-2}\oplus\left(\boldsymbol{2},\boldsymbol{1},\overline{\boldsymbol{21}}\right)x^{-10}yz^{-20}a^{-3}\nonumber\\
&&\oplus\left(\boldsymbol{1},\boldsymbol{10},\boldsymbol{1}\right)x^{14}y^{-2}z^{4}a^{-12}\oplus\left(\boldsymbol{4},\boldsymbol{1},\boldsymbol{1}\right)y^{3}a^{-9}\oplus\left(\boldsymbol{1},\boldsymbol{1},\boldsymbol{1}\right)y^{-6}a^{-12}\nonumber\\
1st & \to & \underline{\left(\boldsymbol{1},\boldsymbol{1}\right)z^{15}a^{15}}\oplus\underline{\left(\boldsymbol{24}\oplus\boldsymbol{1},\boldsymbol{1}\right)x^{2}z^{-3}ac^{-2}}\oplus\left(\boldsymbol{5},\boldsymbol{2}\right)x^{2}z^{-3}ac^{5}\nonumber\\
&&\oplus\underline{\left(\boldsymbol{1},\boldsymbol{1}\right)y^{-1}z^{5}a^{-2}\left(\underline{b}+b^{-1}\right)}\oplus\left(\overline{\boldsymbol{10}},\boldsymbol{1}\right)x^{-10}yz^{-20}a^{-3}\left(b+b^{-1}\right)c^{-4}\nonumber\\
&&\oplus\left(\overline{\boldsymbol{5}},\boldsymbol{2}\right)x^{-10}yz^{-20}a^{-3}\left(b+b^{-1}\right)c^{3}\oplus\underline{\left(\boldsymbol{1},\boldsymbol{1}\right)x^{-10}yz^{-20}a^{-3}\left(b+\underline{b^{-1}}\right)c^{10}}\nonumber\\
&&\oplus\left(\boldsymbol{10},\boldsymbol{1}\right)x^{14}y^{-2}z^{4}a^{-12}\oplus\underline{\left(\boldsymbol{1},\boldsymbol{1}\right)y^{3}a^{-9}\left(b^{3}+b+b^{-1}+\underline{b^{-3}}\right)}\oplus\left(\boldsymbol{1},\boldsymbol{1}\right)y^{-6}a^{-12}\, , \\
2nd & \to & \underline{\left(\boldsymbol{1},\boldsymbol{1}\right)z^{15}a^{15}}\oplus\left(\boldsymbol{4},\boldsymbol{6}\right)x^{2}z^{-3}acd^{-1}\oplus\left(\boldsymbol{4},\boldsymbol{1}\right)x^{2}z^{-3}acd^{6}\oplus\left(\boldsymbol{1},\boldsymbol{6}\right)x^{2}z^{-3}ac^{-4}d^{-1}\nonumber\\
&&\oplus\underline{\left(\boldsymbol{1},\boldsymbol{1}\right)x^{2}z^{-3}ac^{-4}d^{6}}\oplus\underline{\left(\boldsymbol{1},\boldsymbol{1}\right)y^{-1}z^{5}a^{-2}\left(\underline{b}+b^{-1}\right)}\nonumber\\
&&\oplus\underline{\left(\boldsymbol{1},\boldsymbol{14}\oplus\boldsymbol{1}\right)x^{-10}yz^{-20}a^{-3}\left(b+\underline{b^{-1}}\right)d^{-2}}\oplus\left(\boldsymbol{1},\boldsymbol{6}\right)x^{-10}yz^{-20}a^{-3}\left(b+b^{-1}\right)d^{5}\nonumber\\
&&\oplus\left(\boldsymbol{4},\boldsymbol{1}\right)x^{14}y^{-2}z^{4}a^{-12}c^{-3}\oplus\underline{\left(\boldsymbol{5}\oplus\boldsymbol{1},\boldsymbol{1}\right)x^{14}y^{-2}z^{4}a^{-12}c^{2}}\nonumber\\
&&\oplus\underline{\left(\boldsymbol{1},\boldsymbol{1}\right)y^{3}a^{-9}\left(b^{3}+\underline{b}+b^{-1}+b^{-3}\right)}\oplus\left(\boldsymbol{1},\boldsymbol{1}\right)y^{-6}a^{-12}\, ,
\ee
where we underlined the operators containing singlets under the non-abelian symmetries. These are designated to be the marginal operators inserted in the superpotential, and as such have a quotient under the abelian groups, in accordance with the invariant operators we found. Here, in the first case we have used the breaking $\overline{\boldsymbol{7}}_{SU(7)}\rightarrow \frac{1}{c^2}\left(\boldsymbol{5},\boldsymbol{1}\right) + c^5\left(\boldsymbol{1},\boldsymbol{2}\right)$, $\boldsymbol{2}_{SU(7)}\rightarrow b + \frac{1}{b}$ while in the second we have used the breaking $\overline{\boldsymbol{7}}_{SU(7)}\rightarrow \frac{1}{d}\left(\boldsymbol{1},\boldsymbol{6}\right) + d^6\left(\boldsymbol{1},\boldsymbol{1}\right)$, $\boldsymbol{5}_{SU(5)}\rightarrow c\left(\boldsymbol{4},\boldsymbol{1}\right) + \frac{1}{c^4}\left(\boldsymbol{1},\boldsymbol{1}\right)$, $\boldsymbol{2}_{SU(7)}\rightarrow b + \frac{1}{b}$. Finally, we need to identify some of the abelian symmetries with one another so that the underlined operators become true singlets. This will give us the abelian symmetry preserved by these operators. For the first operator $AB^{5}C^{4}DG$ we find the identifications
\be \label{Idn}
& z^{15}a^{15}=x^{2}z^{-3}ac^{-2}=y^{-1}z^{5}a^{-2}b=x^{-10}yz^{-20}a^{-3}b^{-1}c^{10}=y^{3}a^{-9}b^{-3}=1\nonumber\\
& \Rightarrow z=a=1\ ,b=y\ ,c=x\, .
\ee
For the second operator $A^{3}BC^{2}D^{3}E^{2}G$ we find the identifications
\be
& z^{15}a^{15}=x^{2}z^{-3}ac^{-4}d^{6}=y^{-1}z^{5}a^{-2}b=x^{-10}yz^{-20}a^{-3}b^{-1}d^{-2}=x^{14}y^{-2}z^{4}a^{-12}c^{2}=y^{3}a^{-9}b=1\nonumber\\
& \Rightarrow z=y^{-2}\ ,a=y^{2}\ ,b=y^{15}\ ,c=x^{-7}y^{17}\ ,d=x^{-5}y^{10}\, .
\ee
Breaking the symmetry further according to the identification above, the marginal operators are
\be
1st & \to & \left(\boldsymbol{1},\boldsymbol{1}\right)\left(y^{6}+y^{4}+2y^{2}+5+y^{-2}+y^{-6}\right)\oplus\left(\boldsymbol{24},\boldsymbol{1}\right)\oplus\left(\boldsymbol{5},\boldsymbol{2}\right)x^{7}\nonumber\\
&&\oplus\left(\overline{\boldsymbol{5}},\boldsymbol{2}\right)x^{-7}\left(y^{2}+1\right)\oplus\left(\boldsymbol{10},\boldsymbol{1}\right)x^{14}y^{-2}\oplus\left(\overline{\boldsymbol{10}},\boldsymbol{1}\right)x^{-14}\left(y^{2}+1\right)\,,\\
2nd & \to & \left(\boldsymbol{1},\boldsymbol{1}\right)\left(2y^{30}+6+3y^{-30}+y^{-60}\right)\oplus\left(\boldsymbol{4},\boldsymbol{6}\right)y^{15}\oplus\left(\boldsymbol{1},\boldsymbol{14}\right)\left(y^{30}+1\right)\oplus\left(\boldsymbol{5},\boldsymbol{1}\right)\nonumber\\
&&\oplus\left(\boldsymbol{4},\boldsymbol{1}\right)\left(x^{-35}y^{85}+x^{35}y^{-85}\right)\oplus\left(\boldsymbol{1},\boldsymbol{6}\right)\left(x^{35}y^{-70}+x^{-35}y^{100}+x^{-35}y^{70}\right)\,.
\ee

Next we write down the broken currents
\be
b.c.^{1st} & \to & \left(\boldsymbol{1},\boldsymbol{1}\right)\left(y^{2}+4+y^{-2}\right)\oplus\left(\boldsymbol{24},\boldsymbol{1}\right)\oplus\left(\boldsymbol{5},\boldsymbol{2}\right)x^{7}\oplus\left(\overline{\boldsymbol{5}},\boldsymbol{2}\right)x^{-7}\,,\\
b.c.^{2nd} & \to & \left(\boldsymbol{1},\boldsymbol{1}\right)\left(y^{30}+5+y^{-30}\right)\oplus\left(\boldsymbol{5},\boldsymbol{1}\right)\oplus\left(\boldsymbol{1},\boldsymbol{14}\right)\nonumber\\
&&\oplus\left(\boldsymbol{4},\boldsymbol{1}\right)\left(x^{-35}y^{85}+x^{35}y^{-85}\right)\oplus\left(\boldsymbol{1},\boldsymbol{6}\right)\left(x^{-35}y^{70}+x^{35}y^{-70}\right)\, .
\ee
The remaining marginal operators after being eaten by the broken currents
\be
1st & \to & \left(\boldsymbol{1},\boldsymbol{1}\right)\oplus\left(\boldsymbol{1},\boldsymbol{1}\right)\left(y^{6}+y^{4}+y^{2}+y^{-6}\right)\oplus\left(\overline{\boldsymbol{5}},\boldsymbol{2}\right)x^{-7}y^{2}\nonumber\\
&&\oplus\left(\boldsymbol{10},\boldsymbol{1}\right)x^{14}y^{-2}\oplus\left(\overline{\boldsymbol{10}},\boldsymbol{1}\right)x^{-14}\left(y^{2}+1\right)\, ,\\
2nd & \to & \left(\boldsymbol{1},\boldsymbol{1}\right)\oplus\left(\boldsymbol{1},\boldsymbol{1}\right)\left(y^{30}+2y^{-30}+y^{-60}\right)\nonumber\\
&&\oplus\left(\boldsymbol{4},\boldsymbol{6}\right)y^{15}\oplus\left(\boldsymbol{1},\boldsymbol{14}\right)y^{30}\oplus\left(\boldsymbol{1},\boldsymbol{6}\right)x^{-35}y^{100}\, .
\ee
We see that we are left with one singlet in each case, meaning one exactly marginal operator, and thus there is a $1d$ subspace on the conformal manifold preserving $U(1)^{2}\times SU(2)\times SU(5)$ symmetry and a  $1d$ subspace preserving $U(1)^{2}\times USp(4)\times USp(6)$ symmetry. 

Looking at the remaining operators of the first case one can easily see that we can generate another singlet by combining the two operators $\left(\boldsymbol{1},\boldsymbol{1}\right)y^{6}$ and $\left(\boldsymbol{1},\boldsymbol{1}\right)y^{-6}$ breaking $U(1)_y$. The remaining operators after the symmetry breaking and the subtraction of the broken current are
\be
4\left(\boldsymbol{1},\boldsymbol{1}\right)\oplus\left(\overline{\boldsymbol{5}},\boldsymbol{2}\right)x^{-7}\oplus2\left(\overline{\boldsymbol{10}},\boldsymbol{1}\right)x^{-14}\oplus\left(\boldsymbol{10},\boldsymbol{1}\right)x^{14}.
\ee
Thus, we have additional three directions preserving $U(1)\times SU(2)\times SU(5)$ together with the one we already found preserving a higher amount of symmetry.

Next, we can break $U(1)\times SU(2)\times SU(5)\to U(1)\times SU(2)\times\left(USp(4)\times U(1)_{a'}\right)$ and identify $a'=x^{-7}$  in a similar procedure to the one we did before to find
\be
6\left(\boldsymbol{1},\boldsymbol{1}\right)\oplus2\left(\boldsymbol{5},\boldsymbol{1}\right)\oplus\left(\boldsymbol{4},\boldsymbol{2}\right)\oplus\left(\boldsymbol{4},\boldsymbol{1}\right)x^{-35}\oplus\left(\boldsymbol{1},\boldsymbol{2}\right)x^{-35},
\ee
giving two additional directions preserving $U(1)\times SU(2)\times USp(4)$. One can continue breaking the symmetry as follows $U(1)\times SU(2)\times USp(4)\to U(1)\times SU(2)\times SU(2)^2 \to U(1)\times SU(2)^3  \to U(1)\times SU(2)^2 \to U(1)\times SU(2) \to U(1)^2$ and find the remaining operators
\be \label{eofo}
12 \oplus 3x^{-35}\left(y'+y'^{-1}\right)\,.
\ee
Thus, we find a $12$ dimensional conformal manifold preserving $U(1)^2$ on a general point. We cannot break the symmetry any more because we are left only with exactly marginal operators and operators that cannot combine to form a singlet of the flavor symmetry. The latter ones are then marginally irrelevant operators. We note that the $U(1)$ symmetry responsible for the irrelevancy is a non-trivial combination of the intrinsic $U(1)$ groups and Cartan elements in the non-abelian groups, given by the identifications in \eqref{Idn} and $a'=x^{-7}$. The presence of such symmetries may not be immediately apparent, but can be uncovered through the process of successive breakings as done here. Finally, we note that we can get to the same end result, \eqref{eofo}, also by successive breakings from the second $1d$ subspace we found.  

\subsubsection{Cases with a complex subspace structure}

We have seen by now various cases defined for a group with variable rank where the existence and behavior of the conformal manifold depends on the rank. Usually, this is due to the existence of additional operators at special ranks. Occasionally, this is due to some marginal operators becoming singlets under some symmetries at special ranks. Interestingly, there are also cases where there is a conformal manifold for every rank, sometimes also with the same dimension and generically preserved symmetry for every rank, but the subspace structure of the conformal manifold is sensitive to the rank of the gauge group. This occurs where the identity of the independent invariants, but not necessarily their number, depends on the rank. The existence of these subspaces is important when considering quiver theories, as then it might be necessary to know which symmetries can be gauged without completely lifting the conformal manifold, and so it is useful to also consider these details.

As an example we consider case 9 in table \ref{USpGen2}. This is a theory of a $USp(2N)$ gauge theory with three antisymmetric traceless chiral fields, and twelve fundamental chiral fields. The symmetry of the free point is $U(1)_c\times SU(3)\times SU(12)$, and we assign charges under the non-anomalous $U(1)_c$ group as follows. We take the antisymmetric chiral fields to have charge $2$, and the fundamental chiral fields to have charge $-(N-1)$. 

We then have the following marginal operators
\be
A=AS^{3} & \sim & \left(\boldsymbol{10},\boldsymbol{1}\right)c^{6}\nonumber\\
B=AS\times F^{2}&\sim&\left(\boldsymbol{3},\boldsymbol{66}\right)c^{-2(N-2)}\, .
\ee
Here, we have assumed that $N>2$. The case of $N=2$ is special as then the operator $A$ does not exist, and we will not consider it in detail here.

As there is only one $U(1)$ group, it is straightforward to find abelian invariants. The more interesting question, is finding the non-abelian invariants. From $A$, we can build the two $SU(3)$ invariants $A^4$ and $A^6$. These are positively charged under $U(1)_c$, and as $B$ is negatively charged under $U(1)_c$, if we can find a non-abelian singlet made just from it then there is a quotient. Indeed, the combination $B^{12}$ contains a singlet, where the $SU(3)$ indices are contracted with the epsilon tensor to form an $SU(3)$ invariant, and the $SU(6)$ indices are contracted in a mixed symmetry product given by the Young diagram made from three rows of four boxes each. This ensures that the full product is symmetric. We then see that there is indeed a quotient, and using similar methods as in the previous sections, one can show that there is a $56$ dimensional conformal manifold, on a generic point of which the global symmetry is completely broken, regardless of $N$.

An interesting question then is what symmetries can be preserved on special subspaces. As we are inserting a marginal operator in the $\boldsymbol{10}$ of $SU(3)$, it should be broken at least to its Cartan. The effect of $B$ is subtler. Generically, a single operator in the antisymmetric of an $SU(2k)$ group breaks that group to $USp(2k)$. If there are two independent operators in the antisymmetric representation of $SU(2k)$ group, then that group is broken to $USp(2n)\times USp(2k-2n)$, and so forth. In this case, we needed to use a third rank antisymmetric contraction for the $SU(6)$ indices in $B$, due to the need to make the operator $SU(3)$ invariant. This suggests that the breaking pattern is $SU(12)\rightarrow USp(2k)\times USp(2l)\times USp(12-2k-2l)$ for some numbers $k$ and $l$. However, we can ask whether it is possible to preserve more symmetry, specifically, whether we can preserve the full $USp(12)$ subgroup of $SU(12)$. 

In order to have such a subspace, we need to form the quotient using the totally symmetric, determinant type, invariant for the antisymmetric of $SU(6)$. We then consider the operator $B^6$, and contract the $SU(6)$ indices fully symmetrically. This then forces us to contract also the $SU(3)$ indices in the fully symmetric matter, which leads to the six index symmetric tensor representation of $SU(3)$. To form an $SU(3)$ invariant, we multiply this with the operator $A^6$, which contains the conjugate six index symmetric tensor representation. The resulting operator has the $U(1)_c$ charge $-12(N-4)$. We then have three possible behaviors depending on the value of $N$. If $N>4$ then the operator we found has negative charge and together with the $SU(3)$ invariant $A^3$, we have a quotient. In these cases there is a subspace of the conformal manifold preserving the $USp(12)$ subgroup of $SU(12)$. If $N=4$ then it is invariant under the full global symmetry, and gives a quotient by itself. We then expect a $1d$ subspace, spanned by this combination, preserving $U(1)\times USp(12)$, where the extra $U(1)$ is expected as the quotient requires one operator less and so we expect one less marginally irrelevant operator and one more conserved current. Finally, if $N<4$ then this operator is also positively charged and there is no quotient of this type, implying that we can not preserve the $USp(12)$ subgroup of $SU(12)$ on any subspace on the conformal manifold.

Finally, it is possible to show that we can get a quotient fixing two independent antisymmetric matrices for any $N>2$, so in those cases the minimal number of $USp$ subgroups of $SU(12)$ that can be preserved is two. For $N=2$, as we do not have the operator $A$, we must make an invariant from $B$ alone and so the minimal number of $USp$ subgroups of $SU(12)$ that can be preserved is three. The $N=2$ case also has various other special subspaces preserving various subgroups of $SU(3)\times SU(12)$, see \cite{Razamat:2020gcc} for a detailed example of one such subspace.    

\section{Lists of models with vanishing beta function}\label{app:betavanish}

In this appendix we summarize all possible strictly ${\cal N}=1$ solutions of conditions (\ref{AnomalyC}) and (\ref{BetaZero}). For the complete list, these should be supplemented by the ${\cal N}=2$ and ${\cal N}=4$ solutions in \cite{Bhardwaj:2013qia}. The list is organized by gauge group and matter representation, in the same vein as done in the bulk of the paper.

\subsection{Special unitary groups}

Here we summarize solutions when the gauge group is $SU(N)$.

\subsubsection{Cases with two adjoints}\label{appendix:wtwoadj}

Here we summarize solutions when the matter content contains two chiral fields in the adjoint representation. The possible solutions to conditions (\ref{AnomalyC}) and (\ref{BetaZero}) for generic $N$ are:

\begin{enumerate}

\item - $N_F=N_{\overline{F}}=N$

\item - $N_{AS}=1, \quad N_F=3 , \quad N_{\overline{F}}=N-1$

\item - $N_{AS}=N_{\overline{AS}}=1, \quad N_F=N_{\overline{F}}=2$ 

\end{enumerate}

Here we have suppressed the two adjoints for brevity, but we remind the reader that they are also part of the matter content. Also, for chiral choices there are two possibilities given by complex conjugation and we have only written one, as they are physically the same differing merely by redefining the $SU(N)$ generators.   

Besides these, there are several solutions that exist only for small $N$:

\begin{enumerate}

\item - $N_{AS}=2, \quad N_F=6-N , \quad N_{\overline{F}}=N-2, \quad N=5,6$

\item - $N_{AS}=3, \quad N_F=N_{\overline{F}}=1, \quad N=4$

\item - $N_{AS}=4, \quad N=4$

\item - $N_{AS}=2, \quad N_{\overline{AS}}=1, \quad N_{\overline{F}}=1, \quad N=5$

\end{enumerate}

Here we have only listed cases that do not reduce to one of the generic families.

\subsubsection{Cases with a single adjoint}\label{appendix:woneadj}

Here we summarize solutions when the matter content contains one chiral field in the adjoint representation, but that only have $\mathcal{N}=1$ supersymmetry. The possible solutions to conditions (\ref{AnomalyC}) and (\ref{BetaZero}) for generic $N$ are then:

\begin{enumerate}

\item - $N_S=N_{\overline{S}}=1, \quad N_{AS}=1, \quad N_F=1, \quad N_{\overline{F}}=N-3$

\item - $N_{S}=1, \quad N_{AS}=1, \quad N_{\overline{F}}=2N$

\item - $N_{S}=1, \quad N_{\overline{AS}}=1, \quad N_{F}=N-4, \quad N_{\overline{F}}=N+4$

\item - $N_{S}=1, \quad N_{\overline{AS}}=2, \quad N_{F}=N-5, \quad N_{\overline{F}}=7$

\item - $N_{S}=1, \quad N_{F}=N-3, \quad N_{\overline{F}}=2N+1$

\item - $N_{AS}=2, \quad N_{\overline{AS}}=1, \quad N_F=5 , \quad N_{\overline{F}}=N+1$

\item - $N_{AS}=2, \quad N_F=6 , \quad N_{\overline{F}}=2N-2$

\item - $N_{AS}=1, \quad N_F=N+3 , \quad N_{\overline{F}}=2N-1$ 

\end{enumerate}

Here we have suppressed the adjoint for brevity, but we remind the reader that it is also part of the matter content. Also, for chiral choices there are two possibilities given by complex conjugation and we have only written one, as they are physically the same differing merely by a redefining the $SU(N)$ generators. We also note that while for generic $N$ these models have only $\mathcal{N}=1$ supersymmetry, for some this is enhanced to $\mathcal{N}=2$ for low values of $N$. 

Besides these, there are several solutions that exist only for small $N$:

\begin{enumerate}

\item - $N_{S}=1, \quad N_{\overline{AS}}=3, \quad N_F=N-6 , \quad N_{\overline{F}}=10-N, \quad N=6, 7, 8, 9, 10$

\item - $N_{AS}=7, \quad N_F=N_{\overline{F}}=1, \quad N=4$

\item - $N_{AS}=5, \quad N_F=15-3N , \quad N_{\overline{F}}=2N-5, \quad N=4, 5$

\item - $N_{AS}=4, \quad N_{\overline{AS}}=2, \quad N_{\overline{F}}=2, \quad N=5$

\item - $N_{AS}=4, \quad N_{\overline{AS}}=1,  \quad N_F=1 , \quad N_{\overline{F}}=4, \quad N=5$

\item - $N_{AS}=4, \quad N_F=12-2N , \quad N_{\overline{F}}=2N-4, \quad N=5, 6$

\item - $N_{AS}=3, \quad N_{\overline{AS}}=2, \quad N_F=7-N , \quad N_{\overline{F}}=3, \quad N=5, 6, 7$

\item - $N_{AS}=3, \quad N_{\overline{AS}}=1, \quad N_F=8-N , \quad N_{\overline{F}}=N, \quad N=5, 6, 7, 8$

\item - $N_{AS}=3, \quad N_F=9-N , \quad N_{\overline{F}}=2N-3, \quad N=5, 6, 7, 8, 9$

\end{enumerate}

Here we have only listed cases that do not reduce to one of the generic families.

\subsubsection{No adjoints, with symmetrics}\label{appendix:symmnoadj}

Here we summarize solutions when the matter content contains chiral fields in the symmetric representation, but no chiral fields in the adjoint representation. The possible solutions to conditions (\ref{AnomalyC}) and (\ref{BetaZero}) for generic $N$ are:

\begin{enumerate}

\item - $N_{S}=2, \quad N_{\overline{S}}=2, \quad N_F=N-4, \quad N_{\overline{F}}=N-4$

\item - $N_{S}=2, \quad N_{\overline{S}}=1, \quad N_{\overline{AS}}=2, \quad N_F=N-7, \quad N_{\overline{F}}=5$

\item - $N_{S}=2, \quad N_{\overline{S}}=1, \quad N_{\overline{AS}}=1, \quad N_F=N-6, \quad N_{\overline{F}}=N+2$

\item - $N_{S}=2, \quad N_{\overline{S}}=1, \quad N_F=N-5, \quad N_{\overline{F}}=2N-1$

\item - $N_{S}=2, \quad N_{\overline{AS}}=3, \quad N_F=N-9, \quad N_{\overline{F}}=11$

\item - $N_{S}=2, \quad N_{\overline{AS}}=2, \quad N_F=N-8, \quad N_{\overline{F}}=N+8$

\item - $N_{S}=2, \quad N_{\overline{AS}}=1, \quad N_F=N-7, \quad N_{\overline{F}}=2N+5$

\item - $N_{S}=2, \quad N_F=N-6, \quad N_{\overline{F}}=3N+2$

\item - $N_{S}=1, \quad N_{\overline{S}}=1, \quad N_{AS}=2, \quad N_{\overline{AS}}=2, \quad N_F=2 , \quad N_{\overline{F}}=2$

\item - $N_{S}=1, \quad N_{\overline{S}}=1, \quad N_{AS}=2, \quad N_{\overline{AS}}=1, \quad N_F=3 , \quad N_{\overline{F}}=N-1$

\item - $N_{S}=1, \quad N_{\overline{S}}=1, \quad N_{AS}=2, \quad N_F=4 , \quad N_{\overline{F}}=2N-4$

\item - $N_{S}=1, \quad N_{\overline{S}}=1, \quad N_{AS}=1, \quad N_{\overline{AS}}=1, \quad N_F=N , \quad N_{\overline{F}}=N$

\item - $N_{S}=1, \quad N_{\overline{S}}=1, \quad N_{AS}=1, \quad N_F=N+1 , \quad N_{\overline{F}}=2N-3$

\item - $N_{S}=1, \quad N_{\overline{S}}=1, \quad N_F=2N-2, \quad N_{\overline{F}}=2N-2$

\item - $N_{S}=1, \quad N_{AS}=2, \quad N_{\overline{AS}}=3, \quad N_{\overline{F}}=8$

\item - $N_{S}=1, \quad N_{AS}=2, \quad N_{\overline{AS}}=2, \quad N_F=1, \quad N_{\overline{F}}=N+5$

\item - $N_{S}=1, \quad N_{AS}=2, \quad N_{\overline{AS}}=1, \quad N_F=2, \quad N_{\overline{F}}=2N+2$

\item - $N_{S}=1, \quad N_{AS}=2, \quad N_F=3, \quad N_{\overline{F}}=3N-1$

\item - $N_{S}=1, \quad N_{AS}=1, \quad N_{\overline{AS}}=3, \quad N_F=N-3, \quad N_{\overline{F}}=9$

\item - $N_{S}=1, \quad N_{AS}=1, \quad N_{\overline{AS}}=2, \quad N_F=N-2, \quad N_{\overline{F}}=N+6$

\item - $N_{S}=1, \quad N_{AS}=1, \quad N_{\overline{AS}}=1, \quad N_F=N-1, \quad N_{\overline{F}}=2N+3$

\item - $N_{S}=1, \quad N_{AS}=1, \quad N_F=N, \quad N_{\overline{F}}=3N$

\item - $N_{S}=1, \quad N_{\overline{AS}}=3, \quad N_F=2N-6, \quad N_{\overline{F}}=10$

\item - $N_{S}=1, \quad N_{\overline{AS}}=2, \quad N_F=2N-5, \quad N_{\overline{F}}=N+7$

\item - $N_{S}=1, \quad N_{\overline{AS}}=1, \quad N_F=2N-4, \quad N_{\overline{F}}=2N+4$

\item - $N_{S}=1, \quad N_F=2N-3, \quad N_{\overline{F}}=3N+1$

\end{enumerate}

As previously, for chiral choices there are two possibilities given by complex conjugation and we have only written one, as they are physically the same, differing merely by redefining the $SU(N)$ generators. 

Besides these, there are several solutions that exist only for small $N$:

\begin{enumerate}

\item - $N_{S}=2, \quad N_{\overline{AS}}=4, \quad N_F=N-10, \quad N_{\overline{F}}=14-N, \quad N=10, 11, 12, 13, 14$

\item - $N_{S}=2, \quad N_{\overline{S}}=1, \quad N_{\overline{AS}}=3, \quad N=8$

\item - $N_{S}=1, \quad N_{\overline{S}}=1, \quad N_{AS}=3, \quad N_F=7-N , \quad N_{\overline{F}}=2N-5, \quad N=5,6,7$

\item - $N_{S}=1, \quad N_{\overline{S}}=1, \quad N_{AS}=4, \quad N_{\overline{F}}=4, \quad N=5$

\item - $N_{S}=1, \quad N_{\overline{S}}=1, \quad N_{AS}=5, \quad N_F=1 , \quad N_{\overline{F}}=1, \quad N=4$

\item - $N_{S}=1, \quad N_{\overline{S}}=1, \quad N_{AS}=6, \quad N=4$

\item - $N_{S}=1, \quad N_{\overline{S}}=1, \quad N_{AS}=3, \quad N_{\overline{AS}}=1, \quad N_F=6-N, \quad N_{\overline{F}}=N-2, \quad N=5,6$

\item - $N_{S}=1, \quad N_{\overline{S}}=1, \quad N_{AS}=3, \quad N_{\overline{AS}}=2, \quad N_{\overline{F}}=1, \quad N=5$

\item - $N_{S}=1, \quad N_{AS}=3, \quad N_F=6-N, \quad N_{\overline{F}}=3N-2, \quad N=5,6$

\item - $N_{S}=1, \quad N_{AS}=3, \quad N_{\overline{AS}}=1, \quad N_{\overline{F}}=11, \quad N=5$

\item - $N_{S}=1, \quad N_{AS}=1, \quad N_{\overline{AS}}=4, \quad N_F=N-4, \quad N_{\overline{F}}=12-N, \quad N=5,6,7,8,9,10,11,12$

\item - $N_{S}=1, \quad N_{AS}=1, \quad N_{\overline{AS}}=5, \quad N_F=N-5, \quad N_{\overline{F}}=15-2N, \quad N=5,6,7$

\item - $N_{S}=1, \quad N_{AS}=1, \quad N_{\overline{AS}}=6, \quad N=6$

\item - $N_{S}=1, \quad N_{\overline{AS}}=4, \quad N_F=2N-7, \quad N_{\overline{F}}=13-N, \quad N=5,6,7,8,9,10,11,12,13$

\item - $N_{S}=1, \quad N_{\overline{AS}}=5, \quad N_F=2N-8, \quad N_{\overline{F}}=16-2N, \quad N=5,6,7,8$

\item - $N_{S}=1, \quad N_{\overline{AS}}=6, \quad N_F=2N-9, \quad N_{\overline{F}}=19-3N, \quad N=5,6$

\item - $N_{S}=1, \quad N_{\overline{AS}}=7, \quad N_{\overline{F}}=2, \quad N=5$

\end{enumerate}

Here we have only listed cases that do not reduce to one of the generic families.

\subsubsection{No adjoints, no symmetrics, but antisymmetrics and fundamentals}\label{appendix:asymmnoadjnosymm}

Here we summarize solutions when the matter content contains chiral fields in the fundamental and antisymmetric representations (and their conjugates), but no chiral fields in any other representation. The possible solutions to conditions (\ref{AnomalyC}) and (\ref{BetaZero}) for generic $N$ are:

\begin{enumerate}

\item - $N_{AS}=N_{\overline{AS}}=3, \quad N_F=N_{\overline{F}}=6$

\item - $N_{AS}=3, \quad N_{\overline{AS}}=2, \quad N_F=7, \quad N_{\overline{F}}=N+3$

\item - $N_{AS}=N_{\overline{AS}}=2, \quad N_F=N_{\overline{F}}=N+4$

\item - $N_{AS}=3, \quad N_{\overline{AS}}=1, \quad N_F=8, \quad N_{\overline{F}}=2N$

\item - $N_{AS}=3, \quad N_F=9, \quad N_{\overline{F}}=3N-3$

\item - $N_{AS}=2, \quad N_{\overline{AS}}=1, \quad N_F=N+5, \quad N_{\overline{F}}=2N+1$

\item - $N_{AS}=2, \quad N_F=N+6, \quad N_{\overline{F}}=3N-2$

\item - $N_{AS}=N_{\overline{AS}}=1, \quad N_F=N_{\overline{F}}=2N+2$

\item - $N_{AS}=1, \quad N_F=2N+3, \quad N_{\overline{F}}=3N-1$

\item - $N_F=N_{\overline{F}}=3N$

\end{enumerate}

As previously,for chiral choices there are two possibilities given by complex conjugation and we have only written one, as they are physically the same differing merely by a redefining the $SU(N)$ generators. 

Besides these, there are several solutions that exist only for small $N$:

\begin{enumerate}

\item - $N_{AS}=12, \quad N=4$

\item - $N_{AS}=11, \quad N_F=N_{\overline{F}}=1, \quad N=4$

\item - $N_{AS}=N_{\overline{AS}}=5, \quad N_F=N_{\overline{F}}=10-2N, \quad N=4,5$

\item - $N_{AS}=5, \quad N_{\overline{AS}}=4, \quad N_F=11-2N, \quad N_{\overline{F}}=7-N, \quad N=4,5$

\item - $N_{AS}=6, \quad N_{\overline{AS}}=3, \quad N_{\overline{F}}=3, \quad N=5$

\item - $N_{AS}=N_{\overline{AS}}=4, \quad N_F=N_{\overline{F}}=8-N, \quad N=4,5,6,7,8$

\item - $N_{AS}=5, \quad N_{\overline{AS}}=3, \quad N_F=12-2N, \quad N_{\overline{F}}=4, \quad N=5,6$

\item - $N_{AS}=6, \quad N_{\overline{AS}}=2, \quad N_F=1, \quad N_{\overline{F}}=5, \quad N=5$

\item - $N_{AS}=7, \quad N_{\overline{AS}}=1, \quad N_{\overline{F}}=6, \quad N=5$

\item - $N_{AS}=7, \quad N_F=21-4N, \quad N_{\overline{F}}=3N-7, \quad N=4,5$

\item - $N_{AS}=6, \quad N_{\overline{AS}}=1, \quad N_F=2, \quad N_{\overline{F}}=7, \quad N=5$

\item - $N_{AS}=5, \quad N_{\overline{AS}}=2, \quad N_F=13-2N, \quad N_{\overline{F}}=N+1, \quad N=5,6$

\item - $N_{AS}=4, \quad N_{\overline{AS}}=3, \quad N_F=9-N, \quad N_{\overline{F}}=5, \quad N=5,6,7,8,9$

\item - $N_{AS}=6, \quad N_F=18-3N, \quad N_{\overline{F}}=3N-6, \quad N=5,6$

\item - $N_{AS}=5, \quad N_{\overline{AS}}=1, \quad N_F=14-2N, \quad N_{\overline{F}}=2N-2, \quad N=5,6,7$

\item - $N_{AS}=4, \quad N_{\overline{AS}}=2, \quad N_F=10-N, \quad N_{\overline{F}}=N+2, \quad N=5,6,7,8,9,10$

\item - $N_{AS}=5, \quad N_F=15-2N, \quad N_{\overline{F}}=3N-5, \quad N=5,6,7$

\item - $N_{AS}=4, \quad N_{\overline{AS}}=1, \quad N_F=11-N, \quad N_{\overline{F}}=2N-1, \quad N=5,6,7,8,9,10,11$

\item - $N_{AS}=4, \quad N_F=12-N, \quad N_{\overline{F}}=3N-4, \quad N=5,6,7,8,9,10,11,12$

\end{enumerate}

Here we have only listed cases that do not reduce to one of the generic families.

\subsubsection{Exotic cases}\label{appendix:exotics}

Here we summarize solutions when the matter content contains chiral fields in at least one of the exotic representations in table \ref{SUExRep}. The possible solution for equations (\ref{BetaZero}) and (\ref{AnomalyC}) in this case are: 

\subsection*{$SU(4)$}

\begin{enumerate}

\item - $N_{\bold{20}}=1, \quad N_{\overline{S}}=1, \quad N_{AS}=2, \quad N_F=1$

\item - $N_{\bold{20}}=1, \quad N_{\overline{S}}=1, \quad N_{AS}=1, \quad N_F=2, \quad N_{\overline{F}}=1$

\item - $N_{\bold{20}}=1, \quad N_{\overline{S}}=1, \quad N_F=3, \quad N_{\overline{F}}=2$

\item - $N_{\bold{20}}=1, \quad N_{AS}=2, \quad N_{\overline{F}}=7$

\item - $N_{\bold{20}}=1, \quad N_{AS}=1, \quad N_F=1, \quad N_{\overline{F}}=8$

\item - $N_{\bold{20}}=1, \quad N_F=2, \quad N_{\overline{F}}=9$

\item - $N_{\bold{20'}}=1, \quad N_{Ad}=1$

\item - $N_{\bold{20'}}=1, \quad N_{AS}=4$

\item - $N_{\bold{20'}}=1, \quad N_{AS}=3, \quad N_F=N_{\overline{F}}=1$

\item - $N_{\bold{20'}}=1, \quad N_{AS}=2, \quad N_F=N_{\overline{F}}=2$

\item - $N_{\bold{20'}}=1, \quad N_{AS}=1, \quad N_F=N_{\overline{F}}=3$

\item - $N_{\bold{20'}}=1, \quad N_F=N_{\overline{F}}=4$

\end{enumerate}

\subsection*{$SU(5)$}

\begin{enumerate}

\item - $N_{\bold{45}}=1, \quad N_{\overline{F}}=6$

\end{enumerate}

\subsection*{$SU(6)$}

\begin{enumerate}

\item - $N_{\bold{20}}=6$

\item - $N_{\bold{20}}=5, \quad N_{AS}=1, \quad N_{\overline{F}}=2$

\item - $N_{\bold{20}}=5, \quad N_F=N_{\overline{F}}=3$

\item - $N_{\bold{20}}=4, \quad N_{AS}=2, \quad N_{\overline{F}}=4$

\item - $N_{\bold{20}}=4, \quad N_{AS}=N_{\overline{AS}}=1, \quad N_F=N_{\overline{F}}=2$

\item - $N_{\bold{20}}=4, \quad N_{AS}=1, \quad N_{F}=3, \quad N_{\overline{F}}=5$

\item - $N_{\bold{20}}=4, \quad N_F=N_{\overline{F}}=6$

\item - $N_{\bold{20}}=3, \quad N_{Ad}=1, \quad N_{AS}=1, \quad N_{\overline{F}}=2$

\item - $N_{\bold{20}}=3, \quad N_{S}=N_{\overline{S}}=1, \quad N_{F}=N_{\overline{F}}=1$

\item - $N_{\bold{20}}=3, \quad N_{S}=1, \quad N_{\overline{F}}=10$

\item - $N_{\bold{20}}=3, \quad N_{AS}=N_{\overline{AS}}=2, \quad N_F=N_{\overline{F}}=1$

\item - $N_{\bold{20}}=3, \quad N_{AS}=3, \quad N_{\overline{F}}=6$

\item - $N_{\bold{20}}=3, \quad N_{AS}=2, \quad N_{\overline{AS}}=1, \quad N_{F}=2, \quad N_{\overline{F}}=4$

\item - $N_{\bold{20}}=3, \quad N_{AS}=2, \quad N_{F}=3, \quad N_{\overline{F}}=7$

\item - $N_{\bold{20}}=3, \quad N_{AS}=N_{\overline{AS}}=1, \quad N_F=N_{\overline{F}}=5$

\item - $N_{\bold{20}}=3, \quad N_{AS}=1, \quad N_{F}=6, \quad N_{\overline{F}}=8$

\item - $N_{\bold{20}}=3, \quad N_F=N_{\overline{F}}=9$

\item - $N_{\bold{20}}=2, \quad N_{Ad}=2$

\item - $N_{\bold{20}}=2, \quad N_{Ad}=1, \quad N_{AS}=2, \quad N_{\overline{F}}=4$

\item - $N_{\bold{20}}=2, \quad N_{Ad}=1, \quad N_{AS}=1, \quad N_{F}=3, \quad N_{\overline{F}}=5$

\item - $N_{\bold{20}}=2, \quad N_{S}=N_{\overline{S}}=1, \quad N_{AS}=N_{\overline{AS}}=1$

\item - $N_{\bold{20}}=2, \quad N_{S}=N_{\overline{S}}=1, \quad N_{AS}=1, \quad N_{F}=1, \quad N_{\overline{F}}=3$

\item - $N_{\bold{20}}=2, \quad N_{S}=N_{\overline{S}}=1, \quad N_F=N_{\overline{F}}=4$

\item - $N_{\bold{20}}=2, \quad N_{S}=1, \quad N_{AS}=1, \quad N_{\overline{F}}=12$

\item - $N_{\bold{20}}=2, \quad N_{S}=1, \quad N_{\overline{AS}}=3, \quad N_{\overline{F}}=4$

\item - $N_{\bold{20}}=2, \quad N_{S}=1, \quad N_{\overline{AS}}=2, \quad N_{F}=1, \quad N_{\overline{F}}=7$

\item - $N_{\bold{20}}=2, \quad N_{S}=1, \quad N_{\overline{AS}}=1, \quad N_{F}=2, \quad N_{\overline{F}}=10$

\item - $N_{\bold{20}}=2, \quad N_{S}=1, \quad N_{F}=3, \quad N_{\overline{F}}=13$

\item - $N_{\bold{20}}=2, \quad N_{AS}=x, \quad N_{\overline{AS}}=3, \quad N_{F}=9-3x, \quad N_{\overline{F}}=3-x, \quad x=0,1,2,3$

\item - $N_{\bold{20}}=2, \quad N_{AS}=x, \quad N_{\overline{AS}}=2, \quad N_{F}=10-3x, \quad N_{\overline{F}}=6-x, \quad x=0,1,2$

\item - $N_{\bold{20}}=2, \quad N_{AS}=x, \quad N_{\overline{AS}}=1, \quad N_{F}=11-3x, \quad N_{\overline{F}}=9-x, \quad x=0,1$

\item - $N_{\bold{20}}=2, \quad N_{AS}=4, \quad N_{\overline{F}}=8$

\item - $N_{\bold{20}}=2, \quad N_{F}=N_{\overline{F}}=12$

\item - $N_{\bold{20}}=1, \quad N_{Ad}=2, \quad N_{AS}=1, \quad N_{\overline{F}}=2$

\item - $N_{\bold{20}}=1, \quad N_{Ad}=2, \quad N_F=N_{\overline{F}}=3$

\item - $N_{\bold{20}}=1, \quad N_{Ad}=1, \quad N_{S}=1, \quad N_{\overline{F}}=10$

\item - $N_{\bold{20}}=1, \quad N_{Ad}=1, \quad N_{AS}=x, \quad N_{\overline{AS}}=2, \quad N_{F}=7-3x, \quad N_{\overline{F}}=3-x, \quad x=0,1$

\item - $N_{\bold{20}}=1, \quad N_{Ad}=1, \quad N_{AS}=3, \quad N_{\overline{F}}=6$

\item - $N_{\bold{20}}=1, \quad N_{Ad}=1, \quad N_{AS}=1, \quad N_{F}=6, \quad N_{\overline{F}}=8$

\item - $N_{\bold{20}}=1, \quad N_{S}=N_{\overline{S}}=1, \quad N_{AS}=2, \quad N_{\overline{AS}}=1, \quad N_{\overline{F}}=2$

\item - $N_{\bold{20}}=1, \quad N_{S}=N_{\overline{S}}=1, \quad N_{AS}=2, \quad N_{F}=1, \quad N_{\overline{F}}=5$

\item - $N_{\bold{20}}=1, \quad N_{S}=N_{\overline{S}}=1, \quad N_{AS}=N_{\overline{AS}}=1, \quad N_{F}=N_{\overline{F}}=3$

\item - $N_{\bold{20}}=1, \quad N_{S}=N_{\overline{S}}=1, \quad N_{AS}=1, \quad N_{F}=4, \quad N_{\overline{F}}=6$

\item - $N_{\bold{20}}=1, \quad N_{S}=1, \quad N_{\overline{AS}}=5, \quad N_{F}=N_{\overline{F}}=1$

\item - $N_{\bold{20}}=1, \quad N_{S}=1, \quad N_{\overline{AS}}=4, \quad N_{F}=2, \quad N_{\overline{F}}=4$

\item - $N_{\bold{20}}=1, \quad N_{S}=1, \quad N_{AS}=x, \quad N_{\overline{AS}}=3, \quad N_{F}=3-3x, \quad N_{\overline{F}}=7-x, \quad x=0,1$

\item - $N_{\bold{20}}=1, \quad N_{S}=1, \quad N_{AS}=x, \quad N_{\overline{AS}}=2, \quad N_{F}=4-3x, \quad N_{\overline{F}}=10-x, \quad x=0,1$

\item - $N_{\bold{20}}=1, \quad N_{S}=1, \quad N_{AS}=x, \quad N_{\overline{AS}}=1, \quad N_{F}=5-3x, \quad N_{\overline{F}}=13-x, \quad x=0,1$

\item - $N_{\bold{20}}=1, \quad N_{S}=1, \quad N_{AS}=x, \quad N_{F}=6-3x, \quad N_{\overline{F}}=16-x, \quad x=0,1,2$

\item - $N_{\bold{20}}=1, \quad N_{AS}=x, \quad N_{\overline{AS}}=y, \quad N_{F}=15-3x-y, \quad N_{\overline{F}}=15-x-3y, \quad x+3y\leq 15, 3x+y\leq 15, x\geq y$

\end{enumerate}

\subsection*{$SU(7)$}

\begin{enumerate}

\item - $N_{\bold{35}}=N_{\overline{\bold{35}}}=2, N_{F}=N_{\overline{F}}=1$

\item - $N_{\bold{35}}=3, N_{F}=3, N_{\overline{F}}=9$

\item - $N_{\bold{35}}=3, N_{\overline{AS}}=2, N_{F}=N_{\overline{F}}=1$

\item - $N_{\bold{35}}=3, N_{\overline{AS}}=1, N_{F}=2, N_{\overline{F}}=5$

\item - $N_{\bold{35}}=2, N_{\overline{\bold{35}}}=1, N_{AS}=N_{\overline{AS}}=1, N_{\overline{F}}=2$

\item - $N_{\bold{35}}=2, N_{\overline{\bold{35}}}=1, N_{AS}=1, N_{F}=1, N_{\overline{F}}=6$

\item - $N_{\bold{35}}=2, N_{\overline{\bold{35}}}=1, N_{\overline{AS}}=1, N_{F}=4, N_{\overline{F}}=3$

\item - $N_{\bold{35}}=2, N_{\overline{\bold{35}}}=1, N_{F}=5, N_{\overline{F}}=7$

\item - $N_{\bold{35}}=2, N_{Ad}=1, N_{\overline{AS}}=1, N_{F}=1, N_{\overline{F}}=2$

\item - $N_{\bold{35}}=2, N_{Ad}=1, N_{F}=2, N_{\overline{F}}=6$

\item - $N_{\bold{35}}=2, N_{S}=N_{\overline{S}}=1, N_{\overline{F}}=4$

\item - $N_{\bold{35}}=2, N_{\overline{S}}=1, N_{AS}=2, N_{F}=2, N_{\overline{F}}=1$

\item - $N_{\bold{35}}=2, N_{\overline{S}}=1, N_{AS}=1, N_{F}=6, N_{\overline{F}}=2$

\item - $N_{\bold{35}}=2, N_{\overline{S}}=1, N_{F}=10, N_{\overline{F}}=3$

\item - $N_{\bold{35}}=2, N_{AS}=2, N_{\overline{AS}}=1, N_{\overline{F}}=7$

\item - $N_{\bold{35}}=2, N_{AS}=1, N_{\overline{AS}}=3, N_{F}=2$

\item - $N_{\bold{35}}=2, N_{AS}=1, N_{\overline{AS}}=2, N_{F}=3, N_{\overline{F}}=4$

\item - $N_{\bold{35}}=2, N_{AS}=1, N_{\overline{AS}}=1, N_{F}=4, N_{\overline{F}}=8$

\item - $N_{\bold{35}}=2, N_{\overline{AS}}=3, N_{F}=6, N_{\overline{F}}=1$

\item - $N_{\bold{35}}=2, N_{\overline{AS}}=2, N_{F}=7, N_{\overline{F}}=5$

\item - $N_{\bold{35}}=2, N_{\overline{AS}}=1, N_{F}=8, N_{\overline{F}}=9$

\item - $N_{\bold{35}}=2, N_{AS}=2, N_{F}=1, N_{\overline{F}}=11$

\item - $N_{\bold{35}}=2, N_{AS}=1, N_{F}=5, N_{\overline{F}}=12$

\item - $N_{\bold{35}}=2, N_{F}=9, N_{\overline{F}}=13$

\item - $N_{\bold{35}}=N_{\overline{\bold{35}}}=1, N_{Ad}=1, N_{AS}=1, N_{\overline{F}}=3$

\item - $N_{\bold{35}}=N_{\overline{\bold{35}}}=1, N_{S}=N_{\overline{S}}=1, N_{F}=N_{\overline{F}}=2$

\item - $N_{\bold{35}}=N_{\overline{\bold{35}}}=1, N_{S}=1, N_{\overline{AS}}=1, N_{\overline{F}}=8$

\item - $N_{\bold{35}}=N_{\overline{\bold{35}}}=1, N_{S}=1, N_{F}=1, N_{\overline{F}}=12$

\item - $N_{\bold{35}}=N_{\overline{\bold{35}}}=1, N_{AS}=N_{\overline{AS}}=2, N_{F}=N_{\overline{F}}=1$

\item - $N_{\bold{35}}=N_{\overline{\bold{35}}}=1, N_{AS}=2, N_{\overline{AS}}=1, N_{F}=2, N_{\overline{F}}=5$

\item - $N_{\bold{35}}=N_{\overline{\bold{35}}}=1, N_{AS}=2, N_{F}=3, N_{\overline{F}}=9$

\item - $N_{\bold{35}}=N_{\overline{\bold{35}}}=1, N_{AS}=1, N_{F}=7, N_{\overline{F}}=10$

\item - $N_{\bold{35}}=N_{\overline{\bold{35}}}=1, N_{AS}=N_{\overline{AS}}=1, N_{F}=N_{\overline{F}}=6$

\item - $N_{\bold{35}}=N_{\overline{\bold{35}}}=1, N_{F}=N_{\overline{F}}=11$

\item - $N_{\bold{35}}=1, N_{Ad}=2, N_{F}=1, N_{\overline{F}}=3$

\item - $N_{\bold{35}}=1, N_{Ad}=1, N_{\overline{S}}=1, N_{F}=9$

\item - $N_{\bold{35}}=1, N_{Ad}=1, N_{AS}=1, N_{\overline{AS}}=2, N_{F}=2, N_{\overline{F}}=1$

\item - $N_{\bold{35}}=1, N_{Ad}=1, N_{AS}=2, N_{\overline{F}}=8$

\item - $N_{\bold{35}}=1, N_{Ad}=1, N_{AS}=N_{\overline{AS}}=1, N_{F}=3, N_{\overline{F}}=5$

\item - $N_{\bold{35}}=1, N_{Ad}=1, N_{\overline{AS}}=2, N_{F}=6, N_{\overline{F}}=2$

\item - $N_{\bold{35}}=1, N_{Ad}=1, N_{AS}=1, N_{F}=4, N_{\overline{F}}=9$

\item - $N_{\bold{35}}=1, N_{Ad}=1, N_{\overline{AS}}=1, N_{F}=7, N_{\overline{F}}=6$

\item - $N_{\bold{35}}=1, N_{Ad}=1, N_{F}=8, N_{\overline{F}}=10$

\item - $N_{\bold{35}}=1, N_{S}=N_{\overline{S}}=1, N_{\overline{AS}}=2, N_{F}=4$

\item - $N_{\bold{35}}=1, N_{S}=N_{\overline{S}}=1, N_{AS}=N_{\overline{AS}}=1, N_{F}=1, N_{\overline{F}}=3$

\item - $N_{\bold{35}}=1, N_{S}=N_{\overline{S}}=1, N_{AS}=1, N_{F}=2, N_{\overline{F}}=7$

\item - $N_{\bold{35}}=1, N_{S}=N_{\overline{S}}=1, N_{\overline{AS}}=1, N_{F}=5, N_{\overline{F}}=4$

\item - $N_{\bold{35}}=1, N_{S}=N_{\overline{S}}=1, N_{F}=6, N_{\overline{F}}=8$

\item - $N_{\bold{35}}=1, N_{S}=1, N_{AS}=N_{\overline{AS}}=1, N_{\overline{F}}=13$

\item - $N_{\bold{35}}=1, N_{S}=1, N_{AS}=1, N_{F}=1, N_{\overline{F}}=17$

\item - $N_{\bold{35}}=1, N_{S}=1, N_{\overline{AS}}=4, N_{F}=1, N_{\overline{F}}=2$

\item - $N_{\bold{35}}=1, N_{S}=1, N_{\overline{AS}}=3, N_{F}=2, N_{\overline{F}}=6$

\item - $N_{\bold{35}}=1, N_{S}=1, N_{\overline{AS}}=2, N_{F}=3, N_{\overline{F}}=10$

\item - $N_{\bold{35}}=1, N_{S}=1, N_{\overline{AS}}=1, N_{F}=4, N_{\overline{F}}=14$

\item - $N_{\bold{35}}=1, N_{S}=1, N_{F}=5, N_{\overline{F}}=18$

\item - $N_{\bold{35}}=1, N_{\overline{S}}=1, N_{AS}=4, N_{\overline{F}}=3$

\item - $N_{\bold{35}}=1, N_{\overline{S}}=1, N_{AS}=3, N_{\overline{AS}}=1, N_{F}=3$

\item - $N_{\bold{35}}=1, N_{\overline{S}}=1, N_{AS}=3, N_F=N_{\overline{F}}=4$

\item - $N_{\bold{35}}=1, N_{\overline{S}}=1, N_{AS}=2, N_{\overline{AS}}=1, N_{F}=7, N_{\overline{F}}=1$

\item - $N_{\bold{35}}=1, N_{\overline{S}}=1, N_{AS}=2, N_F=8, N_{\overline{F}}=5$

\item - $N_{\bold{35}}=1, N_{\overline{S}}=1, N_{AS}=N_{\overline{AS}}=1, N_{F}=11, N_{\overline{F}}=2$

\item - $N_{\bold{35}}=1, N_{\overline{S}}=1, N_{AS}=1, N_F=12, N_{\overline{F}}=6$

\item - $N_{\bold{35}}=1, N_{\overline{S}}=1, N_{\overline{AS}}=1, N_{F}=15, N_{\overline{F}}=3$

\item - $N_{\bold{35}}=1, N_{\overline{S}}=1, N_{F}=16, N_{\overline{F}}=7$

\item - $N_{\bold{35}}=1, N_{AS}=x, N_{\overline{AS}}=y, N_{F}=15-4x-y, N_{\overline{F}}=17-x-4y, 4x+y\leq 15, 4y+x\leq 17$

\end{enumerate}

\subsection*{$SU(8)$}

\begin{enumerate}

\item - $N_{\bold{70}}=2, N_{F}=N_{\overline{F}}=4$

\item - $N_{\bold{56}}=2, N_{\overline{S}}=1, N_{AS}=1, N_{\overline{F}}=2$

\item - $N_{\bold{56}}=2, N_{\overline{S}}=1, N_{F}=5, N_{\overline{F}}=3$

\item - $N_{\bold{56}}=2, N_{\overline{AS}}=2, N_{F}=2, N_{\overline{F}}=4$

\item - $N_{\bold{56}}=2, N_{\overline{AS}}=1, N_{F}=3, N_{\overline{F}}=9$

\item - $N_{\bold{56}}=2, N_{F}=4, N_{\overline{F}}=14$

\item - $N_{\bold{56}}=N_{\overline{\bold{56}}}=1, N_{AS}=N_{\overline{AS}}=1, N_{F}=N_{\overline{F}}=3$

\item - $N_{\bold{56}}=N_{\overline{\bold{56}}}=1, N_{AS}=1, N_{F}=4, N_{\overline{F}}=8$

\item - $N_{\bold{56}}=N_{\overline{\bold{56}}}=1, N_{F}=N_{\overline{F}}=9$

\item - $N_{\bold{70}}=1, N_{\bold{56}}=1, N_{\overline{AS}}=1, N_{F}=3, N_{\overline{F}}=4$

\item - $N_{\bold{70}}=1, N_{\bold{56}}=1, N_{F}=4, N_{\overline{F}}=9$

\item - $N_{\bold{70}}=1, N_{Ad}=1, N_{AS}=N_{\overline{AS}}=1$

\item - $N_{\bold{70}}=1, N_{Ad}=1, N_{AS}=1, N_{F}=1, N_{\overline{F}}=5$

\item - $N_{\bold{70}}=1, N_{Ad}=1, N_{F}=N_{\overline{F}}=6$

\item - $N_{\bold{70}}=1, N_{S}=N_{\overline{S}}=1, N_{F}=N_{\overline{F}}=4$

\item - $N_{\bold{70}}=1, N_{S}=1, N_{\overline{AS}}=3$

\item - $N_{\bold{70}}=1, N_{S}=1, N_{\overline{AS}}=2, N_{F}=1, N_{\overline{F}}=5$

\item - $N_{\bold{70}}=1, N_{S}=1, N_{\overline{AS}}=1, N_{F}=2, N_{\overline{F}}=10$

\item - $N_{\bold{70}}=1, N_{S}=1, N_{F}=3, N_{\overline{F}}=15$

\item - $N_{\bold{70}}=1, N_{AS}=N_{\overline{AS}}=2, N_{F}=N_{\overline{F}}=2$

\item - $N_{\bold{70}}=1, N_{AS}=2, N_{\overline{AS}}=1, N_{F}=3, N_{\overline{F}}=7$

\item - $N_{\bold{70}}=1, N_{AS}=2, N_{F}=4, N_{\overline{F}}=12$

\item - $N_{\bold{70}}=1, N_{AS}=N_{\overline{AS}}=1, N_{F}=N_{\overline{F}}=8$

\item - $N_{\bold{70}}=1, N_{AS}=1, N_{F}=9, N_{\overline{F}}=13$

\item - $N_{\bold{70}}=1, N_{F}=N_{\overline{F}}=14$

\item - $N_{\bold{56}}=1, N_{Ad}=1, N_{\overline{S}}=1, N_{F}=7$

\item - $N_{\bold{56}}=1, N_{Ad}=1, N_{AS}=1, N_{F}=1, N_{\overline{F}}=10$

\item - $N_{\bold{56}}=1, N_{Ad}=1, N_{AS}=N_{\overline{AS}}=1, N_{\overline{F}}=5$

\item - $N_{\bold{56}}=1, N_{Ad}=1, N_{\overline{AS}}=2, N_{F}=4, N_{\overline{F}}=1$

\item - $N_{\bold{56}}=1, N_{Ad}=1, N_{\overline{AS}}=1, N_{F}=5, N_{\overline{F}}=6$

\item - $N_{\bold{56}}=1, N_{Ad}=1, N_{F}=6, N_{\overline{F}}=11$

\item - $N_{\bold{56}}=1, N_{S}=N_{\overline{S}}=1, N_{\overline{AS}}=1, N_{F}=3, N_{\overline{F}}=4$

\item - $N_{\bold{56}}=1, N_{S}=N_{\overline{S}}=1, N_{F}=4, N_{\overline{F}}=9$

\item - $N_{\bold{56}}=1, N_{S}=1, N_{\overline{AS}}=3, N_{\overline{F}}=5$

\item - $N_{\bold{56}}=1, N_{S}=1, N_{\overline{AS}}=2, N_{F}=1, N_{\overline{F}}=10$

\item - $N_{\bold{56}}=1, N_{S}=1, N_{\overline{AS}}=1, N_{F}=2, N_{\overline{F}}=15$

\item - $N_{\bold{56}}=1, N_{S}=1, N_{F}=3, N_{\overline{F}}=20$

\item - $N_{\bold{56}}=1, N_{\overline{S}}=1, N_{AS}=3, N_{\overline{F}}=5$

\item - $N_{\bold{56}}=1, N_{\overline{S}}=1, N_{AS}=2, N_{\overline{AS}}=1, N_{F}=4, N_{\overline{F}}=1$

\item - $N_{\bold{56}}=1, N_{\overline{S}}=1, N_{AS}=2, N_{F}=5, N_{\overline{F}}=6$

\item - $N_{\bold{56}}=1, N_{\overline{S}}=1, N_{AS}=1, N_{\overline{AS}}=1, N_{F}=9, N_{\overline{F}}=2$

\item - $N_{\bold{56}}=1, N_{\overline{S}}=1, N_{AS}=1, N_{F}=10, N_{\overline{F}}=7$

\item - $N_{\bold{56}}=1, N_{\overline{S}}=1, N_{\overline{AS}}=1, N_{F}=14, N_{\overline{F}}=3$

\item - $N_{\bold{56}}=1, N_{\overline{S}}=1, N_{F}=15, N_{\overline{F}}=8$

\item - $N_{\bold{56}}=1, N_{AS}=x, N_{\overline{AS}}=y, N_{F}=14-5x-y, N_{\overline{F}}=19-x-5y, 5x+y\leq 14, 5y+x\leq 19$

\end{enumerate}

\subsection*{$SU(9)$}

\begin{enumerate}

\item - $N_{\bold{126}}=1, N_{\overline{S}}=1, N_{F}=8$

\item - $N_{\bold{126}}=1, N_{AS}=N_{\overline{AS}}=1, N_{\overline{F}}=5$

\item - $N_{\bold{126}}=1, N_{AS}=1, N_{F}=1, N_{\overline{F}}=11$

\item - $N_{\bold{126}}=1, N_{\overline{AS}}=2, N_{F}=5$

\item - $N_{\bold{126}}=1, N_{\overline{AS}}=1, N_{F}=N_{\overline{F}}=6$

\item - $N_{\bold{126}}=1, N_{F}=7, N_{\overline{F}}=12$

\item - $N_{\bold{84}}=N_{\overline{\bold{84}}}=1, N_{AS}=1, N_{\overline{F}}=5$

\item - $N_{\bold{84}}=N_{\overline{\bold{84}}}=1, N_{F}=N_{\overline{F}}=6$

\item - $N_{\bold{84}}=1, N_{Ad}=1, N_{\overline{S}}=1, N_{F}=4$

\item - $N_{\bold{84}}=1, N_{Ad}=1, N_{\overline{AS}}=2, N_{F}=1$

\item - $N_{\bold{84}}=1, N_{Ad}=1, N_{\overline{AS}}=1, N_{F}=2, N_{\overline{F}}=6$

\item - $N_{\bold{84}}=1, N_{Ad}=1, N_{F}=3, N_{\overline{F}}=12$

\item - $N_{\bold{84}}=1, N_{S}=1, N_{\overline{F}}=22$

\item - $N_{\bold{84}}=1, N_{S}=N_{\overline{S}}=1, N_{F}=1, N_{\overline{F}}=10$

\item - $N_{\bold{84}}=1, N_{S}=N_{\overline{S}}=1, N_{\overline{AS}}=1, N_{\overline{F}}=4$

\item - $N_{\bold{84}}=1, N_{\overline{S}}=1, N_{AS}=2, N_{\overline{AS}}=1, N_{\overline{F}}=1$

\item - $N_{\bold{84}}=1, N_{\overline{S}}=1, N_{AS}=N_{\overline{AS}}=1, N_F=6, N_{\overline{F}}=2$

\item - $N_{\bold{84}}=1, N_{\overline{S}}=1, N_{AS}=2, N_{F}=1, N_{\overline{F}}=7$

\item - $N_{\bold{84}}=1, N_{\overline{S}}=1, N_{AS}=1, N_{F}=7, N_{\overline{F}}=8$

\item - $N_{\bold{84}}=1, N_{\overline{S}}=1, N_{\overline{AS}}=1, N_{F}=12, N_{\overline{F}}=3$

\item - $N_{\bold{84}}=1, N_{\overline{S}}=1, N_{F}=13, N_{\overline{F}}=9$

\item - $N_{\bold{84}}=1, N_{AS}=x, N_{\overline{AS}}=y, N_{F}=12-6x-y, N_{\overline{F}}=21-x-6y, 6x+y\leq 12, 6y+x\leq 21$

\end{enumerate}

\subsection*{$SU(10)$}

\begin{enumerate}

\item - $N_{\bold{120}}=N_{\overline{\bold{120}}}=1, N_{F}=N_{\overline{F}}=2$

\item - $N_{\bold{120}}=1, N_{\overline{S}}=1, N_{Ad}=1$

\item - $N_{\bold{120}}=1, N_{\overline{S}}=1, N_{AS}=N_{\overline{AS}}=1, N_{F}=N_{\overline{F}}=2$

\item - $N_{\bold{120}}=1, N_{\overline{S}}=1, N_{AS}=1, N_{F}=3, N_{\overline{F}}=9$

\item - $N_{\bold{120}}=1, N_{\overline{S}}=1, N_{\overline{AS}}=1, N_{F}=9, N_{\overline{F}}=3$

\item - $N_{\bold{120}}=1, N_{\overline{S}}=1, N_{F}=N_{\overline{F}}=10$

\item - $N_{\bold{120}}=1, N_{AS}=1, N_{\overline{AS}}=2, N_{\overline{F}}=8$

\item - $N_{\bold{120}}=1, N_{AS}=1, N_{\overline{AS}}=1, N_{F}=1, N_{\overline{F}}=15$

\item - $N_{\bold{120}}=1, N_{AS}=1, N_{F}=2, N_{\overline{F}}=22$

\item - $N_{\bold{120}}=1, N_{\overline{AS}}=3, N_{F}=6, N_{\overline{F}}=2$

\item - $N_{\bold{120}}=1, N_{\overline{AS}}=2, N_{F}=7, N_{\overline{F}}=9$

\item - $N_{\bold{120}}=1, N_{\overline{AS}}=1, N_{F}=8, N_{\overline{F}}=16$

\item - $N_{\bold{120}}=1, N_{F}=9, N_{\overline{F}}=23$

\end{enumerate}

\subsection*{$SU(11)$}

\begin{enumerate}

\item - $N_{\bold{165}}=1, N_{\overline{S}}=1, N_{\overline{AS}}=1, N_{F}=5, N_{\overline{F}}=3$

\item - $N_{\bold{165}}=1, N_{\overline{S}}=1, N_{F}=6, N_{\overline{F}}=11$

\item - $N_{\bold{165}}=1, N_{\overline{AS}}=3, N_{F}=2, N_{\overline{F}}=1$

\item - $N_{\bold{165}}=1, N_{\overline{AS}}=2, N_{F}=3, N_{\overline{F}}=9$

\item - $N_{\bold{165}}=1, N_{\overline{AS}}=1, N_{F}=4, N_{\overline{F}}=17$

\item - $N_{\bold{165}}=1, N_{F}=5, N_{\overline{F}}=25$

\end{enumerate}

\subsection*{$SU(12)$}

\begin{enumerate}

\item - $N_{\bold{220}}=1, N_{\overline{S}}=1, N_{\overline{AS}}=1, N_{\overline{F}}=3$

\item - $N_{\bold{220}}=1, N_{\overline{S}}=1, N_{F}=1, N_{\overline{F}}=12$

\item - $N_{\bold{220}}=1, N_{\overline{F}}=27$

\end{enumerate}

\subsection{Symplectic groups}\label{appendix:sp}

Here we summarize solutions when the gauge group is $USp(2N)$.

\subsubsection{Generic cases}\label{appendix:spgenerics}

Here we list the theories which we dubbed generic. We begin by listing the possible solutions to conditions (\ref{BetaZero}) for generic $N$:

\begin{enumerate}

\item - $N_{S}=2, \quad N_F=2N+2$

\item - $N_{S}=2, \quad N_{AS}=1, \quad N_F=4$

\item - $N_{S}=1, \quad N_{AS}=1, \quad N_F=2N+6$

\item - $N_{AS}=3, \quad N_F=12$

\item - $N_{AS}=2, \quad N_F=2N+10$ 

\item - $N_{AS}=1, \quad N_F=4N+8$

\item - $N_F=6N+6$  

\end{enumerate}

Besides these, there are several solutions that exist only for small $N$:

\begin{enumerate}

\item - $N_{S}=2, \quad N_{AS}=3, \quad N=2$

\item - $N_{S}=2, \quad N_{AS}=2, \quad N_F=2(3-N), \quad N=2,3$

\item - $N_{S}=1, \quad N_{AS}=5, \quad N_F=2, \quad N=2$

\item - $N_{S}=1, \quad N_{AS}=3, \quad N_F=2(5-N), \quad N=2,3,4,5$

\item - $N_{AS}=9, \quad N=2$

\item - $N_{AS}=8, \quad N_F=2, \quad N=2$

\item - $N_{AS}=7, \quad N_F=4, \quad N=2$

\item - $N_{AS}=6, \quad N_F=6(3-N), \quad N=2,3$

\item - $N_{AS}=5, \quad N_F=4(4-N), \quad N=2,3,4$

\item - $N_{AS}=4, \quad N_F=2(7-N), \quad N=2,3,4,5,6,7$

\end{enumerate}

\subsubsection{Exotic cases}\label{appendix:spexotics}

Here we list the theories with vanishing beta functions which we dubbed exotic.

\subsection*{$USp(4)$}

\begin{enumerate}

\item - $N_{\bold{14}}=1, \quad N_{AS}=2$

\item - $N_{\bold{14}}=1, \quad N_{AS}=1, \quad N_{F} = 2$

\item - $N_{\bold{14}}=1, \quad N_{F} = 4$

\item - $N_{\bold{16}}=1, \quad N_{AS}=3$

\item - $N_{\bold{16}}=1, \quad N_{AS}=2, \quad N_{F}=2$

\item - $N_{\bold{16}}=1, \quad N_{AS}=1, \quad N_{F}=4$

\item - $N_{\bold{16}}=1, \quad N_{F}=6$

\end{enumerate}

\subsection*{$USp(6)$}

\begin{enumerate}

\item - $N_{\bold{14'}}=1, \quad N_{S}=2, \quad N_{F}=3$

\item - $N_{\bold{14'}}=2, \quad N_{S}=1, \quad N_{AS}=1, \quad N_{F} = 2$

\item - $N_{\bold{14'}}=1, \quad N_{S}=1, \quad N_{AS}=1, \quad N_{F} = 7$

\item - $N_{\bold{14'}}=4, \quad N_{AS} = 1$

\item - $N_{\bold{14'}}=4, \quad N_{F} = 4$

\item - $N_{\bold{14'}}=3, \quad N_{AS}=2, \quad N_{F} = 1$

\item - $N_{\bold{14'}}=3, \quad N_{AS}=1, \quad N_{F} = 5$

\item - $N_{\bold{14'}}=3, \quad N_{F} = 9$

\item - $N_{\bold{14'}}=2, \quad N_{AS}=3, \quad N_{F} = 2$

\item - $N_{\bold{14'}}=2, \quad N_{AS}=2, \quad N_{F} = 6$

\item - $N_{\bold{14'}}=2, \quad N_{AS}=1, \quad N_{F} = 10$

\item - $N_{\bold{14'}}=2, \quad N_{F} = 14$

\item - $N_{\bold{14'}}=1, \quad N_{AS}=4, \quad N_{F} = 3$

\item - $N_{\bold{14'}}=1, \quad N_{AS}=3, \quad N_{F} = 7$

\item - $N_{\bold{14'}}=1, \quad N_{AS}=2, \quad N_{F} = 11$

\item - $N_{\bold{14'}}=1, \quad N_{AS}=1, \quad N_{F} = 15$

\item - $N_{\bold{14'}}=1, \quad N_{F} = 19$

\end{enumerate}

\subsection*{$USp(8)$}

\begin{enumerate}

\item - $N_{\bold{48}}=1, \quad N_{S}=1, \quad N_{AS}=1$

\item - $N_{\bold{42}}=1, \quad N_{S}=1, \quad N_{AS}=1$

\item - $N_{\bold{42}}=1, \quad N_{S}=1, \quad N_{F}=6$

\item - $N_{\bold{48}}=2, \quad N_{F}=2$

\item - $N_{\bold{48}}=1, \quad N_{\bold{42}}=1, \quad N_{F}=2$

\item - $N_{\bold{42}}=2, \quad N_{F}=2$

\item - $N_{\bold{48}}=1, \quad N_{AS}=2, \quad N_{F}=4$

\item - $N_{\bold{48}}=1, \quad N_{AS}=1, \quad N_{F}=10$

\item - $N_{\bold{48}}=1, \quad N_{F}=16$

\item - $N_{\bold{42}}=1, \quad N_{AS}=2, \quad N_{F}=4$

\item - $N_{\bold{42}}=1, \quad N_{AS}=1, \quad N_{F}=10$

\item - $N_{\bold{42}}=1, \quad N_{F}=16$

\end{enumerate}

\subsection*{$USp(10)$}

\begin{enumerate}

\item - $N_{\bold{110}}=1, \quad N_{AS}=1, \quad N_{F}=1$

\item - $N_{\bold{110}}=1, \quad N_{F}=9$

\end{enumerate}

\subsection{Orthogonal groups}\label{appendix:so}

Here we summarize solutions when the gauge group is $SO(N)$.

\subsubsection{Generic cases}\label{appendix:sogenerics}

Here we list the possible generic solutions for equation (\ref{BetaZero}).

\begin{enumerate}

\item - $N_{S}=2, \quad N_V=N-10$

\item - $N_{S}=1, \quad N_{AS}=1, \quad N_V=N-6$

\item - $N_{S}=1, \quad N_V=2N-8$

\item - $N_{AS}=2, \quad N_V=N-2$

\item - $N_V=3N-6$  

\end{enumerate}

\subsubsection{Exotic cases}\label{appendix:soexotics}

Here we list the possible exotic solutions for equation (\ref{BetaZero}). Here we write cases up to the outer automorphism of the selected $SO$ group. 

\subsection*{$SO(7)$}

\begin{enumerate}

\item - $N_{\bold{8}}=M, \quad N_{V}=15-M$

\item - $N_{AS}=1, \quad N_{\bold{8}}=M, \quad N_{V} = 10-M, \quad \text{$M$ odd}$

\item - $N_{AS}=2, \quad N_{\bold{8}}=M, \quad N_{V} = 5-M$

\item - $N_{S}=1, \quad N_{AS}=1, \quad N_{\bold{8}}=1$

\item - $N_{S}=1, \quad N_{\bold{8}}=M, \quad N_{V} = 6-M$

\item - $N_{AS}=1, \quad N_{\bold{35}}=1$

\item - $N_{\bold{35}}=1, \quad N_{\bold{8}}=M, \quad N_{V} = 5-M$

\item - $N_{\bold{48}}=1, \quad N_{\bold{8}}=1$

\item - $N_{\bold{48}}=1, \quad N_{V} = 1$

\end{enumerate}

\subsection*{$SO(8)$}

\begin{enumerate}

\item - $N_{\bold{8}_S}=M, \quad N_{\bold{8}_C}=L, \quad N_{V}=18-M-L , \quad M\geq L\geq 18-M-L$

\item - $N_{AS}=1, \quad N_{\bold{8}_S}=M, \quad N_{\bold{8}_C}=L, \quad N_{V}=12-M-L , \quad M\geq L\geq 12-M-L, \text{$M$ and $L$ not both even}$

\item - $N_{AS}=2, \quad N_{\bold{8}_S}=M, \quad N_{\bold{8}_C}=L, \quad N_{V}=6-M-L , \quad M\geq L\geq 6-M-L$

\item - $N_{S}=1, \quad N_{AS}=1, \quad N_{\bold{8}_S}=2$

\item - $N_{S}=1, \quad N_{AS}=1, \quad N_{\bold{8}_S}=1, \quad N_{V}=1$

\item - $N_{S}=1, \quad N_{AS}=1, \quad N_{\bold{8}_S}=1, \quad N_{\bold{8}_C}=1$

\item - $N_{S}=1, \quad N_{\bold{8}_S}=M, \quad N_{\bold{8}_C}=L, \quad N_{V}=8-M-L, \quad M\geq L$

\item - $N_{\bold{56}_v}=1, \quad N_{V}=3$

\item - $N_{\bold{56}_v}=1, \quad N_{V}=2, \quad N_{\bold{8}_S}=1$

\item - $N_{\bold{56}_v}=1, \quad N_{V}=1, \quad N_{\bold{8}_S}=2$

\item - $N_{\bold{56}_v}=1, \quad N_{V}=1, \quad N_{\bold{8}_S}=1, \quad N_{\bold{8}_C}=1$

\item - $N_{\bold{56}_v}=1, \quad N_{\bold{8}_S}=3$

\item - $N_{\bold{56}_v}=1, \quad N_{\bold{8}_S}=2, \quad N_{\bold{8}_C}=1$

\end{enumerate}

\subsection*{$SO(9)$}

\begin{enumerate}

\item - $N_{\bold{16}}=M, \quad N_{V}=21-2M$

\item - $N_{AS}=1, \quad N_{\bold{16}}=M, \quad N_{V}=14-2M, \quad \text{$M$ odd}$

\item - $N_{AS}=2, \quad N_{\bold{16}}=M, \quad N_{V}=7-2M$

\item - $N_{S}=1, \quad N_{AS}=1, \quad N_{\bold{16}}=1, \quad N_{V}=1$

\item - $N_{S}=1, \quad N_{\bold{16}}=M, \quad N_{V}=10-2M$

\item - $N_{\bold{84}}=1$

\end{enumerate}

\subsection*{$SO(10)$}

\begin{enumerate}

\item - $N_{\bold{16}}=M, \quad N_{\bold{\overline{16}}}=L, \quad N_{V}=24-2M-2L, M\geq L$

\item - $N_{AS}=1, \quad N_{\bold{16}}=M, \quad N_{\bold{\overline{16}}}=L, \quad N_{V}=16-2M-2L, M>L$

\item - $N_{AS}=2, \quad N_{\bold{16}}=M, \quad N_{\bold{\overline{16}}}=L, \quad N_{V}=8-2M-2L, M\geq L$

\item - $N_{S}=1, \quad N_{AS}=1, \quad N_{\bold{16}}=2$

\item - $N_{S}=1, \quad N_{AS}=1, \quad N_{\bold{16}}=1, \quad N_{\bold{\overline{16}}}=1$

\item - $N_{S}=1, \quad N_{AS}=1, \quad N_{\bold{16}}=1, \quad N_{V}=2$

\item - $N_{S}=1, \quad N_{\bold{16}}=M, \quad N_{\bold{\overline{16}}}=L, \quad N_{V}=12-2M-2L, M\geq L$

\end{enumerate}

\subsection*{$SO(11)$}

\begin{enumerate}

\item - $N_{\bold{32}}=M, \quad N_{V}=27-4M$

\item - $N_{AS}=2, \quad N_{\bold{32}}=2, \quad N_{V}=1$

\item - $N_{AS}=2, \quad N_{\bold{32}}=1, \quad N_{V}=5$

\item - $N_{S}=1, \quad N_{AS}=1, \quad N_{\bold{32}}=1, \quad N_{V}=1$

\item - $N_{S}=1, \quad N_{\bold{32}}=M, \quad N_{V}=14-4M$

\end{enumerate}

\subsection*{$SO(12)$}

\begin{enumerate}

\item - $N_{\bold{32}}=M, \quad N_{\bold{32'}}=L, \quad N_{V}=30-4M-4L, \quad M\geq L$

\item - $N_{AS}=2, \quad N_{\bold{32}}=2, \quad N_{V}=2$

\item - $N_{AS}=2, \quad N_{\bold{32}}=1, \quad N_{\bold{32'}}=1, \quad N_{V}=2$

\item - $N_{AS}=2, \quad N_{\bold{32}}=1, \quad N_{V}=6$

\item - $N_{S}=1, \quad N_{AS}=1, \quad N_{\bold{32}}=1, \quad N_{V}=2$

\item - $N_{S}=1, \quad N_{\bold{32}}=M, \quad N_{\bold{32'}}=L, \quad N_{V}=16-4M-4L, \quad M\geq L$

\end{enumerate}

\subsection*{$SO(13)$}

\begin{enumerate}

\item - $N_{\bold{64}}=M, \quad N_{V}=33-8M$

\item - $N_{AS}=2, \quad N_{\bold{64}}=1, \quad N_{V}=3$

\item - $N_{S}=1, \quad N_{\bold{64}}=2, \quad N_{V}=2$ 

\item - $N_{S}=1, \quad N_{\bold{64}}=1, \quad N_{V}=10$ 

\end{enumerate}

\subsection*{$SO(14)$}

\begin{enumerate}

\item - $N_{\bold{64}}=M, \quad N_{\bold{\overline{64}}}=L, \quad N_{V}=36-8M-8L, M\geq L$

\item - $N_{AS}=1, \quad N_{\bold{64}}=M, \quad N_{\bold{\overline{64}}}=L, \quad N_{V}=24-8M-8L, M>L$

\item - $N_{AS}=2, \quad N_{\bold{64}}=1, \quad N_{V}=4$

\item - $N_{S}=1, \quad N_{AS}=1, \quad N_{\bold{64}}=1$

\item - $N_{S}=1, \quad N_{\bold{64}}=2, \quad N_{V}=4$

\item - $N_{S}=1, \quad N_{\bold{64}}=1, \quad N_{\bold{\overline{64}}}=1, \quad N_{V}=4$

\item - $N_{S}=1, \quad N_{\bold{64}}=1,  \quad N_{V}=12$

\end{enumerate}

\subsection*{$SO(15)$}

\begin{enumerate}

\item - $N_{\bold{128}}=2, \quad N_{V}=7$

\item - $N_{\bold{128}}=1, \quad N_{V}=23$

\item - $N_{AS}=1, \quad N_{\bold{128}}=1, \quad N_{V}=10$

\item - $N_{S}=1, \quad N_{\bold{128}}=1, \quad N_{V}=6$

\end{enumerate}

\subsection*{$SO(16)$}

\begin{enumerate}

\item - $N_{\bold{128}}=2, \quad N_{V}=10$

\item - $N_{\bold{128}}=1, \quad N_{\bold{128'}}=1, \quad N_{V}=10$

\item - $N_{\bold{128}}=1, \quad N_{V}=26$

\item - $N_{AS}=1, \quad N_{\bold{128}}=1, \quad N_{V}=12$

\item - $N_{S}=1, \quad N_{\bold{128}}=1, \quad N_{V}=8$

\end{enumerate}

\subsection*{$SO(17)$}

\begin{enumerate}

\item - $N_{\bold{256}}=1, \quad N_{V}=13$

\end{enumerate}

\subsection*{$SO(18)$}

\begin{enumerate}

\item - $N_{\bold{256}}=1, \quad N_{V}=16$

\item - $N_{AS}=1, \quad N_{\bold{256}}=1$

\end{enumerate}

\subsection{Exceptional groups}\label{appendix:exc}

Here we summarize the possible solutions for equation (\ref{BetaZero}) when the gauge group is one of the exceptional groups.

\subsection*{$G_2$}

\begin{enumerate}

\item - $N_{\bold{14}}=2, \quad N_{\bold{7}}=4$

\item - $N_{\bold{7}}=12$

\item - $N_{\bold{27}}=1, \quad N_{\bold{7}}=3$

\end{enumerate}

\subsection*{$F_4$}

\begin{enumerate}

\item - $N_{\bold{52}}=2, \quad N_{\bold{26}}=3$

\item - $N_{\bold{26}}=9$

\end{enumerate}

\subsection*{$E_6$}

\begin{enumerate}

\item - $N_{\bold{78}}=2, \quad N_{\bold{27}}=N_{\overline{\bold{27}}}=2$

\item - $N_{\bold{78}}=2, \quad N_{\bold{27}}=3 , \quad N_{\overline{\bold{27}}}=1$

\item - $N_{\bold{78}}=2, \quad N_{\bold{27}}=4$

\item - $N_{\bold{78}}=1, \quad N_{\bold{27}}=5 , \quad N_{\overline{\bold{27}}}=3$

\item - $N_{\bold{78}}=1, \quad N_{\bold{27}}=6 , \quad N_{\overline{\bold{27}}}=2$

\item - $N_{\bold{78}}=1, \quad N_{\bold{27}}=7 , \quad N_{\overline{\bold{27}}}=1$

\item - $N_{\bold{78}}=1, \quad N_{\bold{27}}=8$

\item - $N_{\bold{27}}=N_{\overline{\bold{27}}}=6$

\item - $N_{\bold{27}}=7, \quad N_{\overline{\bold{27}}}=5$

\item - $N_{\bold{27}}=8, \quad N_{\overline{\bold{27}}}=4$

\item - $N_{\bold{27}}=9, \quad N_{\overline{\bold{27}}}=3$

\item - $N_{\bold{27}}=10, \quad N_{\overline{\bold{27}}}=2$

\item - $N_{\bold{27}}=11, \quad N_{\overline{\bold{27}}}=1$

\item - $N_{\bold{27}}=12$

\end{enumerate}

\subsection*{$E_7$}

\begin{enumerate}

\item - $N_{\bold{133}}=2, \quad N_{\bold{56}}=3$

\item - $N_{\bold{56}}=9$

\end{enumerate}   

\bibliographystyle{ytphys}
\bibliography{refs}

\end{document}